\documentclass[a4paper,11pt]{article}
\pdfoutput=1 % if your are submitting a pdflatex (i.e. if you have
\usepackage{jcappub} % for details on the use of the package, please
\usepackage[T1]{fontenc} % if needed
\usepackage{appendix}
\usepackage{bm}
\usepackage{makecell}
\usepackage{tablefootnote}

\newcommand\footnoteref[1]{\protected@xdef\@thefnmark{\ref{#1}}\@footnotemark}

\title{\boldmath Introducing \textsc{saguaro} - Simulating IGM Evolution and Environments At High Resolution: Setup and First Results}

\author[a]{Christopher Cain\note{Corresponding author.}, }
\author[a]{Aloha Das, }
\author[b]{Anson D'Aloisio, }
\author[c]{Simon Foreman, }
\author[a]{Evan Scannapieco, }
\author[a]{Esteban Moreno,}
\author[b]{Matthew Lugatiman,}
\author[a]{Joshua Cohon,}
\author[d]{Hurum Maksora Tohfa,}
\author[e,f]{and Hy Trac}

\affiliation[a]{School of Earth and Space Exploration, Arizona State University, Tempe, AZ 85281, USA}
\affiliation[b]{Department of Physics and Astronomy, University of California, Riverside, CA 92521, USA}
\affiliation[c]{Department of Physics, Arizona State University, Tempe, AZ 85287, USA}
\affiliation[d]{Department of Astronomy, University of Washington, Seattle, WA 98195-1580, USA}
\affiliation[e]{McWilliams Center for Cosmology and Astrophysics, Department of Physics, Carnegie Mellon University, Pittsburgh, PA 15213, USA}
\affiliation[f]{NSF AI Planning Institute for Physics of the Future, Carnegie Mellon University, Pittsburgh, PA 15213, USA}

\emailAdd{clcain3@asu.edu}
\emailAdd{adas106@asu.edu}
\emailAdd{ansond@ucr.edu}
\emailAdd{simon.Foreman@asu.edu}
\emailAdd{evan.scannapieco@asu.edu}
\emailAdd{emoren47@asu.edu}
\emailAdd{matthew.lugatiman@email.ucr.edu}
\emailAdd{jacohon@asu.edu}
\emailAdd{htohfa@uw.edu}
\emailAdd{hytrac@andrew.cmu.edu}

\abstract{ 
    Small-scale physics in the intergalactic medium (IGM) plays a crucial role in shaping the progress of cosmic reionization and several high-redshift observables that probe this period. Several recent studies have characterized the complex, dynamical response of the IGM to reionization down to kilo-parsec scales, including its effect on observables such as the Ly$\alpha$ forest. However, there has been no concentrated attempt to simulate and characterize these effects across the full parameter space of realistic large-scale IGM environments during reionization. To meet this need, we introduce the \textsc{saguaro} simulation suite, sub-titled ``Simulating IGM Evolution and Environments At High Resolution''. \textsc{saguaro} is a suite of over two hundred high-resolution, coupled radiative-hydrodynamics simulations of IGM gas dynamics during and after reionization. The suite spans a grid of photoionization rates, redshifts of reionization, and box-scale densities. We also simulate other physical effects, such as X-ray pre-heating, recombination radiation, baryon–dark matter free-streaming, and alternative dark matter cosmologies. Our suite includes box sizes of $2$ and $0.25$ $h^{-1}$Mpc, extending to volumes large enough to begin capturing halos above the atomic cooling limit and resolutions high enough to fully resolve the IGM Jeans scale in the cold, neutral universe.  We present a detailed description of the setup and first results from \textsc{saguaro}, descriptions of the IGM gas dynamics and thermal structure, opacity, self-shielding properties, the effect of the IGM on the reionization photon budget, and the halo mass function, and Ly$\alpha$ transmission properties.  \textsc{saguaro} will help facilitate detailed studies of small-scale IGM structure and its effects that will help inform the next generation of reionization simulations and data interpretation.  
}

\begin{document}
\maketitle
\flushbottom

\section{Introduction}
\label{sec:intro}

The Epoch of Reionization (EoR) is a crucial period in the history of the universe, during which ionizing photons from the first galaxies and active galactic nuclei (AGN) ionized most of the neutral hydrogen in the intergalactic medium (IGM).  The EoR is directly probed by several observables.  These include the Ly$\alpha$ forest of high-redshift quasars~\citep{Becker2015,Kulkarni2019,Keating2020,Nasir2020,Bosman2021,Zhu2021,Jin2023,Qin2024,Becker2024,Zhu2024,Becker2024}, Ly$\alpha$ damping wings in quasar~\citep{Davies2018,Wang2020,Durovcikova2024,Hennawi2025}, and galaxy~\citep{Mason2018,Umeda2023,Bunker2023,Tang2024b} spectra, the statistics and ionizing properties of high-redshift galaxies~\citep{Adams2024,Donnan2024,Finkelstein2024,Simmonds2024,Jaskot2024a,Jaskot2024b,Papovich2025}, the CMB~\citep{Planck2018,Reichardt2020,deBelsunce2021,Beringue2025}, and forthcoming observations of the average cosmic 21 cm signal and its spatial fluctuations~\citep{Bowman2018,Singh2018a,HERA2021a,HERA2021b,Berkhout2024,Nunhokee2025}.  These observables are sensitive to the global reionization history and its spatial morphology, which in turn depend on the sources that drove it.  Mapping the properties of these first sources to EoR observables (a chief goal of the field) requires a complete theoretical picture of how galaxy and IGM physics affects the EoR.  

Modeling the EoR is challenging in large part because of its multi-scale nature.  The spatial morphology of the EoR is set mainly by the clustering of ionizing sources on tens to hundreds of megaparsec scales.  The opacity of the IGM to ionizing photons is set by the clumpiness and dynamics of IGM gas down to scales as small as a kilo-parsec~\citep{Gnedin2000}.  The production and escape of these photons from galaxies is set by processes on even smaller scales, down to the sub-parsec scales at play in the circumgalactic interstellar media of galaxies, which cannot be fully resolved even by state-of-the-art reionization simulations~\citep[e.g.][]{Gnedin2014,Ocvirk2018,Rosdahl2018,Kannan2022}.  This ``dynamic range problem'' is worsened by two factors that make the EoR an even more computationally daunting challenge. First, modeling the topology of reionization requires solving the radiative transfer equation in large volumes that include millions of ionizing sources~\citep{Abel2002,Trac2007,Gnedin2014}. Second, the relevant parameter space is broad, reflecting the poorly constrained properties of reionization sources~\citep{Qin2024} and the structure of the IGM~\citep{Choudhury2025}.  Adequately searching this parameter space requires thousands to millions of simulations.  

The focus of this work is the middle tier of the hierarchy of scales described above - namely, the kpc-scale physics that sets the recombination rate and ionizing photon opacity in the ionized IGM.  The dynamics of mini-halos and filaments in dark matter structures below the atomic cooling limit ($\lesssim 10^8$ $h^{-1}M_{\odot}$) is often unresolved in simulations, but can still have important consequences for the reionization process.  These include setting the mean free path (MFP) to ionizing photons~\citep{Prochaska2009,Worseck2014,Becker2021,Cain2021,Zhu2023,Gaikwad2023}, shaping the large-scale spatial morphology of ionized and neutral regions~\citep{McQuinn2007,Mao2019,Bianco2021,Cain2022b}, and determining the number of ionizing photons that sources must produce to complete reionization~\citep[][]{Madau1999,Pawlik2009,Davies2021b,Munoz2024,Davies2024c}.  Structure at these scales can also affect the accuracy of forward-modeling EoR observables, including the Ly$\alpha$ forest~\citep{Doughty2023} and damping wing absorption towards high-redshift Ly$\alpha$ emitting galaxies and quasars~\citep{Park2021b}.  

Studying the dynamics and effects of IGM small-scale structure has been the goal of several studies over the past two decades.  The first detailed numerical simulations along these lines characterized the photo-evaporation of mini-halos by cosmological ionization fronts (I-fronts) under idealized conditions~\citep{Iliev2003,Shapiro2004,Iliev2005}. These results were then combined with analytic estimates to develop a subgrid model to account for the impact of mini-halos on the overall process of reionization \citep{Ciardi2006,Yue2009}.
However, due to computational limitations, these works could not simulate this process fully self-consistently within its larger cosmological context.  
As computational capabilities improved, subsequent works began to model small-scale structure directly in cosmological simulations.  Ref.~\cite{McQuinn2011} exposed small-volume, high-resolution hydrodynamical simulations to ionizing radiation to study the HI column density distribution in the limit that the IGM has responded dynamically to reionization heating.  Ref.~\cite{Emberson2013} performed a similar study, but focused on the limit in which reionization has occurred very recently and the IGM has not yet dynamically responded to heating.  These works reached somewhat different conclusions about the importance of small-scale structure, suggesting that the interplay between radiative transfer (RT) and hydrodynamics is important.  

The interplay between reionization heating and small-scale structure was studied by Ref.~\cite{Park2016} using an approximation to RT (see also Ref.~\cite{Hirata2018}).  They found that small-scale structure is most abundant in regions of the IGM that have been re-ionized very recently, and have not had time to respond dynamically to photo-heating.  However, over a time-scale of a few hundred million years, most structures at mass scales $\lesssim 10^8$ $M_{\odot}$ are destroyed as the gas responds to pressure disequilibrium after ionization, termed ``pressure smoothing''.  Ref.~\cite{DAloisio2020} studied these effects in greater detail using a self-consistent RT+Eulerian hydrodynamics treatment in larger volumes, and explored their implications for reionization. Ref.~\cite{Nasir2021} further studied the column-density distribution and dynamics of self-shielding structures using the same simulation suite.  Most recently, Ref.~\cite{Chan2023} ran complementary simulations using an RT+Lagrangian hydrodynamics code and conducted a focused study of mini-halo evaporation in a cosmological context.  

The aforementioned works focused on relatively small cosmological boxes ($
\lesssim 2$ Mpc on a side) that could directly resolve the processes under study.  Unfortunately, directly resolving small-scale IGM physics in computational volumes that capture the large-scale distribution of ionized and neutral regions ($10$s to $100$s of Mpc) is very computationally challenging.  Simulations that come close~\citep{Gnedin2014,Kaurov2015,Rahmati2017,Ocvirk2018,Rosdahl2018,Kannan2022,Puchwein2023} are sufficiently expensive (costing in the millions of CPU hours) to preclude extensive studies of the reionization parameter space.  On the other hand, semi-numerical approaches that include subgrid treatments for IGM physics~\citep{Sobacchi2014,Davies2021,Qin2024,Choudhury2025} suffer from large parameter spaces and/or simplifications that limit their accuracy.  

An alternative to these approaches is to run a set of radiative-hydrodynamic simulations in small volumes that allow the relevant physics to be directly resolved, and vary a set of ``box-scale'' parameters that dictate the large-scale environment in which the gas resides (such as over-density, ionizing background strength, and the local timing of reionization).  This approach allows the small-scale physics to be self-consistently characterized across a diversity of large-scale IGM environments.  One can then use these results to integrate over distributions that describe these large-scale fluctuations, either analytically or in the form of a sub-grid model in larger-volume simulations.  This approach has been successfully employed by many of the aforementioned studies~\citep{Emberson2013,Park2016,Hirata2018,DAloisio2020,Nasir2021,Cain2022a,Theuns2023}, and recently was used by Refs.~\cite{Cain2021,Cain2022b} to calibrate a subgrid opacity model for use in large-scale simulations.  However, most of these studies have been confined to a relatively small number of simulations ($\sim 10-25$) and, thus, have been limited in their ability to search the parameter space of IGM environments and physical parameters (see Ref.~\cite{Gnedin2024} for further discussion).  

This work presents \textsc{saguaro} (Simulating IGM Evolution and Environments At High Resolution), a new suite of small-scale IGM simulations run using an upgraded version of the setup employed in Ref.~\cite{DAloisio2020}.  \textsc{saguaro} is designed to comprehensively survey the parameter space of large-scale IGM environments.  These simulations capture the coupling of IGM gas dynamics and RT in the wake of ionization fronts (I-fronts), and track these dynamics until $z = 4$, well after reionization has ended~\citep{Bosman2021}.  Our suite includes two box sizes, such that its dynamic range spans nearly $4$ orders of magnitude in spatial scale.  The most important parameters that we focus on are the local redshift of reionization, the ionizing background strength, and the matter over-density at the box scale.  Our suite also includes variations of other potentially important physical parameters, including X-Ray pre-heating of the neutral IGM~\citep{Furlanetto2006,Fialkov2014c}, dark matter free-streaming on small scales~\citep{Cain2022a,Davies2023}, the baryon-dark matter streaming velocity~\citep{Tseliakhovich2010}, and the effects of recombination radiation~\citep{FaucherGiguere2009}. 

The goal of this work is to describe the numerical setup and scope of \textsc{saguaro}, and to explore a number of interesting first results that will serve to motivate further studies.  
This work is organized as follows.  In \S\ref{sec:methods}, we describe the code and simulation setup (\S\ref{subsec:code}-\S\ref{subsec:changes}), the physical parameters covered by the suite (\S\ref{subsec:additional_parameters}-\ref{subsec:box_params}), and the data products output by the simulations (\S\ref{subsec:otf}).  \S\ref{sec:vis_IGM} visualizes and characterizes the gas dynamics in the suite, including the density and thermal structure of the IGM.  \S\ref{sec:opacity} presents a preliminary study of the self-shielding properties of the gas, the HI column density distribution, and the evolution of the IGM opacity in \textsc{saguaro}.  We investigate the implications of \textsc{saguaro} results for reionization and the ionizing photon budget in \S\ref{sec:global_reion}, study the mass function of dark matter halos and comment on prospects for investigating their HI content in \S\ref{sec:dm_halos}, and describe Ly$\alpha$ transmission properties in \S\ref{sec:Lya}.  We summarize our findings and conclude in \S\ref{sec:conc}.  Throughout, we assume the following cosmological parameters: $\Omega_m = 0.305$, $\Omega_{\Lambda} = 1 - \Omega_m$, $\Omega_b = 0.048$, $h = 0.68$, $n_s = 0.9667$ and $\sigma_8 = 0.82$, consistent with Ref.~\cite{Planck2018} results. Distances are in co-moving units unless otherwise specified. 

\section{Numerical Methods}
\label{sec:methods}

\subsection{The Code \& Basic Setup}
\label{subsec:code}

\textsc{saguaro} was run using a modified version of the RadHydro code of~\citep{Trac2004,Trac2007}.  RadHydro uses ray tracing to model the radiation field and tracks RT and hydrodynamics in a fully coupled fashion on a uniform Eulerian grid.  In our setup, the RT and hydro grids have the same number of cells, $N_{\rm gas} = N_{\rm RT} = 1024^3$.  We use two box sizes -- $L_{\rm box} = 2$ $h^{-1}$Mpc and $0.25$ $h^{-1}$Mpc -- for spatial resolutions of $\Delta x_{\rm cell} = 1.95$ $h^{-1}$kpc and $244$ $h^{-1}$pc, respectively.  Dark matter (DM) is modeled using a particle-mesh gravity scheme with $N_{\rm DM} = 1024^3$ particles.  Gravitational forces are calculated by smoothing the DM particles onto a uniform mesh using a triangular-shaped-cloud (TSC) kernel, and solving Poisson's equation for the gravitational potential, which is then used to update the DM and gas velocities.  All simulations are initialized at $z = 300$ (except those with stream velocities, which are initialized at $z = 1080$) with separate baryon and DM transfer functions computed with CAMB~\citep{Lewis2000}.  They are run down to $z = 4$, and all outputs are saved down to that redshift (see \S\ref{subsec:otf}).  The setup is similar to that of Ref.~\cite{DAloisio2020}, with several important upgrades that are outlined here and in \S\ref{subsec:changes}.  

In \textsc{saguaro}, there is no explicit treatment of star formation or other processes internal to galaxies.  Instead, our focus is on modeling the dynamics of IGM gas in response to reionization.  The simulations evolve without radiation (gravity and hydrodynamics only) until the reionization redshift, $z_{\rm re}$.  At this time, plane-parallel rays are cast from the boundaries of $N_{\rm dom}^3$ regular cubical sub-domains.  For most of our runs\footnote{The only exception is for our runs with the lowest value of the ionizing background, for which we use $N_{\rm dom} = 32$.  This is because the mean free path must be much larger than $L_{\rm dom}$ to avoid unwanted attenuation of $\Gamma_{\rm HI}$ across the domain length.  Since the mean free path is shortest in our runs with low $\Gamma_{-12}$, we increase $N_{\rm dom}$ to mitigate such attenuation. } in $2$ $h^{-1}$Mpc boxes, we set $N_{\rm dom} = 16$ for a domain length of $L_{\rm dom} = 125$ $h^{-1}$kpc, and for our $0.25$ $h^{-1}$Mpc boxes, we set $N_{\rm dom} = 4$ ($L_{\rm dom} = 62.5$ $h^{-1}$kpc).  In this setup, we treat the radiation field everywhere as if it were produced by ionizing sources external to the RT domain.  This allows us to leave the detailed internal structure of galaxies unresolved and focus instead on how the radiation emitted by galaxies interacts with the IGM.  Unlike in Ref.~\cite{DAloisio2020}, in which plane parallel rays were sent from only two faces of each domain, we send rays from all $N_{\rm face} = 6$ faces.\footnote{The plane-parallel RT causes in sharp, persistent shadows behind self-shielding mini-halos if rays are sent along only one or two directions, which can result in some low-density gas remaining artificially neutral (see Ref.~\cite{Nasir2021}).  Although some shadowing effects are expected in the real IGM, they should generally be small because of the effects of recombination radiation (which we do not model here) and the fact that regions that have been ionized for a while should be ``visible'' to ionizing sources from many directions.  }  The RT time step ($\Delta t_{\rm RT}$) sub-cycles under the hydrodynamical one ($\Delta t_{\rm hydro}$), whenever it is smaller, to save computation time.  

The spectrum of ionizing radiation is modeled as a power law in frequency of the form $J_{\nu} \propto \nu^{-\alpha}$, with $\alpha = 1.5$ being our fiducial value, typical of commonly-assumed stellar population synthesis models (e.g. see \S 3.1 of \cite{DAloisio2019}, and Refs. \cite{Bressan2012,Choi2017}).  We note that our results are reasonably insensitive to this choice, since IGM photo-heating (within I-fronts and in optically thin gas) is only weakly sensitive to the spectral index~\citep{McQuinn2016, DAloisio2019}.  However, we include several runs with varying values of $\alpha$ to quantify this dependence.  We discretize the radiation into five frequency bins spaced between $1$ and $4$ Ryd, spanning the range of frequencies that ionized HI but not HeII.  The ionizing flux at the source planes is set to produce an approximately spatially constant HI photo-ionization rate, $\Gamma_{\rm HI}$, in highly ionized (optically thin) gas, parameterized by $\Gamma_{-12} \equiv \Gamma_{\rm HI}/[10^{-12}{\rm s}^{-1}]$.  Gas dense enough to self-shield will naturally have $\Gamma_{\rm HI} < \Gamma_{-12}$ thanks to our ray-tracing RT treatment.  We also model collisional ionization of hydrogen and all ionization states of Helium, which become important in gas with high IGM densities ($\Delta \gg 100$) and/or temperatures on the high end of typical values in photo-ionized IGM gas ($T \gtrsim 3 \times 10^4 $K).  

As in Ref.~\cite{DAloisio2020}, we account for density fluctuations above the box scale using the ``DC mode'' formalism of Ref.~\cite{Gnedin2011} (see also the closely-related ``separate universe'' formalism of Ref.~\cite{Wagner2015}).  The density contrast at the box scale is parameterized in terms of the {\it linear} over-density in units of its standard deviation, $\delta/\sigma$, which is independent of redshift.  The non-linear density contrast at the box scale, $\delta_{\rm NL}$, evolves according to spherical collapse in a matter-dominated universe.  At the $2$ $h^{-1}$Mpc box size, we have considered $5$ values of box-scale density - $\delta/\sigma = 0$ (cosmic mean density), $\pm 1$, and $\pm \sqrt{3}$.  For the $L_{\rm box} = 2$ $h^{-1}$Mpc run with $\delta/\sigma = +\sqrt{3}$, $\delta_{\rm NL}$ reaches ``turnaround'' at $z \approx 4$.  At this point, a structure begins to form approaching the mass scale of the box, and the hydrodynamical time step becomes prohibitively small.  As such, we choose to terminate our simulations at $z = 4$.  Over-dense $L_{\rm box} = 0.25$ $h^{-1}$Mpc runs would pass turnaround well before the end of reionization, and several times more computationally expensive than their $L_{\rm box} = 2$ $h^{-1}$Mpc counterparts.  As such, we only run mean-density simulations at this box size.  

Our highest and lowest values of $\delta/\sigma$, $\pm \sqrt{3}$, are chosen strategically to allow us to estimate the averages of various quantities across the entire universe.  Specifically, the average of some quantity $X$ can be written as the average of that quantity over the PDF of $\delta$, that is, $\langle X \rangle = \int X(\delta) P(\delta) d\delta$.  Since $P(\delta)$ is Gaussian, we can take advantage of an integration trick called Gauss-Hermite quadrature, which allows the average of any polynomial of order $5$ or less to be expressed as a sum over the three roots of the Hermite polynomial $H_3(x)$, which in this case are given by $\delta/\sigma = 0$ and $\pm \sqrt{3}$.  Appendix B of Ref.~\cite{DAloisio2020} provides a complete derivation and more details about this averaging procedure.  
We test the accuracy of this averaging procedure in Appendix~\ref{app:DCmode} by comparing the results of low-resolution $2$ $h^{-1}$Mpc boxes to an $8$ $h^{-1}$Mpc box, and find that the average over DC modes is in reasonable agreement with the larger volume.  However, some quantities can differ by up to $20-30\%$ at the lowest redshifts ($z \lesssim 5$), motivating study of improved approaches to estimating IGM-averaged quantities from small-volume simulations in the future.  

 \subsection{Key upgrades to the Ref.~\citep{DAloisio2020} setup}
\label{subsec:changes}

\subsubsection{I-front passage}
\label{subsubsec:IF}

The first important difference between our setup and that of Ref.~\cite{DAloisio2020} is how we treat the passage of I-fronts across RT domains at $z = z_{\rm re}$.  In the original setup, the source planes at the faces of each RT domain began emitting radiation at $z = z_{\rm re}$, with I-fronts fully crossing the domains after a time interval $\Delta t_{\rm IF} \approx L_{\rm dom}/\overline{v}_{\rm IF}^{\rm src}$, where and $\overline{v}_{\rm IF}^{\rm src}$ is the average speed of a plane-parallel I-front crossing the RT domain.  For simulations with photo-ionization rates close to or above the typical IGM average, $\Delta t_{\rm IF}$ is a few Myr or less, which is small compared to the photo-evaporation times of mini-halos ($10$s of Myr) and the hydrodynamic response time of the gas ($100$s of Myr).  However, in simulations with lower values of $\Gamma_{-12}$ and/or large $\delta/\sigma$, $\Delta t_{\rm IF}$ can be comparable to or longer than one or both of these timescales.  In these cases, interpretation of the results may be complicated by the time-delay of I-front passage\footnote{For example, the subgrid ionizing opacity model of Ref.~\cite{Cain2022b} contains separate treatments of photons absorbed in I-fronts and those absorbed behind I-fronts, where the volume-averaged neutral fraction is $\ll 1$.  This means that if I-fronts have not finished crossing the box at $z < z_{\rm re}$, the absorptions within them would be ``double-counted'' in their scheme.  }.  

To avoid this issue, we have developed a scheme that forces I-fronts to finish crossing the RT domains before the IGM begins responding dynamically to reionization.  We achieve this by pausing (or ``freezing'') all gravity and hydrodynamics at $z = z_{\rm re}$ for a fixed amount of time while I-fronts cross the RT domains.  During this time, we track only processes directly related to the radiation field - namely, ionization rates, the ionization state of the gas, and thermal evolution inside I-fronts.  Once ionized cells reach a neutral fraction of $x_{\rm HI} < 0.05$, we also shut off thermal evolution to prevent the gas from spuriously cooling after I-fronts have passed through.  The time period for which dynamics are frozen is given by three times the time required for an I-front driven by an incident photon number flux $F_{\gamma} \equiv \Gamma_{-12}/\sigma_{\rm HI}$ to cross the domain in mean density gas.  For our setup, this is given by
\begin{equation}
    \label{eq:tfreeze}
    t_{\rm freeze} = 3\frac{L_{\rm dom}}{\overline{v}_{\rm IF}^{\rm src}} + \frac{L_{\rm dom}}{c}
\end{equation}
where the average non-relativistic I-front speed is estimated by~\citep{DAloisio2019},
\begin{equation}
    \label{eq:vIF}
    \overline{v}_{\rm IF}^{\rm src} = \frac{F_{\gamma}}{(1+\chi)n_{\rm H}}
\end{equation}
where  $n_{\rm H}$ is the mean hydrogen number density at $z = z_{\rm re}$, and the factor of $1+\chi$ accounts for single ionization of helium within the I-front.  The second term on the RHS, wherein $c$ is the speed of light, accounts for scenarios in which $F_{\gamma}$ is large enough that the I-front speed becomes relativistic.  This approach guarantees that the whole box is re-ionized at the same redshift, independent of the ionizing background strength, gas density, and choice of $L_{\rm dom}$.  One potential drawback is that higher values of $t_{\rm freeze}$ could allow a larger fraction of the gas within dense, self-shielding gas clumps to become highly ionized by slow-moving ``D-Type''\footnote{That is, sub-sonic ionization fronts. } I-fronts before the clumps can dynamically respond, potentially affecting subsequent dynamics.  We show in Appendix~\ref{app:tfreeze} that within a reasonable range of $t_{\rm freeze}$ values, the subsequent gas dynamics are insensitive to the exact choice.  

We note that our RT domain + freezing procedure is qualitatively similar to a commonly-used self-shielding approximation in the literature - namely, to apply an on-the-fly self-shielding prescription to simulations with an otherwise homogeneous UV background to mimic the effect of RT~\citep[e.g.][]{Rahmati2013,Chardin2018}.  The crucial difference is that, since we use are doing ray-tracing RT, our method self-consistently predicts the self-shielding density rather than taking it as a calibrated input from some other simulation.  Indeed, in future work we plan to use \textsc{saguaro} results to calibrate a self-shielding prescription for use in larger-volume simulations.  

\subsubsection{Heating rates within I-fronts}
\label{subsubsec:heating_Ifronts}

One problem with the RT domain setup is that $L_{\rm dom}$ can be comparable to the widths of the I-fronts, which are commonly $\sim 10 (1+z)$ ckpc in mean-density IGM gas~\citep{DAloisio2019,Wilson2024a}.  As a result, cells closer to domain edges will ``see'' radiation with a spectral shape that matches that of the ionizing source planes, while cells closer to the center of the domains will initially see a much harder spectrum, filtered by the intervening partially neutral cells.  This results in a gradient of I-front heating rates across the domains, causing the gas to be colder near the domain edges than in the center.  It also results in the average post I-front temperature, $T_{\rm reion}$, being too low (see Appendix~\ref{app:treion} for further details and tests).  

We approximately correct for this by modeling the heating rates inside I-fronts as if the gas were locally optically thick - that is, in the limit that the optical depth across each partially neutral cell were infinite.  In this limit, the energy injection per H ionization is given by
\begin{equation}
    \label{eq:Eion}
    E_{\rm ion, \tau \gg 1}^{\rm HI} = \overline{E} - E_0^{\rm HI}
\end{equation}
where $\overline{E}$ is the average photon energy of the ionizing spectrum at the source planes and $E_0^{\rm HI} = 13.6$ eV is the ionization potential of HI.  A similar equation applies for the energy injected per HeI ionization.  Note that this differs from the average energy injection per ionization in the optically thin ($\tau \ll 1$) limit, which is weighted by the HI-ionizing cross-section, $\sigma_{\rm HI}$ and is thus somewhat smaller.  This treatment is applied only to cells with neutral fractions $> 5\%$, and heating rates in highly ionized cells are treated in the optically thin limit.  Note that {\it only the heating rates} are modified in this way - the optically thick limit is not applied to the propagation of the I-fronts (if it were, the gas would never ionize!).  

We find that this procedure eliminates spurious temperature gradients across RT domains, and raises post-reionization gas temperature $T_{\rm reion}$, especially near the edges of RT domains, relative to the original Ref.~\cite{DAloisio2020} setup.  An unavoidable side effect of this procedure is that the cooling rates inside the I-front, which are sensitive to its detailed thermal structure~\citep{DAloisio2019}, are not modeled correctly.  Fortunately, we find that this likely has only a modest ($10\%$ or smaller) effect on $T_{\rm reion}$.  In Appendix~\ref{app:treion}, we compare $T_{\rm reion}$ just after I-front passage in one of our simulations to the prediction of the $T_{\rm reion}$ prescribed in Ref.~\cite{DAloisio2019}, which is based on detailed 1-D RT simulations, and find reasonably good agreement.  This agreement holds at the $\sim 10\%$ or better level across different combinations of parameters used in our simulations, giving us confidence in the accuracy of the thermal history in \textsc{saguaro}.  

\subsubsection{Reduced speed of light}
\label{subsubsec:RSLA}

To speed up computation time, Ref.~\cite{DAloisio2020} applied an adaptive reduced-speed-of-light approximation (RSLA) scheme.\footnote{Specifically, they set the speed of light to $0.1 c_{\rm true}$ until the volume-weighted neutral fraction of the simulation box drops below $1\%$, after which the speed of light is reduced to $0.01 c_{\rm true}$. The rationale is to keep the speed of light significantly above the relevant velocity-scale. While I-fronts are traversing the RT domains, $c$ must be significantly faster than the I-fronts speeds. Afterward, $c$ must be much faster than the sound speed in the ionized gas ($\sim 20$ km/s).}    We optimize their implementation in two key ways.  The first is that during I-front passage (see \S\ref{subsubsec:IF}), we set the speed of light equal to
\begin{equation}
    \label{eq:ctilde_IF}
    \tilde{c}_{\rm IF} = \min(10 \times \overline{v}_{\rm IF}^{\rm src},c)
\end{equation}
which is equivalent to $10\times$ the I-front crossing speed in mean density gas, or the true speed of light, whichever is smaller.  This condition minimizes $\tilde{c}_{\rm IF}$ while ensuring that it is sufficiently large to allow the I-front to correctly capture the speed of the I-front, and thus $T_{\rm reion}$.  We have found that using a $\tilde{c}_{\rm IF}$ a factor of $2$ below Eq.~\ref{eq:ctilde_IF} does not significantly change $T_{\rm reion}$, demonstrating convergence.  

After I-front passage, the only velocity scales that are important are the sound speed in ionized gas, which sets the velocity of the gas, and the speed of D-type I-fronts inside self-shielding gas clumps in the process of photo-evaporation.  The former is $\sim 20$ km/s, or $\approx 10^{-4}c$, and the speed of D-type I-fronts is comparable to this~\citep{Shapiro2004}.  The speed of light in the ``ionized'' phase of the simulation after I-front passage, $\tilde{c}_{\rm ionized}$, need only be significantly larger than both of these velocity scales.  We find that $\tilde{c}_{\rm ionized} = 2 \times 10^{-3} c$ gives reasonably converged results for the ionization state and opacity of the gas (see Appendix~\ref{app:RSLA}), a factor of $5$ smaller than the $10^{-2} c$ value adopted in Ref.~\cite{DAloisio2020}.  We note that using a speed of light this low sometimes results in the light-crossing time of the RT cell, $\Delta t_{\rm RT}$, becoming larger than the hydrodynamical timestep $\Delta t_{\rm hydro}$, in which case we simply set $\Delta t_{\rm RT} = \Delta t_{\rm hydro}$.  In these cases, $\tilde{c}_{\rm ionized}$ effectively varies from one time step to the next, in which case we appropriately re-scale the ionizing photon number density everywhere adaptively at each time step to maintain a fixed incident ionizing flux, $F_{\gamma}$.  We show convergence with respect to our choices of the reduced speed of light in Appendix~\ref{app:RSLA}.  

\subsubsection{Sub-cycling time steps}
\label{subsubsec:timestep}

Because we use the reduced speed of light, some cells may have characteristic chemical or thermal timescales much shorter than $\Delta t_{\rm RT}$, including those with very high densities and/or at the edges of self-shielding regions.  In these cases, using $\Delta t_{\rm RT}$ to evolve the chemical and thermal equations can cause inaccuracies.  Because the thermal and chemical equations are defined locally (that is, they involve no spatial derivatives), we can define a sub-cycling thermo-chemical timestep in each cell given by
\begin{equation}
    \label{eq:deltat_cycle}
    \Delta t_{\rm cycle} = \min(\Delta t_{\rm RT}, 0.1 t_{\rm therm}, 0.01 t_{\rm chem})
\end{equation}
where $t_{\rm therm} \equiv T/|dT/dt|$ is the local thermal evolution timescale\footnote{We use a factor of $0.1$ multiplier instead of $0.01$ for $t_{\rm therm}$ here because $t_{\rm therm}$ can sometimes be extremely short, especially when collisional cooling is occurring at very high densities. }, and $t_{\rm chem} \equiv \min(n_{\rm HI}/|dn_{\rm HI}/dt|)$ is the timescale associated with changes in the neutral hydrogen density.  This procedure does not affect overall computation time significantly, since only a tiny fraction of cells have $\Delta t_{\rm cycle} \ll \Delta t_{\rm RT}$ by this criterion.  However, we do find it produces more accurate and converged results at very high densities and within D-type I-fronts.  

In Appendix~\ref{app:subcycle}, we show convergence tests of our sub-cycling procedure.  We find that the main physical process that causes large inaccuracies when sub-cycling is not used is collisional ionizations.  Ref.~\cite{DAloisio2020} turned off collisional ionizations to avoid these inaccuracies, while in \textsc{saguaro} we include them.  Notably, we find that our sub-cycling procedure results in converged results for the mass-averaged HI fraction, demonstrating that we capture self-shielding correctly.  We also find however, that the sub-cycling criteria given by Eq.~\ref{eq:deltat_cycle} is not stringent enough to fully capture the total recombination rate in all gas, which is highly sensitive to small ionized fractions in gas at very high densities ($\Delta > 10^3$).  We further find that more stringent time stepping criteria can largely, but not completely, mitigate these inaccuracies.  Fortunately, as we show in Appendix~\ref{app:subcycle}, these differences have neglibile effects on the self-shielding properties of the gas and its ionizing photon opacity.  

Whenever the HI (or HeI) fraction of a cell changes significantly during an RT time step, so will the optical depth across the cell.  In such cases, using the HI number density at the beginning of the time step to calculate optical depths will cause absorption rates in the cell to be incorrect.  To mitigate this, for cells with $\Delta t_{\rm cycle} <  \Delta t_{\rm RT}$, we average the HI and HeI number densities across sub-cycled time-steps, and use these to re-calculate the cell-wise optical depths.  This is done iteratively until convergence in the time step-averaged number densities are reached, or until a maximum number of $20$ iterations have elapsed.  This procedure is qualitatively similar to that adopted in the C2-Ray code~\citep{Mellema2006} to properly capture significant changes in HI number density across large RT time steps.  

\subsection{Additional physical parameters}
\label{subsec:additional_parameters}

As described above, the three most important box-scale parameters varied in \textsc{saguaro} are the reionization $z_{\rm re}$, photo-ionization rate $\Gamma_{\rm HI}$, and box-scale density, parameterized by $\delta/\sigma$.  As we describe in \S\ref{subsec:box_params}, these parameters are varied jointly across all possible combinations for our fiducial IGM physics model (see below).  However, we also include a number of simulations with other physical assumptions, usually for one or just a few combinations of $z_{\rm re}$, $\Gamma_{\rm HI}$, and $\delta/\sigma$.  Below, we summarize our fiducial physical assumptions, and describe the additional physical parameters that we vary in \textsc{saguaro}.  

\subsubsection{Fiducial physics}
\label{subsubsec:fiducial}

Our fiducial suite assumes a concordance $\Lambda$CDM cosmology with parameters given in \S\ref{sec:intro}.  We assume that the IGM prior to reionization cools adiabatically, with no external source of pre-heating, until it is re-ionized at $z = z_{\rm re}$.  The ionizing spectrum of the incident radiation is assumed to have a spectral slope of $\alpha = 1.5$.  We assume the case B recombination rate in ionized gas and ignore baryon-dark matter streaming velocities~\citep{Tseliakhovich2010}.  

\subsubsection{X-Ray pre-heating}
\label{subsubsec:xray}

Concordance models of the formation of the first galaxies and AGN predict that the IGM should have been heated above the adiabatic limit by pre-heating from X-rays~\citep{Furlanetto2006,Fialkov2014c}.  Recent upper limits on the 21 cm power spectrum have confirmed this picture~\cite{HERA2021a,HERA2021b,HERA2022,Nunhokee2025}.  Pre-heating by X-rays causes some pressure-smoothing of dense structures prior to reionization, which reduces the clumpiness of the IGM when I-fronts sweep through~\citep{DAloisio2020,Park2021a,Davies2023}.  Following Ref.~\cite{DAloisio2020}, we model this effect by imposing a minimum gas temperature\footnote{Injecting the heat over time rather than all at once at $z = 15$ would reduce the effect of pressure smoothing on the Baryons at later times.  Ref.~\cite{Davies2023} used a more realistic heating treatment and found the difference to be qualitatively small.  } $T_{\min}$ at $z < 15$, around the time when X-ray sources are believed to have begun heating the IGM~\citep{Eide2018}.  Following Ref.~\cite{Park2021a}, we include values of $T_{\min} = 10$, $100$, and $1000$ K, spanning the (wide) range of possible values expected based on concordance theoretical models.  Note that $10$ K is similar to lower limits on IGM gas temperature at $z \sim 8$ from Ref.~\cite{HERA2022}.  

\subsubsection{Ionizing spectral index}
\label{subsubsec:ionSpec}

We can vary the heat input into the IGM {\it during} reionization by changing the spectral index of the ionizing background, $\alpha$.  This is motivated by the fact that the sources of reionization can have a significant range of spectral shapes.  The ionizing spectrum of galaxies themselves is highly uncertain, ranging from $\alpha = 3$ to $0.5$ across reasonable models (see Ref.~\cite{DAloisio2019} and references therein).  Furthermore, a contribution from AGN~\citep{Chardin2015,Madau2024} and spectral filtering by the IGM itself could further harden the radiation field.  In addition to our fiducial $\alpha = 1.5$, we also consider values of $3$, $0.5$, and $-0.5$.   

\subsubsection{Alternative dark matter models}
\label{subsubsec:altDM}

Non-standard models for the physics of DM can produce significantly different IGM small-scale structure than the standard CDM scenario.  Popular alternative DM models include warm dark matter (WDM,~\citep{Viel2005,Viel2013,Irsic2024}), axion DM~\citep{Irsic2019}, and fuzzy DM~\citep{Litz2018}.  Most viable models predict a suppression (or sometimes an enhancement) in the matter power spectrum at small scales relative to the CDM expectation, and agreement with CDM at large scales that observations constrain (see Fig.~10 of Ref.~\cite{Gilman2021}).  Here, we use the WDM scenario as a general proxy for all alternative DM cosmologies that predict a suppression of small-scale power.  We model WDM scenarios in \textsc{saguaro} by modifying initial transfer functions of DM and baryons using the procedure described in Ref.~\citep{Viel2005}, as we did in Ref.~\citep{Cain2022a}.  The minimum wavenumber up to which small-scale power is suppressed is roughly proportional to the WDM particle mass, $m_{\rm X}$.  We include values of $m_{\rm X}$ as small as $3$ keV, at the edge of the range ruled out by the Ly$\alpha$ forest~\cite{Irsic2024}, and up to $20$ keV, well above current lower limits.  

\subsubsection{Baryon-dark matter streaming}
\label{subsubsec:vbc}

At the epoch of recombination ($z \sim 1100$), baryons underwent baryon acoustic oscillations (BAOs), giving them a velocity offset relative to dark matter, denoted $v_{\rm bc}$, with an rms average value of $\sigma_{\rm bc} = 30 \left(1+z\right)/1100$ km/s~\citep{Tseliakhovich2010}.  This baryon-dark matter ``streaming velocity'' was supersonic by a factor of a few at kinematic decoupling, and suppressed the formation of baryonic structures at mass scales $\lesssim 10^8$ $M_{\odot}$.  The stream velocity affected the formation of Pop III stars~\cite{Naoz2012,OLeary2012}, and black holes~\citep{Tanaka2014}, may have acted as a channel for globular cluster formation~\citep{Naoz2014,Lake2025}, and likely imprints a signature on the 21cm power spectrum from cosmic dawn and the EoR~\citep{McQuinn2012,Munoz2019,Cain2020}.  Most importantly for this work, stream velocities can suppress clumpiness of the IGM prior to reionization, analogously to the effect of WDM~\citep{Hirata2018,Cain2020,Park2021a}.  Since $v_{\rm bc}$ is roughly coherent at scales of a few Mpc, we include it by boosting the gas velocity along one direction of the box, following the procedure in Ref.~\cite{Cain2020}.  As in Ref.~\cite{Cain2020}, simulations with stream velocities are initialized at $z = 1080$ instead of $300$, so that the formation of structures can be self-consistently evaluated from the time the velocity offset is sourced (see also Ref.~\citep{Hirata2018}).  We use values of $v_{\rm bc} = 20$, $41$, and $65$ km/s, following the Gaussian quadrature averaging argument in Ref.~\cite{Cain2020}.  

\subsubsection{Recombination radiation}
\label{subsubsec:recomb_rad}

The last physical parameter we vary is the temperature-dependent recombination coefficient in ionized gas, $\alpha(T)$.  Recombinations directly to the ground state of HI produce a photon with energy $E \gtrsim 13.6$ eV, which itself can ionize HI.  Since explicitly tracking the production and re-absorption of HI-ionizing recombination radiation is often computationally difficult, they are typically treated in one of two limiting cases.  In the ``case B'' limit, these photons are treated as if they are re-absorbed immediately after being produced, and in the ``case A'' limit, they are treated as if they are never re-absorbed.  Our fiducial setup assumes the case B limit, which is motivated by the fact that the majority of recombinations occur in over-dense, ionized gas where the local mean free path to them is very short.  However, it is possible that a significant fraction of these photons escape the dense systems where they are produced, in which case the IGM opacity may be best-described by something in between the case A and B limits~\citep{FaucherGiguere2009}.  To study this, several of our simulations (see below) use the case A recombination coefficient for HII instead of case B when calculating the gas ionization state.  

\subsection{Summary of the suite}
\label{subsec:box_params}

Table~\ref{tab:simulation_summary} summarizes the entire \textsc{saguaro} suite.  The first column gives the name of each subset of simulations, and the second its box size in $h^{-1}$Mpc.  The third, fourth, and fifth columns give the values of $z_{\rm re}$, $\Gamma_{-12} \equiv \Gamma_{\rm HI}/10^{-12} {\rm s}^{-1}$, and $\delta/\sigma$ included in the subset.  When more than one column contains more than one value, that means the subset includes {\it all possible combinations} of those parameters.  For example, the ``Core'' suite includes all possible combinations of $z_{\rm re} \in \{5,7,9,15\}$, $\Gamma_{-12} \in \{0.03,0.3,3,30\}$, and $\delta/\sigma \in \{0, \pm \sqrt{3}\}$, for a total of $48$ runs.  The sixth column lists the physics assumed in the run, and the last column gives the number of runs in the subset.  The full \textsc{saguaro} suite contains $236$ simulations and cost over $5$ million CPU hours to complete.  

\renewcommand{\arraystretch}{1.5}
\begin{table}[h!]
    \centering
    \footnotesize
    \begin{tabular}{|c|c|c|c|c|c|c|}
    \hline \hline
       {\bf Name} & {\bf $L$/[$h^{-1}$Mpc]} & {\bf $\bm{z_{\rm re}}$} & $\bm{\Gamma_{-12}}$ & $\bm{\delta/\sigma}$ & {\bf Physics} & {\bf \#}\\
       \hline \hline
        Core & $2$ & \makecell{$\{5,7, 9,15\}$} & \makecell{$\{0.03,0.3,$ \\ $3,30\}$} & $\{0,\pm \sqrt{3}\}$ & Fiducial & 48\\
        Core-Ex.\tablefootnote{As of the time of writing, $123/139$ (nearly $90\%$) of the Core-Ex. simulations are complete, and computational time has been secured to run the remaining $13$.  } & $2$ & $+\{6,8,12\}$ & $+\{0.1,1\}$ & $+\{\pm 1\}$ & Fiducial & 139\\
        Core-HR & $0.25$ & \makecell{$\{5,7, 9,15\}$} & \makecell{$\{0.03,0.3,$ \\ $3,30\}$} & 0 & Fiducial & 16\\
        %Low-$\Gamma_{\rm HI}$ & $2$ $h^{-1}$Mpc & 7 & 0.003 & 0 & Fiducial & 1\\
        X-Ray & $2$ & 7 & 0.3 & $\{0,\pm \sqrt{3}\}$ & $T_{\min} \in \{10,10^2,10^3\}$ K & 7\\
        Ion. Spectrum & $2$ & 7 & 0.3 & 0 & $\alpha \in \{-0.5,0.5,3\}$ & 3\\
        WDM & $2$  & 7 & 0.3 & $\{0,\pm \sqrt{3}\}$ & $m_{\rm X} \in \{3,5\}$ keV & 6\\
         Recomb Rad. & $2$ & 7 & 0.3 & $\{0,\pm \sqrt{3}\}$ & Case A & 3\\
        SV & $2$ & 7 & 0.3 & 0 & $v_{\rm bc} \in \{20,41,65\}$ km/s & 3\\
        X-Ray-HR & $0.25$ & 7 & 0.3 & 0 & $T_{\min} \in \{10,10^2,10^3\}$ K & 3\\
        WDM-HR & $0.25$ & 7 & 0.3 & 0 & $m_{\rm X} \in \{3,5,7,10,20\}$ keV & 5\\
        SV-HR & $0.25$ & 7 & 0.3 & 0 & $v_{\rm bc} \in \{20,41,65\}$ km/s & 3\\
       \hline \hline 
       {\bf Total} & & & & & & 236\\
       \hline \hline
    \end{tabular}
    \caption{Summary of all simulations in the \textsc{saguaro} suite.  The Core suite represents the ``foundation'' of \textsc{saguaro}, and includes all combinations of the parameters listed in the 3rd-5th columns above.  The Core-Ex. suite extends this grid to include several more values of each parameter, reaching a total of over $150$ simulations.  Core-HR uses a smaller box size and includes the same values of $z_{\rm re}$ and $\Gamma_{-12}$ as Core, but only at $\delta/\sigma = 0$ (mean density).  Several smaller subsets of simulations vary the additional parameters, described in \S\ref{subsec:additional_parameters} and briefly summarized in the 6th column.  See text for further details.  }
    \label{tab:simulation_summary}
\end{table}

The ``Core'' suite, which will be the be the main focus of the rest of this work, has a box size of $2$ $h^{-1}$Mpc and includes all combinations of $4$ values each of $z_{\rm re}$ and $\Gamma_{-12}$, and three values of $\delta/\sigma$ (see above).  We are also in the process of running\footnote{As of submission of this paper, the Core-Ex. suite is $\approx 2/3$ complete, with another $2-3$ months of runtime expected.  } an extension of the Core suite, denoted ``Core-Ex.'', that adds $z_{\rm re} \in \{6,8,12\}$, $\Gamma_{-12} \in \{0.1,1\}$ and $\delta/\sigma \in \{\pm 1\}$ to the Core suite.  In total, Core + Core-Ex. includes (almost)\footnote{The Core-Ex. suite does not include combinations of $\Gamma_{-12} = 30$ with any of the other parameters.} all possible combinations of $7$ values of $z_{\rm re}$, $6$ values of $\Gamma_{-12}$, and $5$ values of $\delta/\sigma$, for a total of $187$ simulations.  These runs will serve as the input for an improved version of the subgrid opacity model described in Refs.~\cite{Cain2021,Cain2022b}, which will be presented in a later paper.   We also include a set of simulations with much higher spatial resolution in $0.25$ $h^{-1}$Mpc boxes, which we denote ``Core-HR''.  These include the same combinations of $z_{\rm re}$ and $\Gamma_{-12}$ included in the Core suite, but run at the cosmic mean density only ($\delta/\sigma = 0$).  This suite is intended to help correct quantities inferred from the Core suite for missing spatial resolution, and study processes that require sub-kpc resolution to fully capture, such as turbulence~\citep{Cain2025b} and photo-evaporation of mini-halos~\citep{Shapiro2004,Chan2023}.   

We have run several smaller sets of simulations that vary physical parameters outlined in \S\ref{subsec:additional_parameters}.  In all these cases, we fix $z_{\rm re} = 7$ (motivated by the small Planck optical depth,~\citep{Tristram2024}) and $\Gamma_{-12} = 0.3$ (motivated by typical values measured from the Ly$\alpha$ forest,~\citep{Becker2013,Gaikwad2023}).  For $2$ $h^{-1}$Mpc boxes, we sometimes include all three values of $\delta/\sigma$.  The ``X-Ray'' subset includes simulations with minimum temperatures of $T_{\min} = 10$, $100$, and $1000$ K (see \S\ref{subsubsec:xray}) at the $2$ $h^{-1}$Mpc box size.  For the runs with $T_{\min} = 100$ and $1000$ K, we include $\delta/\sigma = 0$ and $\pm \sqrt{3}$, while the $T_{\min} = 10$ K case includes only $\delta/\sigma = 0$.  We also ran three mean-density runs at the $0.25$ $h^{-1}$Mpc box size, one for each $T_{\rm min}$ value, which we denote ``X-Ray-HR''.  These runs characterize the effect of X-Ray pre-heating on structure before reionization, which we have already found has a significant effect on small-scale structure and pressure smoothing~\citep{Cain2024a,Cain2025b}.  The Ion. Spectrum runs vary the power law index of the assumed radiation spectrum after reionization, including spectral slopes of $\alpha = -0.5$, $0.5$, and $3$, in addition to the fiducial $\alpha = 1.5$.  We include only $2$ $h^{-1}$Mpc boxes for this parameter.  

\textsc{saguaro} also includes simulations with warm dark matter, streaming velocities, and case A recombination rates.  The WDM suite is made up of  $2$ $h^{-1}$Mpc simulations with $m_{\rm X} = 3$ and $5$ keV, both of which are run for all $\delta/\sigma$ values.  The WDM-HR suite is made up of  $0.25$ $h^{-1}$Mpc simulations with
$m_{\rm X} = 3$, $5$, $7$, $10$, and $20$ at mean density.  The Stream Velocity and Stream Velocity-HR suites each have three values of $v_{\rm bc}$ and are all run at mean density.  Finally, we have run three $2$ $h^{-1}$Mpc simulations with the case A recombination rate, at $\delta/\sigma = 0$ and $\pm \sqrt{3}$.  

\subsection{Simulation outputs}
\label{subsec:otf}

This section describes the various data products produced and saved for each \textsc{saguaro} simulation.  These include full simulation outputs stored at a small number of time steps, and quantities calculated ``on-the-fly'' and saved at every hydrodynamical time step.  

\subsubsection{Full Snapshots}
\label{subsubsec:grids}

Full outputs of DM, gas, and RT data are saved at discrete redshift snapshots.  These are saved at highest temporal fidelity shortly after $z_{\rm re}$, when the gas is evolving most rapidly, and less frequently thereafter.  For a given $z_{\rm re}$, the first three snapshots are saved at redshift intervals of $\Delta z = 0.1$\footnote{The exception to this is $z_{\rm re} = 5$, for which we also save $z = 4.95$.  }, and a fourth at $z_{\rm re} - 0.5$.  Afterwards, snapshots are saved in redshift intervals of $\Delta z = 0.5$ when $z < 10$, and $\Delta z = 1$ when $z > 10$.   For example, for $z_{\rm re} = 7$, we save full outputs at $z = 6.9$, $6.8$, $6.7$, $6.5$, $6$, $5.5$, $5$, $4.5$, and $4$.  A total of between $7$ and $20$ snapshots are saved per simulation, depending on the value of $z_{\rm re}$.  We save the following quantities in each snapshot: 
\begin{itemize}

    \item {\bf Dark Matter Particles}: $x,$ $y,$ and $z$ positions of all dark matter particles, and their velocities in the $x,$ $y,$ and $z$ directions.  

    \item {\bf Gas Cells}: density, velocities in the $x,$ $y,$ and $z$ directions, temperature, and ionization state of HI, HeI, and HeII.  
    
    \item {\bf RT Cells}: Photo-ionization rates and heating rates per particle for HI, HeI, and HeII, photon number densities in each frequency bin, and absorption rates of photons in each frequency bin, and cell-wise optical depth in each frequency bin.  
    
\end{itemize}
We save full gas, DM, and RT grid outputs for all sub-sets of simulations except the Core-Ex. set, for which we save only full gas outputs (to save disk space).  Full snapshot data is being saved long-term on the Texas Advanced Computing Center (TACC) Ranch tape-storage.  

We have also saved 2D slices through the simulation volume at every hydrodynamical timestep to facilitate visualizations, including movies, of various quantities.  We save slices of the DM density (mapped to a grid with TSC), gas density, gas temperature, HI, HeI, and HeII fractions, photo-ionization rates, heating rates, and cell-wise redshift of reionization.  Lastly, we have saved 2D maps of the Lyman Limit ($912\text{\AA}$) optical depth, $\tau_{912}$, obtained from integrating along one axis of the box, which allows us to conveniently visualize the self-shielding properties of the gas.  

\subsubsection{Box-averaged quantities}
\label{subsubsec:otf}

At the end of every hydro time-step, we save the following box-averaged quantities: 

\begin{itemize}

    \item Mean positions and velocities of all DM particles, and average velocities in all gas cells.  

    \item Volume and mass-weighted averages of the neutral fraction, temperature, sound speed, and photo-ionization rate.  

    \item Volume-averaged ionizing photon absorption rates for HI and HeI in each frequency bin.  
    
    \item Volume-averaged recombination rates for HII and HeII.  

\end{itemize}

We also estimate the number of recombinations and absorptions in the box in several different ways.  Following Ref.~\cite{DAloisio2020}, average recombination rates are calculated in $5$ different ways: (i) including all cells, (ii) including only cells with $\Gamma_{\rm HI} > 0.01 \langle \Gamma_{\rm HI} \rangle$, where the angle brackets denote a volume average, (iii) cells with $\Gamma_{\rm HI} > 0.001 \langle \Gamma_{\rm HI} \rangle$, (iv) cells with $\Gamma_{\rm HI} > 0$, and (v) cell with $\Delta > 200$, where $\Delta$ is the gas density in units of the cosmic mean.  We use the same thresholds to calculate and save {\it absorption} rates in only the lowest energy bin ($14.48$ eV) and summed over all frequency bins.  Absorption rates calculated for other individual frequency bins average all cells in the volume.  

We also output several definitions of the clumping factor, which quantifies the boost in the recombination rate relative to an IGM with spatially uniform density~\citep{Pawlik2009,Kaurov2015}.  The clumping factor can be defined directly in terms of the density field as
\begin{equation}
    \label{eq:clump_density}
    C_{\Delta < \Delta_0} = \langle \Delta^2 \rangle_{\Delta < \Delta_0}
\end{equation}
where the average runs over the whole volume and counts only cells with a given threshold $\Delta < \Delta_0$.  We output $C_{\Delta < \Delta_0}$ for $\Delta_0 = 50$, $100$, $200$, and $\infty$.  Additional definitions of the clumping factor are possible that use the recombination and ionization rates that we also calculate - these are discussed in \S\ref{subsec:clumping}.  

\subsubsection{The Mean Free Path}
\label{subsubsec:est_mfp}

A chief goal of \textsc{saguaro} is to facilitate more accurate modeling of the ionizing photon opacity of the IGM, which is quantified by the mean free path (MFP,~\citep{Prochaska2009,Worseck2014,Becker2021,Zhu2023,Gaikwad2023,Davies2024c}).  Toward this end, we estimate the MFP, $\lambda_{\rm mfp}$, in \textsc{saguaro} using three definitions that have been used before in the literature.  We refer to the first of these as the ``flux-based'' definition, given by
\begin{equation}
    \label{eq:lambda_flux_def}
    \lambda_{\rm mfp}^{\rm flux} \equiv \frac{F_{\gamma}}{\langle \Gamma_{\rm HI} n_{\rm HI} \rangle_{\rm V}}
\end{equation}
where $F_{\gamma}$ is the ionizing flux incident at the boundaries of the RT domains and the denominator is the volume-averaged HI absorption rate summed over all frequency bins.  We can define a frequency-dependent version of this definition, $\lambda_{\rm mfp}^{\rm flux}(\nu)$, by replacing the numerator and denominator of Eq.~\ref{eq:lambda_flux_def} with fluxes and absorption rates (respectively) in individual frequency bins.  This definition was first tested in Ref. \cite{DAloisio2020}, used to calibrate the subgrid opacity model defined in Refs.~\cite{Cain2021,Cain2022b}, and is derived in Appendix C of Ref.~\cite{Cain2022b} (see also Appendix A of Ref.~\cite{Cain2023}).  This definition directly relates the MFP, or equivalently the absorption coefficient $\kappa \equiv 1/\lambda_{\rm mfp}$, to the total rate at which ionizing photons are being absorbed in the simulation\footnote{This differs from most definitions in the fact that it must be calculated directly from the radiation field itself, and not in post-processing from the HI number density of the gas.  }.  

We refer to our second estimator as the ``segment-based'' definition, which is given by 
\begin{equation}
    \label{eq:lambda_seg_def}
    \lambda_{\rm mfp}^{\rm seg} \equiv \frac{-L_{\rm seg}}{\ln(\langle f_{\rm out}\rangle)}
\end{equation}
where $L_{\rm seg}$ is some segment length that must satisfy $L_{\rm seg} \ll \lambda_{\rm mfp}^{\rm seg}$, and $\langle f_{\rm out} \rangle \equiv \langle e^{-\tau} \rangle$ is the transmitted flux (at some given frequency) averaged over a large number of randomly positioned and oriented segments of length $L_{\rm seg}$, where $\tau = \int_{0}^{L_{\rm seg}} n_{\rm HI}(x) \sigma_{\rm HI}dx = \sigma_{\rm HI} N_{\rm HI}$, where $N_{\rm HI}$ is the HI column density over the segment.  This definition was first introduced in Ref.~\cite{Emberson2013}, and later in Refs.~\cite{DAloisio2020} and~\cite{Nasir2021}.  It is equivalent to defining the MFP as an integral over the HI column density distribution~\citep{Paresce1980,Becker2013,Rahmati2017} if that distribution is defined as the set of $N_{\rm HI}$ for all segments of length $L_{\rm seg}$ (see Ref.~\cite{Nasir2021}).  Since the choice of $L_{\rm seg}$ is to some degree arbitrary, we evaluate Eq.~\ref{eq:lambda_seg_def} for $L_{\rm seg} = L_{\rm box}/2^n$, where $n$ runs from $0$ to $6$, such that we vary $L_{\rm seg}$ by about $1.5$ orders of magnitude.  Evaluating Eq.~\ref{eq:lambda_seg_def} involves integrating the HI column density distribution, $f(N_{\rm HI})$~\citep{OMeara2007,Prochaska2010}, over a large number of randomly positioned and oriented sightlines of varying lengths.  We bin $N_{\rm HI}$ logarithmically into $1000$ bins between $\log(N_{\rm HI}/[{\rm cm}^{-2}]) = 9$ and $24$, and save the distributions at intervals of $\Delta z = 0.01$.  

Our last definition directly leverages the {\it definition} of the MFP - namely, the average distance traveled by an ionizing photon emitted at a random location and in a random direction in the IGM.  Translating this directly to a mathematical definition gives (see Appendix C of Ref.~\cite{Chardin2015}),
\begin{equation}
    \label{eq:lambda_theory_def}
    \lambda_{\rm mfp}^{\rm def} \equiv \frac{\left\langle \int_{1}^{0} x df\right\rangle}{\left\langle \int_{1}^{0} df \right\rangle}
\end{equation}
Here, $x$ is the distance from the starting point of a randomly-placed sightline, $f(x) \equiv e^{-\tau(x)}$ is the transmitted flux (at some frequency) along that sightline from its start, and $\tau(x) = \int_{0}^{x} dx' n_{\rm HI}(x') \sigma_{\rm HI}$ integrated along the sightline.  The angle brackets in Eq.~\ref{eq:lambda_theory_def} denote an average over a large number of sightlines.  Eq.~\ref{eq:lambda_theory_def} gives the average distance traveled along a sightline weighted by the fraction of photons traveling that distance, which is precisely the definition of the MFP.  

Formally, the integrals in Eq.~\ref{eq:lambda_theory_def} run to $f = 0$ (equivalently, $x = \infty$), so Eq.~\ref{eq:lambda_theory_def} cannot be exactly evaluated numerically.  We approximate $\lambda_{\rm mfp}^{\rm def}$ by terminating sightlines when $f(x) < 10^{-3}$, and find that this gives converged results\footnote{Note that in this limit, the denominator of Eq.~\ref{eq:lambda_flux_def} becomes unity, as pointed out by Ref.~\cite{Chardin2015}.  }.  We also find that using $10,000$ sightlines in the average is sufficient for convergence.  Since both Eq.~\ref{eq:lambda_seg_def} and~\ref{eq:lambda_theory_def} require integrations over large path lengths (especially the latter when $\lambda_{\rm mfp}$ is large), we do not output these at every hydrodynamical time step, but rather only in intervals\footnote{By contrast, Eq.~\ref{eq:lambda_flux_def} is a simple average over the box, so we output it on every time step.} of $\Delta z = 0.01$.  We note that all three of these definitions are equivalent in the limit that the entire universe is filled with highly ionized, optically thin gas.  Differences between them emerge in the presence of optically thick and/or partially neutral systems.  We will defer a detailed exploration of these differences to a future work, and simply comment here that they give similar results across most of our parameter space (see also \S\ref{subsec:mfp}).    

\subsubsection{Sub-volumes}
\label{subsubsec:subvol}

It is interesting to study how much density fluctuations within our $2$ $h^{-1}$Mpc boxes affect the local properties of the IGM.  This variability could e.g. help develop more accurate subgrid treatments of quantities like the MFP at scales smaller than our simulated volumes (see Refs.~\cite{Mao2019,Bianco2021} for an example of such a procedure).  Towards this end, we have calculated a number of quantities in a collection of sub-volumes of side lengths $0.25$, $0.5$, and $1$ $h^{-1}$Mpc.  The number of sub-volumes of each size is $N_{\rm sub} = 8^3$, $8^3$, and $4^3$, respectively\footnote{Note that for the $0.5$ $h^{-1}$Mpc and $1$ $h^{-1}$Mpc, sub-volumes are allowed to overlap.}.  We calculate the following quantities in each sub-volume and output them every $\Delta z = 0.01$: 

\begin{itemize}

    \item Average values of density, temperature, and ionization state of HI and HeI.  

    \item The recombination clumping factor, ionizing photon absorption rates, and recombination rates.  

    \item All three definitions of the MFP, given above.  
    
\end{itemize}

\section{Gas properties in \textsc{saguaro}}
\label{sec:vis_IGM}

\subsection{Visualizing the Core Simulations}
\label{subsec:core}

\subsubsection{Density}
\label{subsubsec:vis_density}

\begin{figure}[h!]
    \centering
    \includegraphics[scale=0.255]{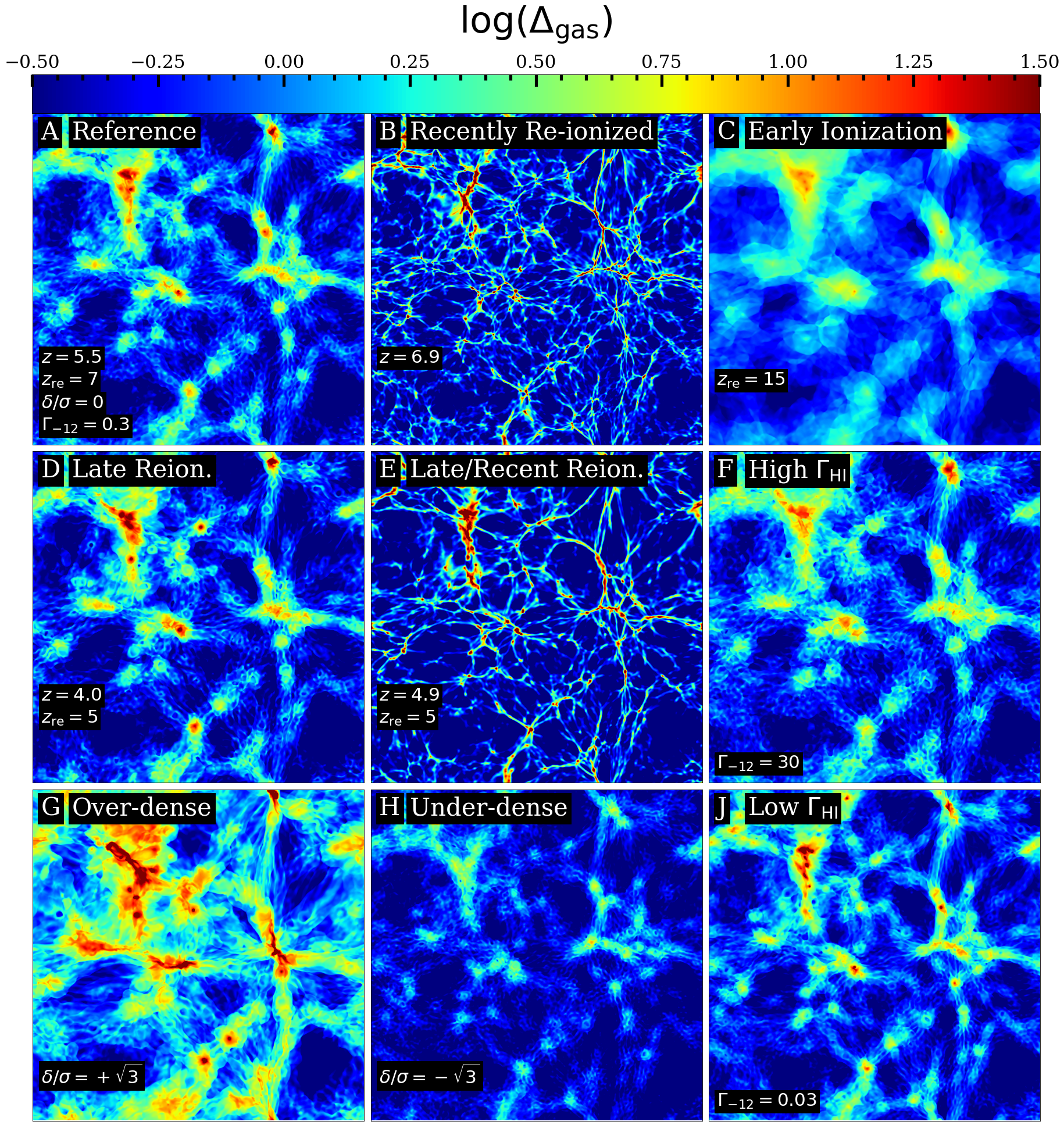}
    \caption{Visualization of the gas density in slices through different simulations and redshifts from the ``Core'' set of \textsc{saguaro} simulations.  The upper left panel (A) shows the simulation with $z_{\rm re} = 7$, $\delta/\sigma = 0$, and $\Gamma_{-12} = 0.3$ at $z = 5.5$, which we use as a reference point for comparison.  Panel B shows the same slice at $z = 6.9$, highlighting the initially clumpy state of the IGM just after reionization, which is followed by considerable pressure smoothing.  This comparison is made again for a lower $z_{\rm re} = 5$ in Panels D and E.  Panel C shows that earlier reionization results in less surviving structure at fixed redshift.  Panels F and J show the effect of increasing and decreasing $\Gamma_{\rm HI}$, respectively.  Higher (lower) $\Gamma_{\rm HI}$ results in less (more) pronounced, compact density peaks due to the interplay between self-shielding and pressure smoothing.  Panels G and H highlight the enhancement of structure formation in over and under-dense boxes, respectively. }
    \label{fig:example_density}
\end{figure}

In this section, we visualize and summarize the important qualitative features of IGM gas dynamics in \textsc{saguaro}.  Figure~\ref{fig:example_density} shows a set of slices (each one cell, or $2$ $h^{-1}$kpc thick) through the density field from different simulations and snapshots in the Core suite.  The panels are indexed by letters in the upper left, and the simulation parameters and redshift are given in the lower right.  Here and in the rest of this section, we will use the simulation with $z_{\rm re} = 7$, $\delta/\sigma = 0$, and $\Gamma_{-12} = 0.3$, at $z = 5.5$ ($\Delta t \sim 300$ Myr after reionization), which is shown in panel A, as our reference point for most comparisons. Other panels show only parameters whose values changed relative to panel A.  The text in the upper left gives a brief qualitative summary of what changed relative to panel A.

We can see how the gas evolves dynamically by comparing panels A and B.  The latter is shown at $z = 6.9$ ($\Delta t \sim 10$ Myr) for the same parameters as panel A.  In B, the most of the gas is concentrated into abundant, small filaments and clumps, reflecting the short IGM Jeans scale prior to reionization heating.  In A, most of these clumps have been destroyed or ``puffed out'', reflecting the much longer Jeans scale of the hot, ionized gas.  A conspicuous interference pattern, pointed out by e.g. Refs.~\cite{DAloisio2020} and~\cite{Puchwein2023}, arises due to the overlap of expanding clumps.  Panels D and E draw the same comparison, but at a somewhat lower redshift in simulations that re-ionize later - $z_{\rm re} = 5$, at $z = 4$ and $4.9$ (respectively).  The difference between snapshots is qualitatively similar, but the $z = 4.9$ case in E shows much more concentrated structure formation than in B due to its lower redshift.  The same difference can be seen between panels D and A, but to a lesser degree.  This ``relaxation'' process has been observed and characterized in a number of previous studies~\citep{Park2016,Hirata2018,DAloisio2020,Nasir2021,Chan2023,Puchwein2023}.  Panel C differs from A only in the reionization redshift, $z_{\rm re} = 15$, and displays considerably less small-scale structure than A.  This is a combination of two effects. First, the gas was re-ionized earlier, before many $\gtrsim 10^8 M_{\odot}$ structures could form, and formation of these structures was subsequently suppressed by reionization heating.  Second, dense structures that were present at $z = 15$ were ionized at a redshift when the dynamical time was shorter~\citep{DAloisio2020} and have had more time to respond to pressure smoothing by $z = 5.5$.  Collectively, panels A-E demonstrate the non-trivial dependence of gas dynamics on $z_{\rm re}$.  

Panels F and J are the same as panel A, but with $100\times$ larger and $10\times$ smaller $\Gamma_{\rm HI}$, respectively. The large-scale features of the gas distribution match well, but some differences are notable at smaller scales.  Panel F displays less concentrated over-densities, especially noticeable in the top left and top right.  This trend is reversed in panel J, which retains many dense clumps with $\Delta \gtrsim 50$.  This owes to the differences in self-shielding properties.  Higher $\Gamma_{\rm HI}$ results in gas at higher densities being ionized and heated, smoothing out density fluctuations that would have otherwise remained cold and compact.  For our lowest $\Gamma_{-12}$, gas with $\Delta < 100$ can remain self-shielded well after reionization, allowing mini-halos to persist for hundreds of Myr~\citep{Chan2023}.  These findings highlight the importance of modeling the interplay between self-shielding and pressure smoothing.  
Lastly, panels G and H show the same as A, but with $\delta/\sigma = +\sqrt{3}$ (over-dense) and $-\sqrt{3}$ (under-dense).  Respectively, these show more and less prominent structure formation, reflecting the conditions in the biased regions they simulate.  In particular, the over-dense run retains a large number of dense structures that can retain gas long after reionization.  

\subsubsection{Temperature}
\label{subsubsec:vis_temperature}

\begin{figure}[h!]
    \centering
    \includegraphics[scale=0.26]{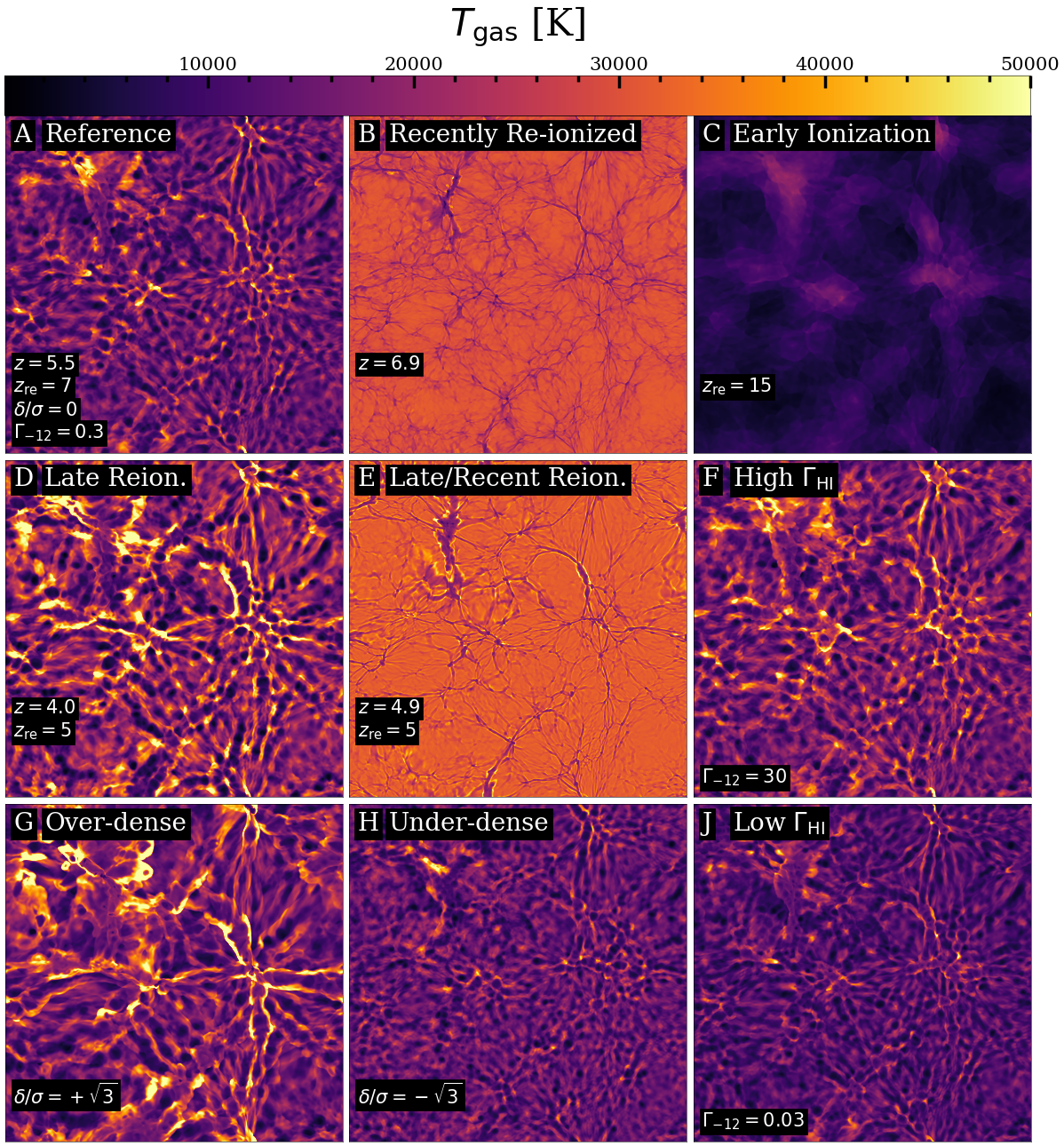}
    \caption{Gas temperature for the same snapshots and layout as in Figure~\ref{fig:example_density}.  The thermal structure proceeds from mostly uniform (B) just after ionization to a complex structure with small-scale fluctuations driven by pressure smoothing several hundred Myr later (A).  These fluctuations disappear if reionization was early and happened long ago (C), and temperatures are higher and fluctuations stronger if reionization occurs later (D and E).  The gas is hotter and displays stronger $T$ fluctuations if $\Gamma_{\rm HI}$ is higher (F and J).  Regions with densities higher (G) and lower (H) than the cosmic mean display stronger and weaker $T$ fluctuations, respectively.  }
    \label{fig:example_temperature}
\end{figure}

Figure~\ref{fig:example_temperature} shows gas temperature for the same snapshots visualized in Figure~\ref{fig:example_density}.  The thermal structure of the IGM varies even more than its density structure across redshifts and simulations. In panel B, the gas is hot ($\sim 30,000$ K) and nearly isothermal, reflecting the roughly uniform $T_{\rm reion}$ heating left behind by fast-moving I-fronts~\citep{DAloisio2019,Zeng2021}.  The over-dense filaments are slightly cooler than the rest of the gas, since I-fronts moved through them more slowly.  By $z = 5.5$ (A), the thermal structure has evolved dramatically.  The dense filaments have since expanded (Figure~\ref{fig:example_density}) and cooled in their centers, whilst gas at lower densities around the edges of filaments is compressed and heated, creating a web of ``hot-spots'', some of which reach up to $\sim 50,000$K.  The gas furthest from the expanding filaments cools steadily as the universe expands.  The resulting thermal structure is very complex, and strongly deviates from the commonly-assumed one-to-one temperature density relation in the post-reionization IGM~\cite{McQuinn2016,Keating2018}.  The physics driving this process in \textsc{saguaro} is explored in more detail in Ref.~\cite{Cain2024a} (see also Ref.~\cite{Hirata2018}).  

In panels C-E, we see how this picture depends on $z$ and $z_{\rm re}$.  Panel C shows that in patches that re-ionize very early, small-scale temperature fluctuations driven by pressure smoothing disappear, and temperature becomes a tight function of density.  We see the same trends in D and E as in A and B, but with some notable differences.  First, the gas is slightly hotter in E than in B, reflecting the lower densities and correspondingly faster I-front speeds at fixed incident ionizing flux (set by $\Gamma_{\rm HI}$).  The gas is also much hotter in D, despite the time interval between $z_{\rm re}$ and $z$ being longer than that in A ($\Delta t = 369$ vs.\ $279$ Myr).  This is in part because the gas is less dense on average and has a longer cooling timescale at lower redshift, and also because more structure has formed in the cold IGM by $z = 5$, resulting in a more ``violent'' dynamical response.  Small-scale temperature fluctuations are more important and persist for longer in patches of the IGM that reionized later.  

In panels F and J, we see that $\Gamma_{\rm HI}$ has a modest impact of temperature fluctuations. Panel F is slightly hotter and has stronger $T$ fluctuations than A, and the opposite is true of panel J.  Two effects are at play here: the speed of I-fronts that sweep through at $z_{\rm re}$ and the degree to which gas is compressed around dense, expanding clumps.  Higher $\Gamma_{\rm HI}$ requires a larger incident ionizing flux, which drives faster I-fronts and hotter $T_{\rm reion}$. In addition, high $\Gamma_{\rm HI}$ heats the gas at higher densities, pushing it out of the potential wells and intensifying the compression of the nearby gas, resulting in higher temperature peaks.  

In panels G and H, we see that the effect of density on the thermal structure is partially degenerate with both $z_{\rm re}$ and $\Gamma_{\rm HI}$.  The under-dense box (H) is colder and displays slightly weaker $T$ fluctuations than A, similar to J.  In this case, the gas cools more rapidly due to adiabatic expansion (since voids expand faster than the average universe), and there are fewer dense structures present to drive small-scale $T$ fluctuations.  The opposite is true in panel G, which most closely resembles panel D.  The denser, more compact structures in the over-dense run reach higher pressures when heated and expand more violently, heating the surrounding gas to higher temperatures than in the mean-density case.  

\subsubsection{Self-shielding}
\label{subsubsec:self_shielding}

\begin{figure}
    \centering
    \includegraphics[scale=0.26]{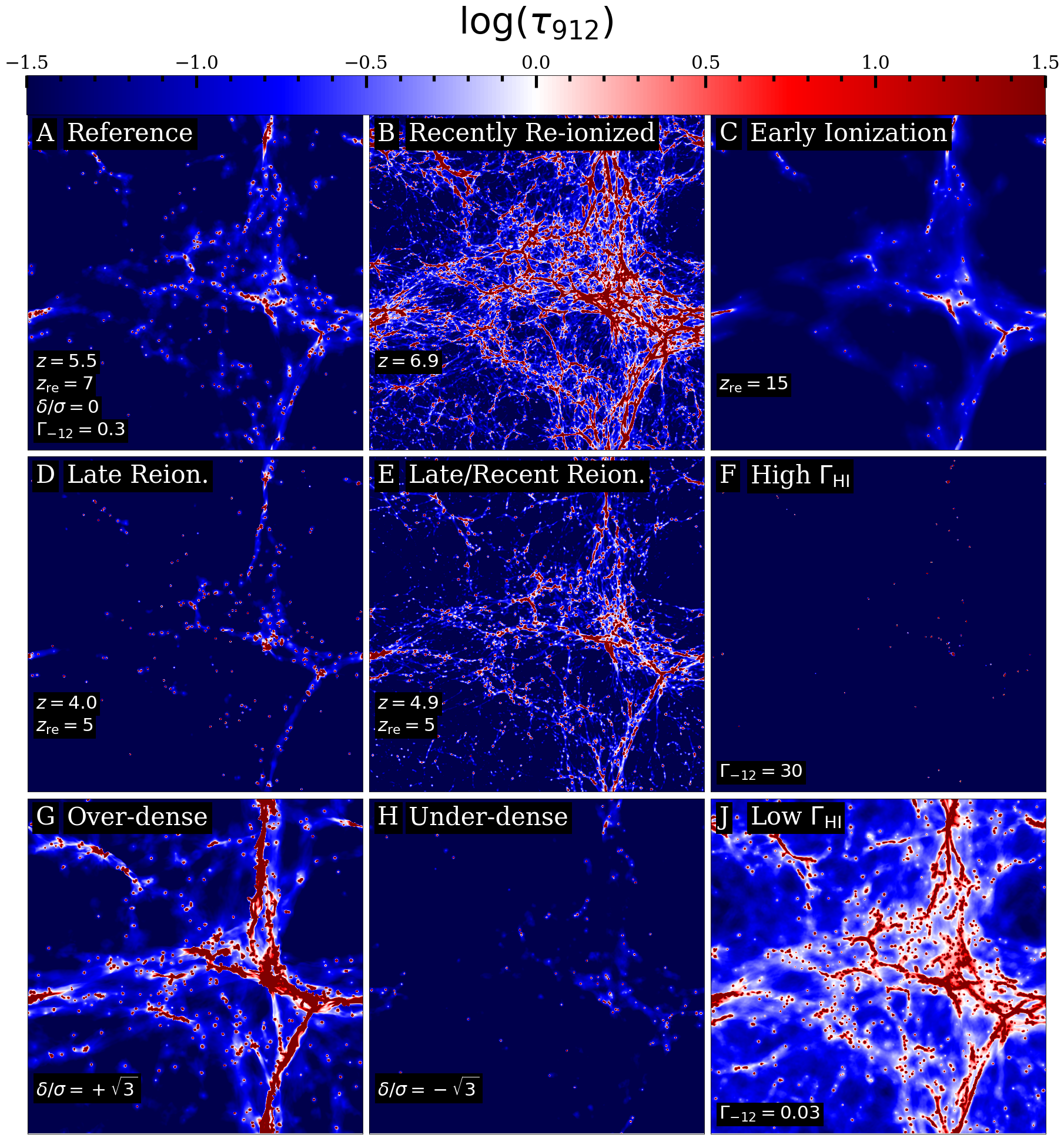}
    \caption{Visualization of self-shielding properties of the Core suite.  The maps show $\tau_{912}$ integrated along one axis of the box, with dark red regions intersecting self-shielding absorbers along the integrated axis.  Self-shielding depends on photo-evaporation of self-shielding systems (A and B), the redshift of reionization (C-E).  The strength of the ionizing background modulates the self-shielding density, with higher $\Gamma_{\rm HI}$ shrinking the sizes of absorbers (F and J).  The abundance of systems massive enough to self-shield increases dramatically with box-scale density (G and H).  }
    \label{fig:example_tau}
\end{figure}

Figure~\ref{fig:example_tau} shows the Lyman limit optical depth, $\tau_{912}$, integrated the length of the box along one axis, in the same format as Figure~\ref{fig:example_density}.  The color map is chosen such that the blue regions denote sightlines with $\tau_{912} < 1$, and the red regions denote sightlines with $\tau_{912} > 1$.  The dark red patches, with $\tau_{912} \gg 1$, intersect self-shielded structures along the integrated axis.  We see that $\sim 10$ Myr after ionization, the IGM is filled with tiny self-shielding structures (B), but most of these are gone a few hundred Myr later (A).  Photo-evaporation of neutral gas in the smallest mini-halos drives this process~\cite{Nasir2021}.  We see in panel C that fewer small self-shielded systems survive to $z = 5.5$ when the universe reionized early, reflecting the relative lack of small-scale structure seen in Figure~\ref{fig:example_density}.  In panel D, self-shielding systems are more compact and  fill a smaller fraction of the projected space than in A, and we see the same difference comparing B to E.  

When $\Gamma_{\rm HI}$ is very high (F), these regions shrink to the highest density peaks in the box, and are barely visible, while in the case with low $\Gamma_{\rm HI}$ (J), they expand considerably and fill a large fraction of the projected 2D space.  This reflects the strong dependence of the self-shielding density threshold on $\Gamma_{\rm HI}$ (see \S\ref{sec:opacity}) and the fact that self-shielding absorbers below this threshold survive much longer when $\Gamma_{\rm HI}$ is low.  The cosmic mean density drops towards lower redshifts, requiring larger over-densities to self-shield, shrinking the sizes of absorbers even as structure formation drives more gas to high densities.  In panels G and H, we see the effect of structure formation on self-shielding.  Higher densities and a larger abundance of massive structures results in abundant self-shielding systems in the over-dense box even hundreds of My after reionization, while the opposite occurs in the under-dense box.  

\subsection{Other parameters}
\label{subsec:other_physics}

\begin{figure}[h!]
    \centering
    \includegraphics[scale=0.25]{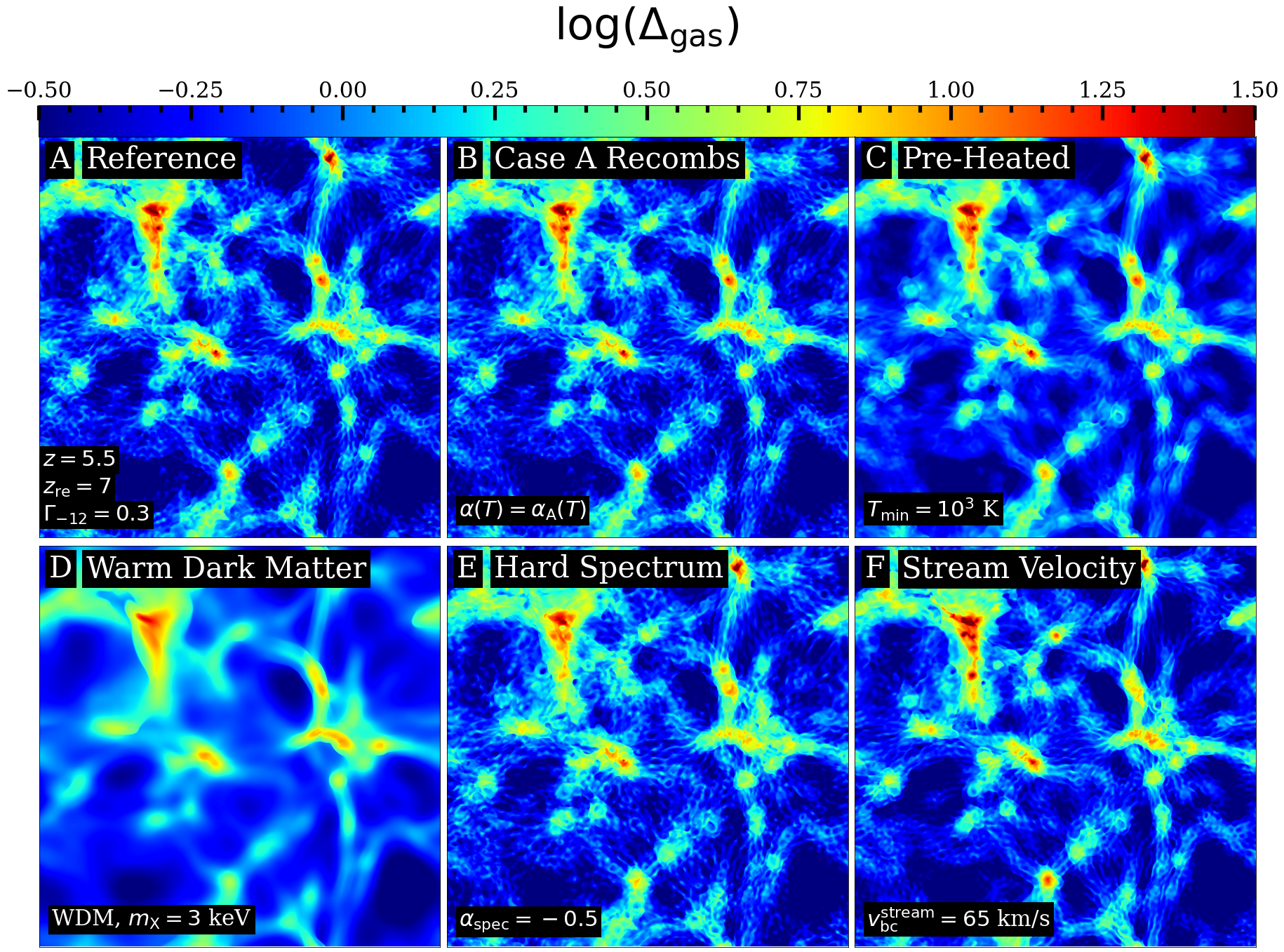}
    \caption{Effect of other physical parameters (rows 4-8 of Table~\ref{tab:simulation_summary}) on the gas density at $z = 5.5$.  Panel A is the same as that in Figure~\ref{fig:example_density}.  Assuming Case A recombinations has almost no effect on the density field (B) as does a non-zero streaming velocity (F).  X-ray pre-heating to $T = 10^3$ K (C) erases the smallest structures before reionization starts, and has a modest effect on the subsequent evolution.  Free-streaming of DM particles (D) prevents these structures from forming in the first place, and has a larger effect than pre-heating for our choice of parameters.  We see that the effects of a harder ionizing spectrum (E) and streaming velocities (F) are both small, similar to panel B.  }
    \label{fig:example_density_phase_2}
\end{figure}

\begin{figure}[h!]
    \centering
    \includegraphics[scale=0.25]{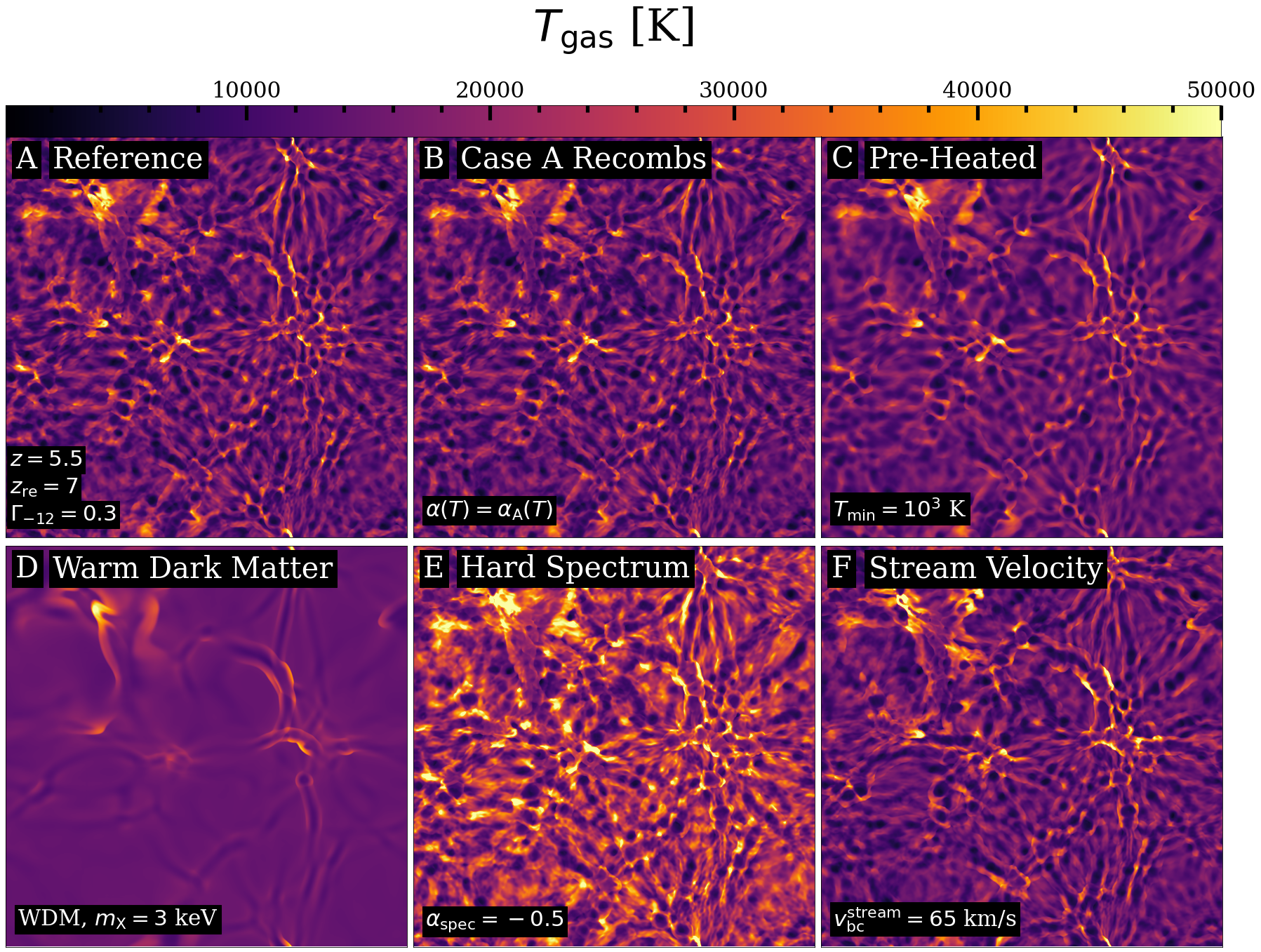}
    \caption{Gas temperature for the same snapshots shown in Figure~\ref{fig:example_density_phase_2}.  As expected, the models with case A recombinations (B) and streaming velocity (F) are very similar to the fiducial case, and even the X-ray heated model (C) displays only modestly less small-scale thermal structure.  The WDM model (D) shows considerably less evidence of compression heating around expanding filaments, indicating that the complex thermal structure in our simulations is a characteristic feature of small-scale DM power.  A harder ionizing spectrum leads to more heating (E), but the small-scale features of the temperature map are similar to the other CDM models.  }
    \label{fig:example_temperature_phase_2}
\end{figure}

\begin{figure}[h!]
    \centering
    \includegraphics[scale=0.25]{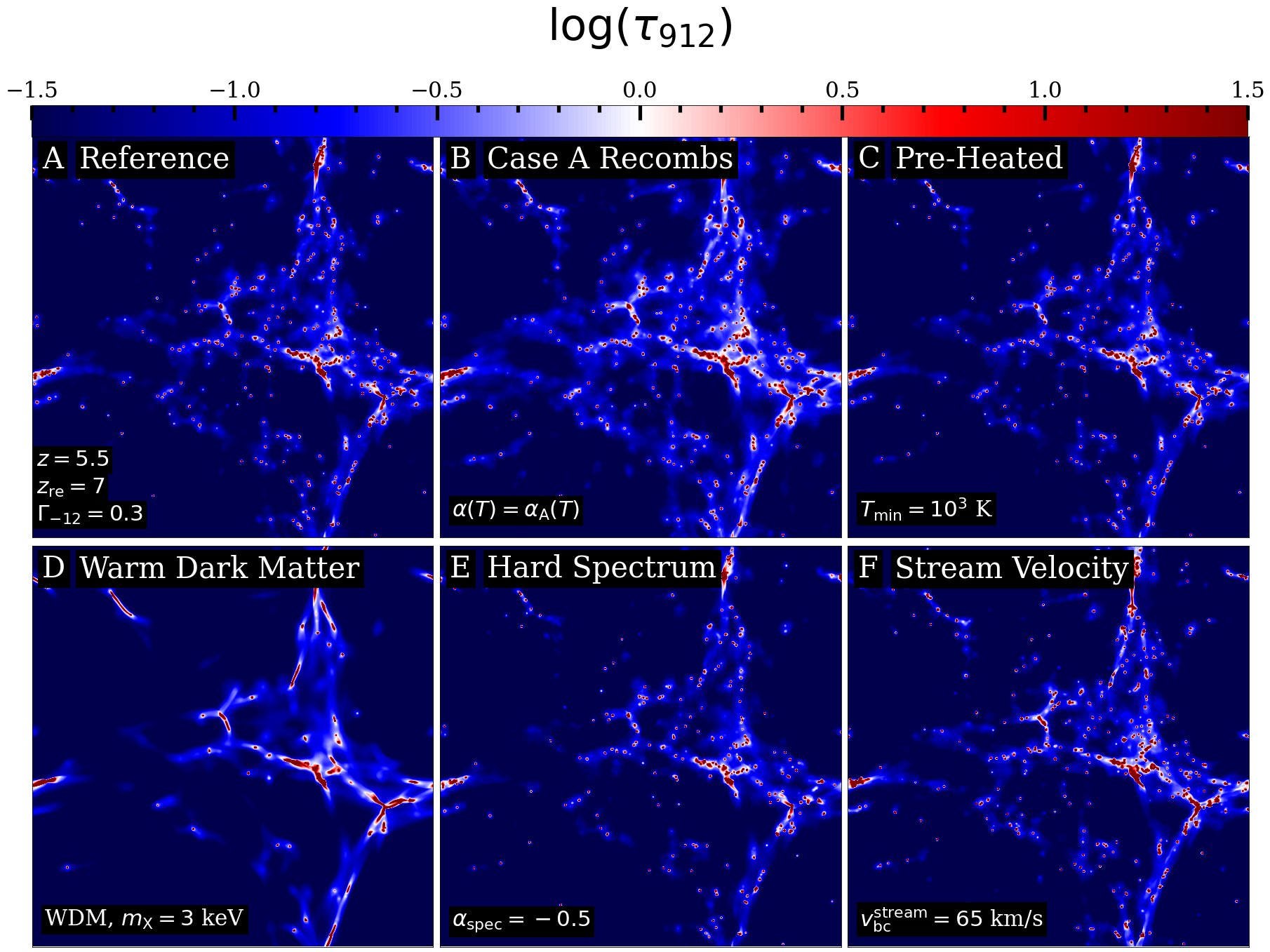}
    \caption{Self-shielding properties of the simulations in Figure~\ref{fig:example_density_phase_2}, visualized in the same way as in Figure~\ref{fig:example_tau}.  See text for details.  }
    \label{fig:example_tau_phase_2}
\end{figure}

In this section, we visualize the effects of other physical parameters in \textsc{saguaro} (see Table~\ref{tab:simulation_summary}) on density, temperature, and self-shielding properties.  Figure~\ref{fig:example_density_phase_2} shows slices through selected $2$ $h^{-1}$Mpc simulations outside the Core suite.  All models have $z_{\rm re} = 7$, $\Gamma_{-12} = 0.3$, $\delta/\sigma = 0$, and are shown at $z = 5.5$.  Panel A is identical to the same panel in Figure~\ref{fig:example_density}.  Respectively, the remaining panels show models that assume the case A recombination rate (B), X-ray heating to $T_{\min} = 10^3$K (C), a WDM particle mass of $m_{\rm X} = 3$ keV (D), an ionizing spectral slope of $\alpha_{\rm spec} = -0.5$ (E), and a stream velocity of $v_{\rm bc} = 65$ km/s (F).  We keep the same layout for Figure~\ref{fig:example_temperature_phase_2} (for temperature) and Figure~\ref{fig:example_tau_phase_2} (for self-shielding).  

The effects of assuming the case A recombination rate on the density field are almost negligible.  Only on close inspection can differences be seen between panels A and B - in the latter, the densest clumps are slightly more compact. This is due to the slightly higher self-shielding threshold when assuming case A.  However, the effects of order-of-magnitude changes in self-shielding properties on the density field were already seen to be modest (panels F and J in Figure~\ref{fig:example_density}), and the difference here is only a factor of $\approx 1.6$.  The effects of X-ray heating (C) are more prominent, but still modest.  The density field is smoother at the smallest scales and displays a less prominent interference pattern.  Pre-heating raises the Jeans scale and smooths out small-scale structure before reionization starts, resulting in less pressure smoothing later.  

The effect of DM free-streaming parameterized by our WDM model (panels D) is qualitatively similar to that of X-ray pre-heating, but more extreme given the rather low value of $m_{\rm X} = 3$ keV that we show here.  As in panel C, the large-scale structure of the density field remains the same, but the small-scale interference pattern is absent.  Without any dark matter mini-halos to grow them, baryonic structures below the free-streaming scale never form in the first place, and thus are not destroyed when reionization occurs (see Refs.~\cite{Cain2022a} and~\cite{Davies2023} for similar results).  Panel E shows that the effect of a harder ionizing spectrum has a small effect on the density structure, similar to the effect of assuming the case A recombination rate in panel B.  In panel F, we see that the effect of streaming velocities is small\footnote{Two effects in the $v_{\rm bc}$ simulations obscure the small differences that may be present due to extra pressure smoothing.  First, these simulations are initialized at $z = 1080$ rather than $300$, producing a small difference in early structure formation.  Second, the slice is cut normal to the direction of the stream velocity, such that baryonic structures are slightly shifted relative to the $v_{\rm bc} = 0$ case.  } compared to those of X-ray pre-heating and DM free-streaming, reflecting the findings of Ref.~\cite{Cain2020,Long2022}.  Structures small enough to be affected by $v_{\rm bc}$ are destroyed before $z = 5.5$.  

Figure~\ref{fig:example_temperature_phase_2} shows temperature for the same simulations as Figure~\ref{fig:example_density_phase_2}. As we might expect, panels A, B, and F are all very similar, since case A recombinations and streaming velocities have little effect on the gas dynamics (at least, several hundred Myr after $z_{\rm re}$).  As with density, X-ray pre-heating (C) results in a modest difference in the amount of small-scale structure in the $T$ field, with the difference between compression-heated and expansion-cooled regions being less extreme~\citep{Cain2024a}.  This effect is more dramatic in panel D, more so than one might expect from the differences in density structure in Figure~\ref{fig:example_density_phase_2}.  Indeed, there is comparatively little small-scale thermal structure in D.  The mini-halos and filaments that source small-scale $T$ fluctuations in response to reionization heating are suppressed in the WDM case.  Their absence means that small-scale $T$ fluctuations do not form after reionization, rendering temperature a much more well-behaved function of density~\citep{McQuinn2016}.  These findings suggest that the complex thermal structure seen in our fiducial simulations is a direct consequence of small-scale DM structure present in the CDM cosmology.  Modeling this accurately could enhance the utility of the Ly$\alpha$ forest, which is sensitive to small-scale thermal structure, as a means of constraining alternative DM cosmologies.  In Panel E, we see (as expected) that a harder ionizing spectrum results in a significantly hotter IGM, although the small-scale features of the thermal structure remain similar to those in the other CDM models.  

In Figure~\ref{fig:example_tau_phase_2}, we see that the self-shielding properties of the IGM are reasonably similar in all these scenarios.  We see the effect of case A recombinations (panel B) most clearly.  Self-shielded structures are slightly larger than they are in panel A, reflecting the higher self-shielding density in the Case A run.  Interestingly, there is little difference between C and A, suggesting that self-shielding systems massive enough to survive pressure smoothing are those that are unaffected by X-ray pre-heating.  The same seems to be true of streaming velocities in panel F.  The WDM model (D) differs only in the abundance of the smallest self-shielders, reflecting the similarity of their density fields on large scales.  The $\alpha_{\rm spec} = -0.5$ model (E) has slightly lower opacity than the others, owing to the increased gas temperature.  

\subsection{Visualizing the HR Simulations}
\label{subsec:HR}

\begin{figure}[h!]
    \centering
    \includegraphics[scale=0.25]{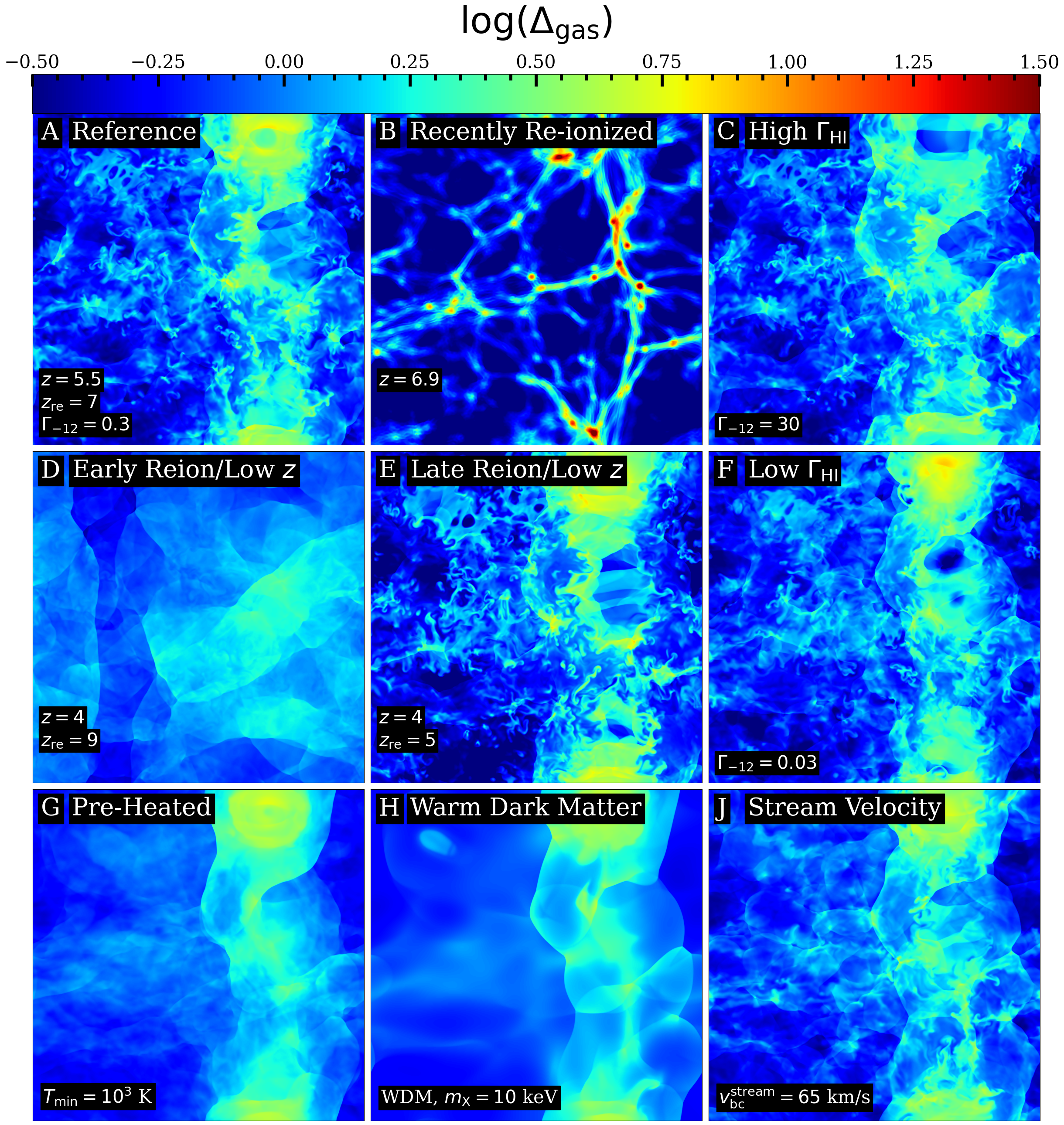}
    \caption{Visualization of the density field in some of the HR simulations.  Pressure smoothing evacuates nearly all the gas from halos and filaments by $z = 5.5$ in our fiducial simulation (panels A and B), and drives the formation of turbulent eddies that fill much of the volume.  Panels C and F show the effect of $\Gamma_{\rm HI}$, and panels D and E compare the gas at $z = 4$ for $z_{\rm re} = 9$ and $5$, respectively.  The turbulent eddies have almost completely dissipated in the former, since the force driving them has disappeared, are still very prominent in panel E ($z_{\rm re} = 5$).  Panels G, H, and J show the effects of X-ray pre-heating, warm dark matter, and streaming velocities, respectively.  All of these have the effect of erasing or suppressing turbulence.  }
    \label{fig:example_density_small_boxes}
\end{figure}

In this section, we will visualize the gas properties in the ``HR'' simulations (rows 3 and 9-11 of Table~\ref{tab:simulation_summary}).  These have $8^3$ higher spatial resolution and volumes the same factor smaller than the larger \textsc{saguaro} boxes, and are all run at the mean density ($\delta/\sigma = 0$).  Figure~\ref{fig:example_density_small_boxes} shows the density field for a representative subset of these.  

As in the previous sections, panels A and B show the density field at $z = 5.5$ and $6.9$ in the reference simulation.  Even at $z = 6.9$, we can see an interference pattern starting to form around the filaments and mini-halos as their gas begins expanding.  The more rapid appearance of this pattern reflects the smaller distances scales being resolved.  In panel A, nearly all the structures in the box have been destroyed by pressure smoothing, since the box is too small to host halos large enough to survive relaxation.  Remarkably, the density field shows distinct small-scale turbulent eddies, with slightly over-dense gas mixing with mildly under-dense gas where the filaments used to be.  This result is unexpected because standard means of generating turbulence in the IGM require effects related to galaxy formation and evolution~\citep[e.g.][]{Balsara2004,Kulsrud2008}.  However, in our case, the effect is driven entirely by the back-reaction of IGM small-scale structure to reionization heating.  The eddies are formed due to the pressure imbalance between hot, slightly over-dense and cold, under-dense gas as filaments expand and over-lap with each other.  We observe them only in the HR simulations, which have marginally high enough resolution to begin to capture them.  The physics driving the formation of these eddies and their potential implications are explored in detail in Ref.~\cite{Cain2025b}.  

%[\chris{paragraph about the gamma dependence in C and F}]\\
As in Figure~\ref{fig:example_density}, we show the effects of increasing and decreasing $\Gamma_{\rm HI}$ in panels C and F, respectively.  The gas distribution is mostly similar,  with modest differences in the large filamentary structure filling much of the right half of the slice.  The over-density in the top right of each panel is more (less) diffuse in the run with higher (lower) $\Gamma_{\rm HI}$.  There is also a ``hole'' in the density field slightly below this in panel F that is absent in the other panels.  These differences correspond to variation in the evaporation rates of the dense clumps (visible in panel B) that merge to form the diffuse filament.  This shows that self-shielding has an important effect on gas dynamics at small scales near evaporating mini-halos, echoing previous findings~\citep{Shapiro2004,Iliev2005,Chan2023}.  

\begin{figure}[h!]
    \centering
    \includegraphics[scale=0.25]{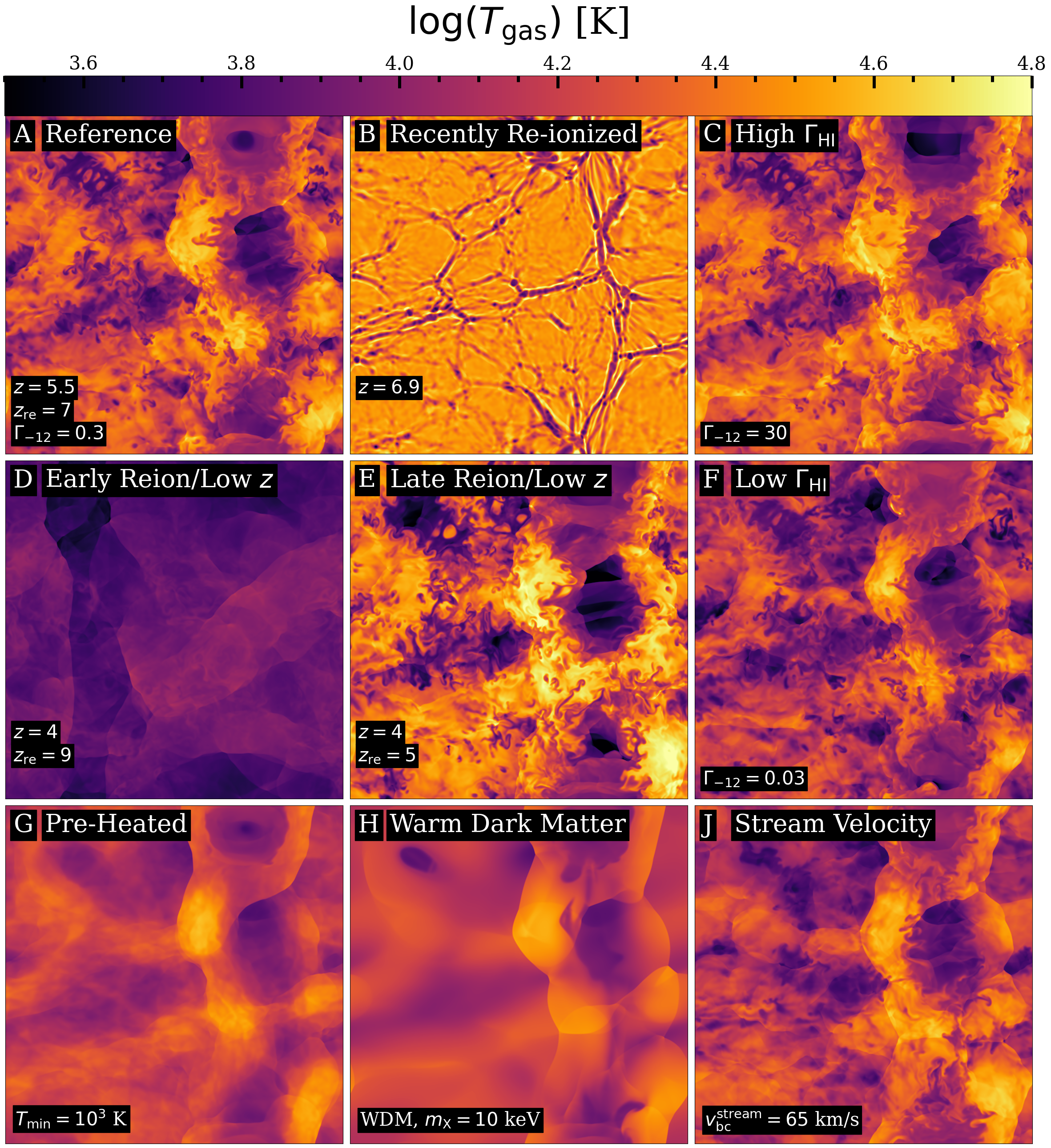}
    \caption{Visualization of the temperature for the same HR snapshots in Figure~\ref{fig:example_density_small_boxes}.  In all cases, the turbulent eddies form along the boundary between gas phases that are different by a factor of a few in temperature.  Hotter (colder) gas corresponds to under (over)-dense gas, respectively, in Figure~\ref{fig:example_density_small_boxes}.  } 
    \label{fig:example_temperature_small_boxes}
\end{figure}

Panels D and E compare the fate of turbulent eddies at $z = 4$, the lowest redshift we simulated, in runs with $z_{\rm re} = 9$ and $5$, respectively.  In the former, there is no sign of eddies.  Since pressure smoothing is their driving force, the eddies begin to dissipate after the IGM reaches a new pressure equilibrium, eventually fading away as the universe expands.  Since the pressure smoothing timescale shortens with increasing redshift, it's likely that turbulence in regions that re-ionized early does not persist past the end of reionization.  However, for $z_{\rm re} = 5$, the eddies are more prominent than in panel A, suggesting that (all else being the same) turbulence may persist the longest in the last regions to re-ionize.  

\begin{figure}[h!]
    \centering
    \includegraphics[scale=0.25]{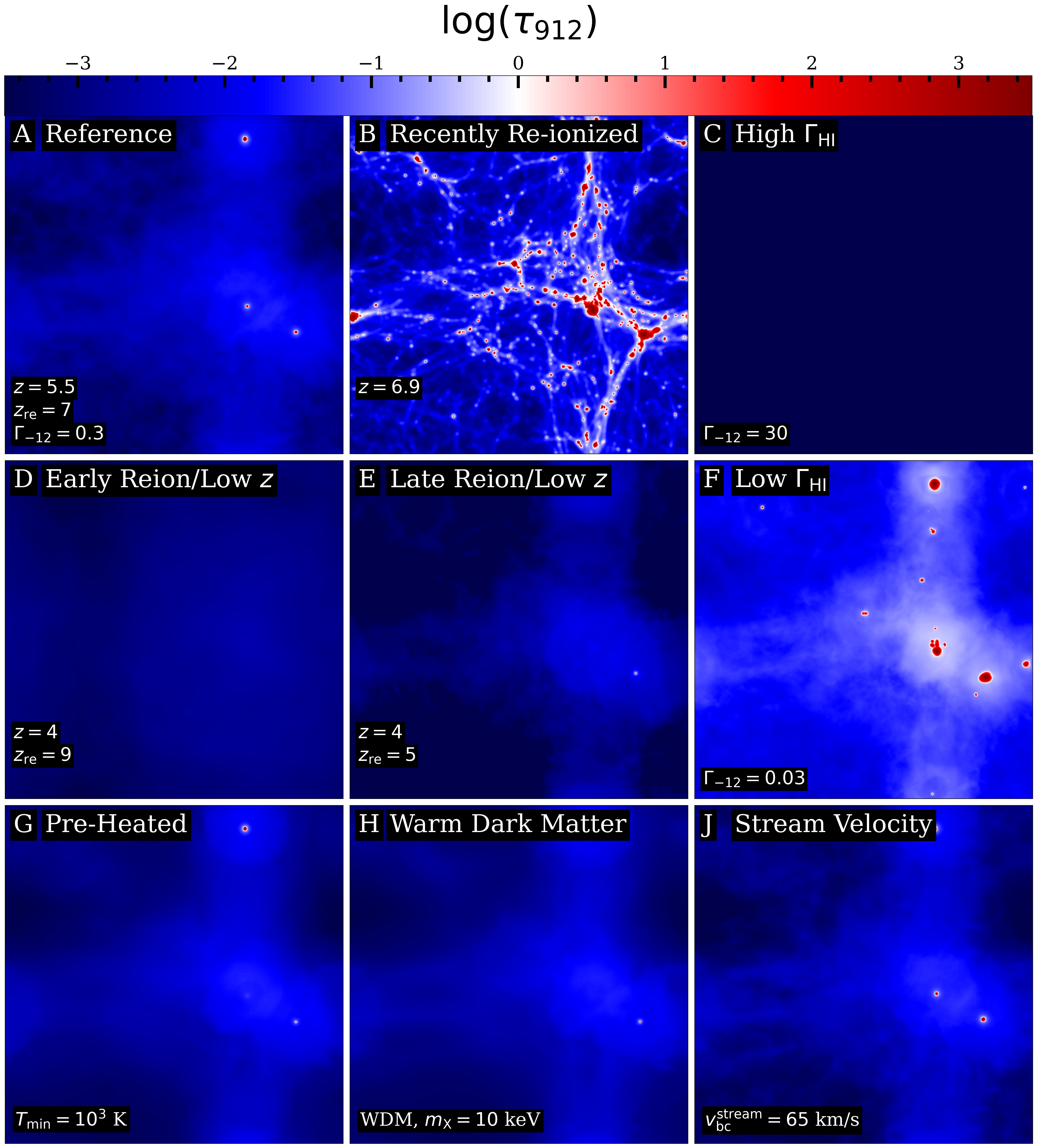}
    \caption{Self-shielding in the HR simulations.  While there is significant self-shielding shortly after $z_{\rm re}$ (panel B), there is almost none in the others except in Panel F, where the ionizing background is low enough to allow a few self-shielding systems to survive for $\sim 300$ Myr.  } 
    \label{fig:example_tau_small_boxes}
\end{figure}

The bottom rows show the same snapshot as A, but with $T_{\min} = 10^3$ K (G), $m_{\rm X} = 10$ keV (H) and $v_{\rm bc}^{\rm stream} = 65$ km/s (J).  We see that all three of these effects suppress or even prevent the formation of turbulence, since they all either prevent the smallest gas structures from forming to begin with (WDM and streaming velocities) or destroy them before $z_{\rm re}$ (pre-heating).  As we noted in Ref.~\cite{Cain2025b}, the formation of turbulence in the diffuse IGM requires DM sub-structure down to sufficiently small scales, a stream velocity close to or below the RMS average, and that the IGM is sufficiently cold prior to reionization.  The last condition is likely to prevent turbulence from forming in a significant fraction of the universe, as much of the neutral IGM likely reaches at least a few hundred K by the time reionization is half over at $z \sim 7-8$.

Figure~\ref{fig:example_temperature_small_boxes} shows the temperature in the HR simulations.  We see in many of the panels that the turbulence corresponds to mixing of gas phases a factor of a few different in temperature, with the slightly cooler gas corresponding to the denser gas in Figure~\ref{fig:example_density_small_boxes}.  This shows that turbulent features correspond to mixing of recently compression-heated low-density gas at the edges of expanding filaments with expansion-cooled low-density gas within the filaments.  Thus, it seems that turbulent eddies form only when expanding filaments overlap, at least for the spatial resolution we are able to reach.  This explains why $\Gamma_{\rm HI}$ has only a modest impact on the turbulence (C and F) and lower $z_{\rm re}$ generates the most prominent turbulence (E).  The pressure smoothing of filaments and clumps, except those dense enough to self-shield, proceeds the same way in runs with different $\Gamma_{\rm HI}$.  At lower $z_{\rm re}$, more dense structures form in the cold, neutral IGM, driving more violent pressure smoothing.  In the runs that are missing the smallest structures (G-J), the structures that are present are too far apart to overlap with each other before pressure-smoothing is complete, preventing eddies from forming.

Finally, Figure~\ref{fig:example_tau_small_boxes} shows $\tau_{912}$ through the HR boxes.  In contrast to Figure~\ref{fig:example_tau}, in most of the panels we see little or no self-shielding, despite the presence of such systems in panel B.  The only exception is panel F, which has the lowest $\Gamma_{\rm HI}$ and allows a few of the most massive clumps to remain self-shielded for several hundred Myr.  In all other cases, none of the halos in the box are massive enough to hold onto dense gas after reionization.  We also see that the turbulent features visible in the Figure~\ref{fig:example_density_small_boxes}-\ref{fig:example_temperature_small_boxes} slices are absent from the integrated $\tau_{912}$ plots, which shows that the small density fluctuations that drive turbulence do not contribute meaningfully to the IGM opacity.  

\subsection{Thermodynamics of the IGM}
\label{subsec:eq_of_state}

\subsubsection{Global averages}
\label{subsubsec:global_averages}

\begin{figure}[h!]
    \centering
    \includegraphics[scale=0.15]{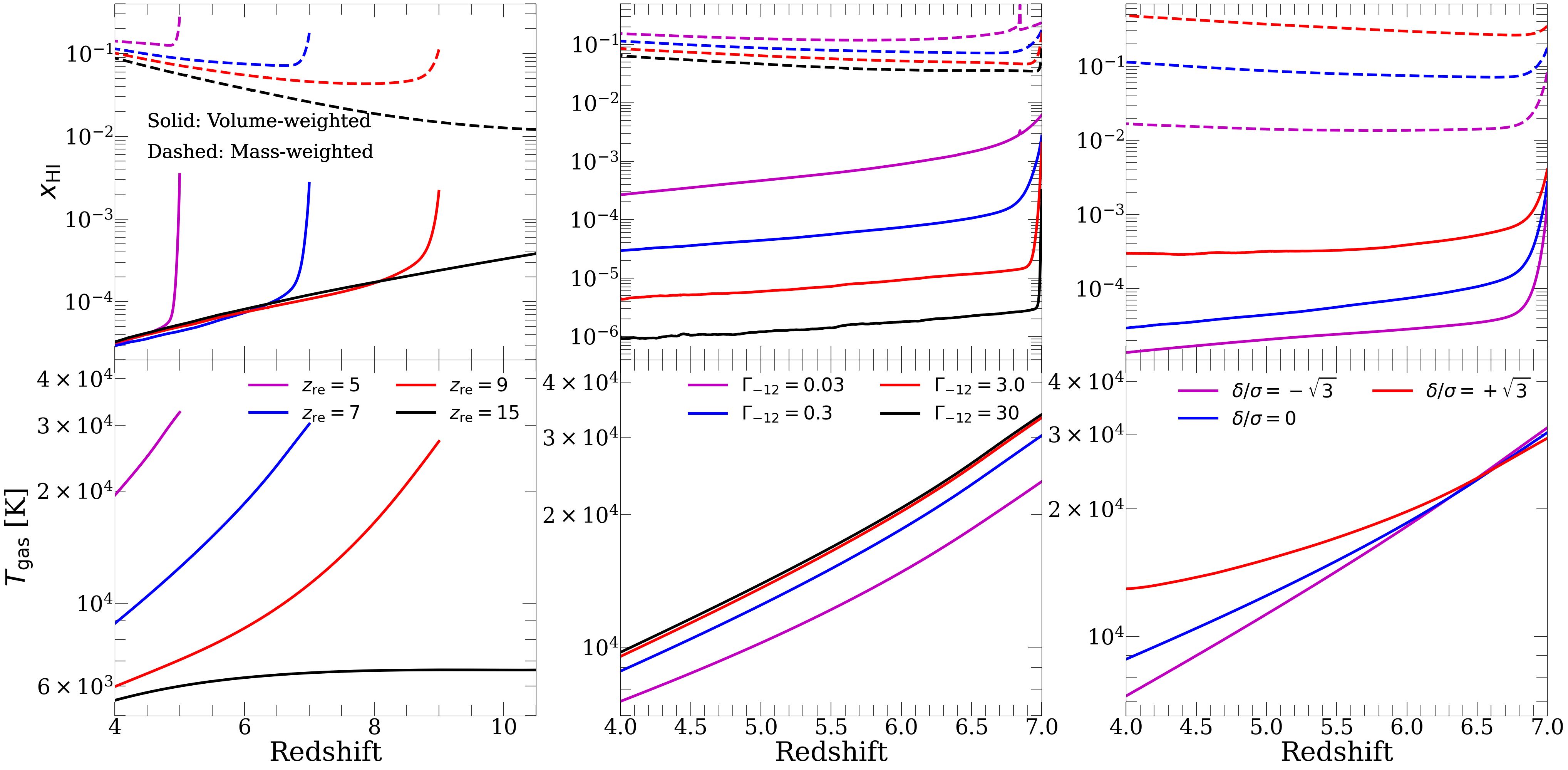}
    \caption{Evolution of box-averaged quantities in the Core suite.  The top and middle rows show $x_{\rm HI}$ and $T_{\rm gas}$, respectively.  From left to right, the columns vary $z_{\rm re}$, $\Gamma_{\rm HI}$, and $\delta/\sigma$ one at a time (see text).  Solid and dashed curves denote volume and mass-weighted averages, respectively.  The volume-weighted average $x_{\rm HI}$ is mostly insensitive to $z_{\rm re}$, but is affected significantly by both $\Gamma_{\rm HI}$ and $\delta/\sigma$.  The mass-weighted $x_{\rm HI}$, which is dominated by gas in self-shielded systems, is mildly sensitive to both $z_{\rm re}$ and $\Gamma_{\rm HI}$, but depends most strongly on $\delta/\sigma$.  The volume-averaged $T_{\rm gas}$ is most sensitive (at fixed $z$) to $z_{\rm re}$, since this determines how long the gas has had to cool after being heated by I-fronts.  It is modestly sensitive to $\Gamma_{\rm HI}$, which sets $T_{\rm reion}$, and is sensitive to $\delta/\sigma$ at a similar level.  }
    \label{fig:fractions_T}
\end{figure}

In this section, we quantify the evolution of the temperature and ionization state of the IGM.  We start with several globally-averaged quantities in Figure~\ref{fig:fractions_T}.  The top row shows the mean neutral fraction vs. redshift, with solid and dashed lines denoting volume and mass-weighted averages, respectively.  We show the volume-averaged $T$ in the same format in the bottom row\footnote{We omit the mass-averaged temperature here because it is very noisy - see Appendices~\ref{app:RSLA}-\ref{app:subcycle} for further details on the reasons for this.  }.  Our reference simulation has $z_{\rm re} = 7$, $\Gamma_{-12} = 0.3$ and $\delta/\sigma = 0$ (the blue curve in every panel), and we vary each parameter one at a time going from left to right.  

In the top-left panel, we see that the volume-averaged neutral fraction evolves rapidly just after $z_{\rm re}$, and quickly settles on the steady $\propto (1+z)^3$ evolution expected from the expansion rate, independent of $z_{\rm re}$.  The mass-averaged $x_{\rm HI}$, however, is much higher and retains memory of $z_{\rm re}$ even down to $z = 4$.  It also gradually increases towards lower redshifts after an initial decline near $z_{\rm re}$ as more gas accretes onto the most massive halos in the box, which are able to self-shield well after reionization.  As Figure~\ref{fig:example_density} showed, more structures form and survive to lower redshift when the universe re-ionizes later, and these hold on to more self-shielded neutral gas.  In the top-middle panel, we see that the volume-average $x_{\rm HI}$ scales strongly - almost linearly - with $\Gamma_{\rm HI}$, but the dependence of the mass-averaged $x_{\rm HI}$ is much weaker.  This is because a significant fraction of the self-shielded mass is retained at extremely high densities that can remain neutral even in the presence of a very strong ionizing background.  The top right shows that $x_{\rm HI}$ depends most strongly on the local over-density.  Around $30-40\%$ of the mass in the over-dense box is neutral, most of which is concentrated in a small number of the most massive halos.  We note that the mass-weighted HI fraction would behave somewhat differently if galaxy formation were included in our simulations, since the most massive halos in our box would be partially self-ionized by internal stellar radiation.  

The volume-weighted $T$ is most sensitive to $z_{\rm re}$ (bottom left) and modestly sensitive to $\Gamma_{\rm HI}$ and $\delta/\sigma$.  The gas is impulsively heated to $T_{\rm reion}$ at $z_{\rm re}$, and cools afterwards mostly via cosmic expansion and Compton cooling~\citep{McQuinn2016}.  At a fixed $z$, $T$ is higher the more recently the gas was ionized.  Simulations with higher $\Gamma_{\rm HI}$ have higher $T_{\rm reion}$ owing to the faster speeds of the I-fronts, but this effect seems to saturate at our highest $\Gamma_{\rm HI}$ values.  Simulations with different densities start out at similar $T_{\rm reion}$, but cool more slowly (quickly) due to the slower (faster) local expansion rate in over-densities (voids).  Note that we do not show the mass-weighted temperature here because it is noisy, likely due to the effects of shock-heating and collisional ionizations in a tiny fraction of cells with extremely high over-densities (see Appendix~\ref{app:subcycle}).

\subsubsection{Temperature-density relation}
\label{subsubsec:TDR}

\begin{figure*}[hbt!] %0.185
    \centering
    \includegraphics[scale=0.175]{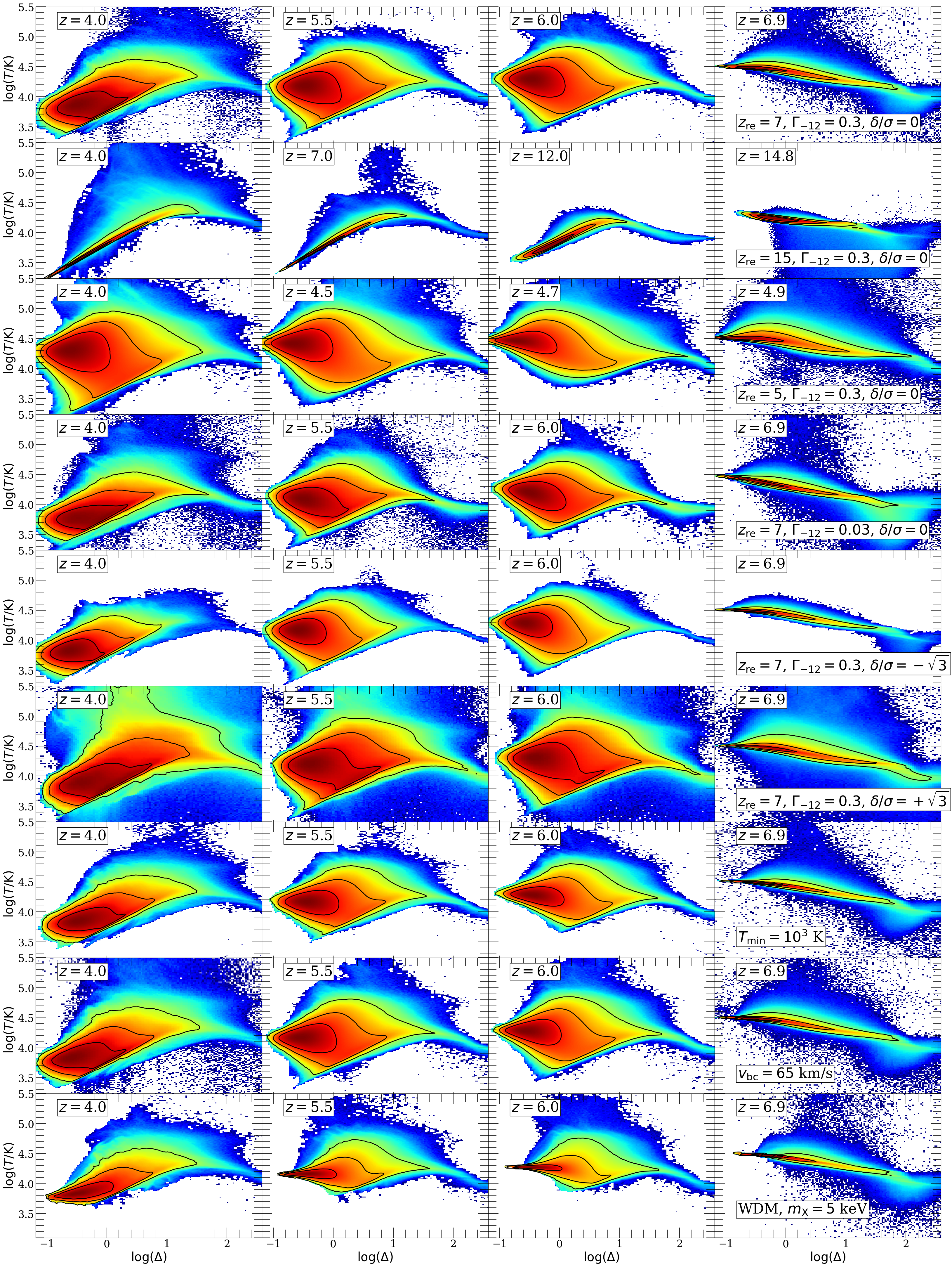}
    \caption{Temperature-density phase diagrams for a subset of our $2$ $h^{-1}$Mpc simulations.  Each row shows a different simulation and the columns show different redshifts, increasing from left to right.  The black lines denote $1$, $2$, and $3\sigma$ contours.  The simulation parameters for each row are denoted in the right-most column.  See text for details.  }
    \label{fig:phase_diagrams_L2}
\end{figure*}

We have already studied the temperature-density relation (TDR) in several \textsc{saguaro} simulations in Ref.~\cite{Cain2024a} - in this section, we extend that study and include our HR simulations.  Figure~\ref{fig:phase_diagrams_L2} shows the TDR for a subset of $2$ $h^{-1}$Mpc runs.  Each row shows a different simulation, indicated by the labels at the far right, and redshift increases from left to right.  The left-most panels all show $z = 4$, and the right-most show snapshots shortly after $z_{\rm re}$.  Each panel shows $\log(T)$ vs.\ $\log(\Delta)$ - the color-scale is logarithmic, and the concentric black lines denote $1$, $2$, and $3\sigma$ contours.  

The top row is our fiducial Core simulation, which is also the fiducial high-resolution run studied in Ref.~\cite{Cain2024a}.  The gas is almost isothermal at $z = 6.9$ (panel B of Figure~\ref{fig:example_temperature}), but by $z = 6$, the TDR has broadened, spanning an order of magnitude in $T$ close to the mean density.  This broadening arises because the pressure-smoothing of filaments and mini-halos drives adiabatic cooling in over-densities and compression-heating at lower densities, as described in Ref.~\cite{Cain2024a}.  We note that this effect is distinct from the broadening of the global $T-\Delta$ relation due to patchy reionization on much larger scales~\citep[e.g.][]{DAloisio2015,Keating2018}.    A power law shape emerges by $z = 4$, but still with $\approx 0.5$ dex of scatter.  In the 2nd row, we see that if the gas re-ionizes early enough ($z_{\rm re} = 15$), the wide scatter seen in the fiducial case never appears, and the TDR remains a tight power law down to $z = 4$.  For $z_{\rm re}$ this high, the IGM has formed much less small-scale structure by the time it ionizes and the cooling time-scale is very short, so it reaches thermal equilibrium quickly.  The opposite is true for $z_{\rm re} = 5$ (third row) which shows the most scatter at $z = 4$.  In the 4th row, we see that a factor of $10$ lower $\Gamma_{\rm HI}$ has a small effect on the TDR, acting mainly to lower the initial temperature slightly, but without changing the subsequent qualitative behavior.  

The fifth and sixth rows show the under and over-dense simulations, respectively.  The scatter in the TDR increases with $\delta/\sigma$ because more massive structures are being destroyed by pressure smoothing.  At $z = 6$ and $5.5$ in the over-dense run, we see hints of a bimodal distribution, with peaks in the TDR slightly below and above the mean density (at $\sim 20,000$ K and $\lesssim 10,000$ K, respectively).  By $z = 4$, a third ``hot'' phase forms at $T > 10^5$~K and slightly over-dense gas, which might correspond to shock-heated outflows around the most massive structures in the box.  The last three rows show runs with $T_{\min} = 10^3$ K, $v_{\rm bc} = 65$ km/s, and $m_{\rm X} = 5$ keV.  As we showed in Ref.~\cite{Cain2024a}, pre-heating reduces the scatter in the TDR modestly, but preserves the same qualitative behavior.  Streaming velocities have an even smaller effect, as seen in Figure~\ref{fig:example_temperature}.  WDM has a substantial effect, flattening the TDR at the low-density end and reducing the scatter considerably for mildly over-dense gas.  

\begin{figure}[h!]
    \centering
    \includegraphics[scale=0.195]{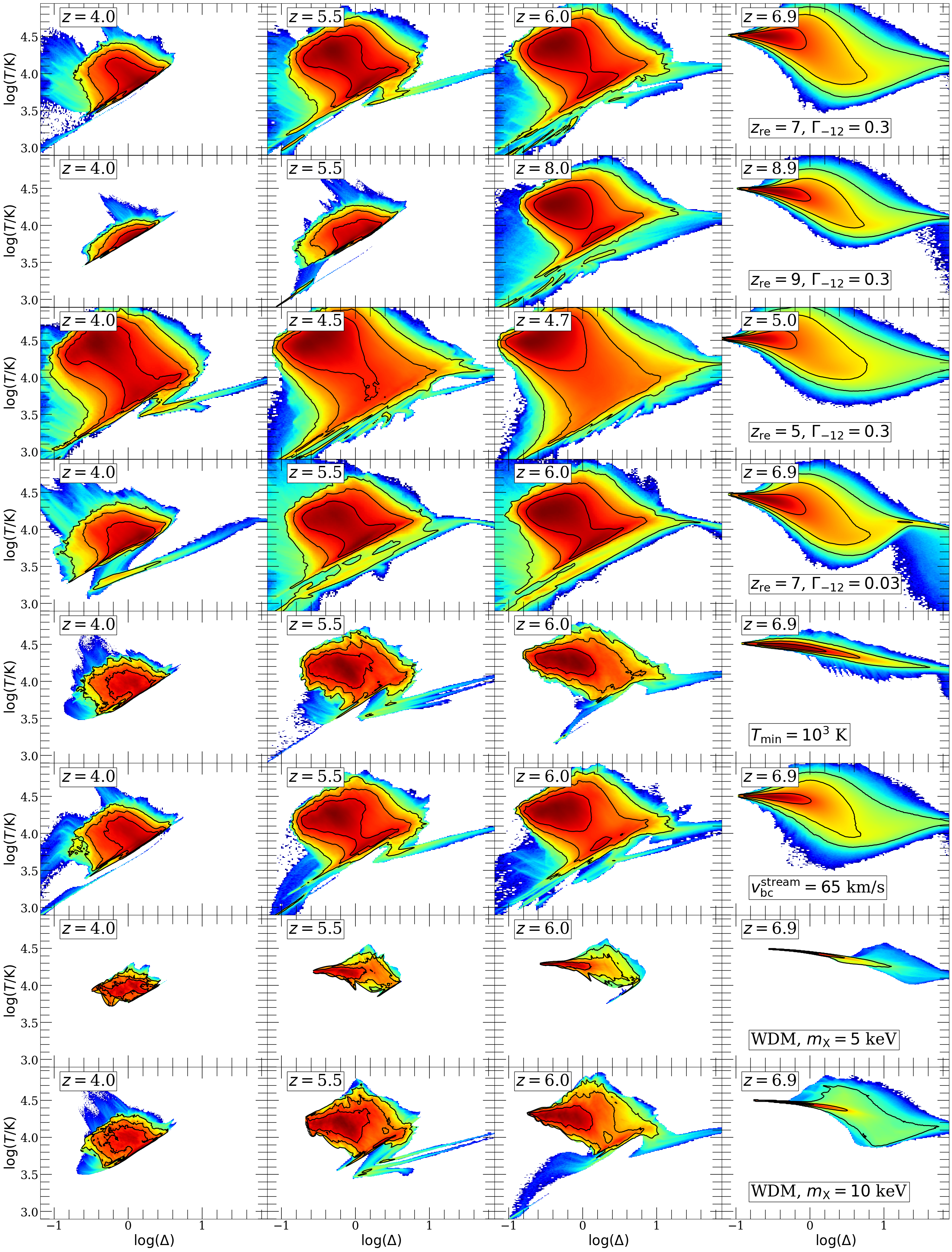}
    \caption{TDR for some of the HR simulations, in the same format as Figure~\ref{fig:phase_diagrams_L2}.  The scatter just after $z_{\rm re}$ is much larger in our fiducial run ($\Gamma_{-12} = 0.3$, $z_{\rm re} = 7$, first row) since the expansion of filaments is resolved at earlier times.  The TDR is noticeably bimodal, resulting in the turbulent mixing of gas phases seen in Figure~\ref{fig:example_temperature_small_boxes}.  Models with X-ray heating, stream velocities, and WDM show less scatter and a lack of (or suppressed) bimodal distribution.  See text for details.  }
    \label{fig:phase_diagrams_L025}
\end{figure}

We show the TDR for a selection of HR simulations in Figure~\ref{fig:phase_diagrams_L025}, with our fiducial Core-HR run in the top row.  Unlike in the $2$ $h^{-1}$Mpc boxes, a large scatter emerges by $z = 6.9$ because higher resolution allows us to resolve expansion cooling of filaments much closer to $z_{\rm re}$.  By $z = 6$ (and $5.5$), the gas has formed a clear bimodal distribution, similar to the trend seen in the over-dense run in Figure~\ref{fig:phase_diagrams_L2}.  Indeed, it is these two distinct gas phases that we see turbulently mixing in Figure~\ref{fig:example_density_small_boxes}-\ref{fig:example_temperature_small_boxes}.  There is also a tail extending down to $\Delta \sim 0.1$ and $T \sim 1000$ K, which may correspond to the extreme under-densities formed in the aftermath of mini-halo photo-evaporation (see Ref.~\cite{Shapiro2004} and Appendix B of Ref.~\cite{Cain2025b}).  By $z = 4$, there are no high over-dense structures left in the box, and the TDR has collapsed to a small region of phase space, but still with significant scatter.  The lack of over-dense structures by this redshift is a limitation of the tiny box size - as we saw in Figure~\ref{fig:phase_diagrams_L2}, larger volumes can capture halos massive enough to hold onto their gas to this redshift.  

The second and third rows show HR run with $z_{\rm re} = 9$ and $5$, respectively.  Both display the same qualitative behavior seen in the top row.  For $z_{\rm re} = 9$, the bimodal structure has vanished by $z = 5.5$, and the TDR has collapsed to a half-dex range in both $\Delta$ and $T$ by $z = 4$.  As shown in Figure~\ref{fig:phase_diagrams_L2}, for $z_{\rm re} = 5$ the scatter in the TDR is larger, and by $z = 4$ a bimodal structure has begun to emerge. In the 4th row, we see that a run with $\Gamma_{-12} = 0.03$ behaves similarly to the fiducial case, with one exception: a small amount of gas remains with $\Delta > 10$ even at $z = 4$, and lies on a fairly tight power-law relation shifted $\approx 0.3$ dex in $T$ below the rest of the gas in the $\log{\Delta}-\log{T}$ plane.  The persistence of over-dense gas for longer is likely due to self-shielding, and reflects the longer timescale for the handful of self-shielding structures in our tiny volume to photo-evaporate.   This gas cooled due to hydrodynamic expansion significantly later than the rest of the dense gas that was originally in filaments, and thus lives in its own part of phase space.  

In the last four rows, we show runs with $T_{\min} = 10^3$K, $v_{\rm bc}^{\rm stream} = 65$ km/s, $m_{\rm X} = 5$ keV, and $m_{\rm X} = 10$ keV.  The differences we see here with the fiducial run are much larger than the ones seen in Figure~\ref{fig:phase_diagrams_L2}.  The HR runs not only resolve tiny structures that are missing in the $2$ $h^{-1}$Mpc boxes, which are more sensitive to these effects, but are also missing most of the large structures that are insensitive to them.  In the $T_{\min} = 10^3$K run, we see no scatter at $z = 6.9$ and no bimodal distribution at later times, consistent with the absence of turbulent mixing in Figures~\ref{fig:example_density_small_boxes}-\ref{fig:example_temperature_small_boxes}.  The run with $v_{\rm bc}^{\rm stream} = 65$ km/s does show scatter and a two-phase structure, but the mildly overdense, cooler gas phase is less prominent than in the fiducial run.  The density field in the $m_{\rm X} = 5$ keV run is sufficiently smooth to prevent any non-trivial behavior in the TDR.  In the $m_{\rm X} = 10$ keV case, the TDR is flat at the low-density end, with some scatter at $\Delta > 1$, but still without a multi-phase structure.    

\section{Self-shielding \& the IGM opacity}
\label{sec:opacity} 

\subsection{The gas density distribution \& self-shielding properties}
\label{subsec:gasPDF}

\begin{figure}[h!]
    \centering
    \includegraphics[scale=0.35]{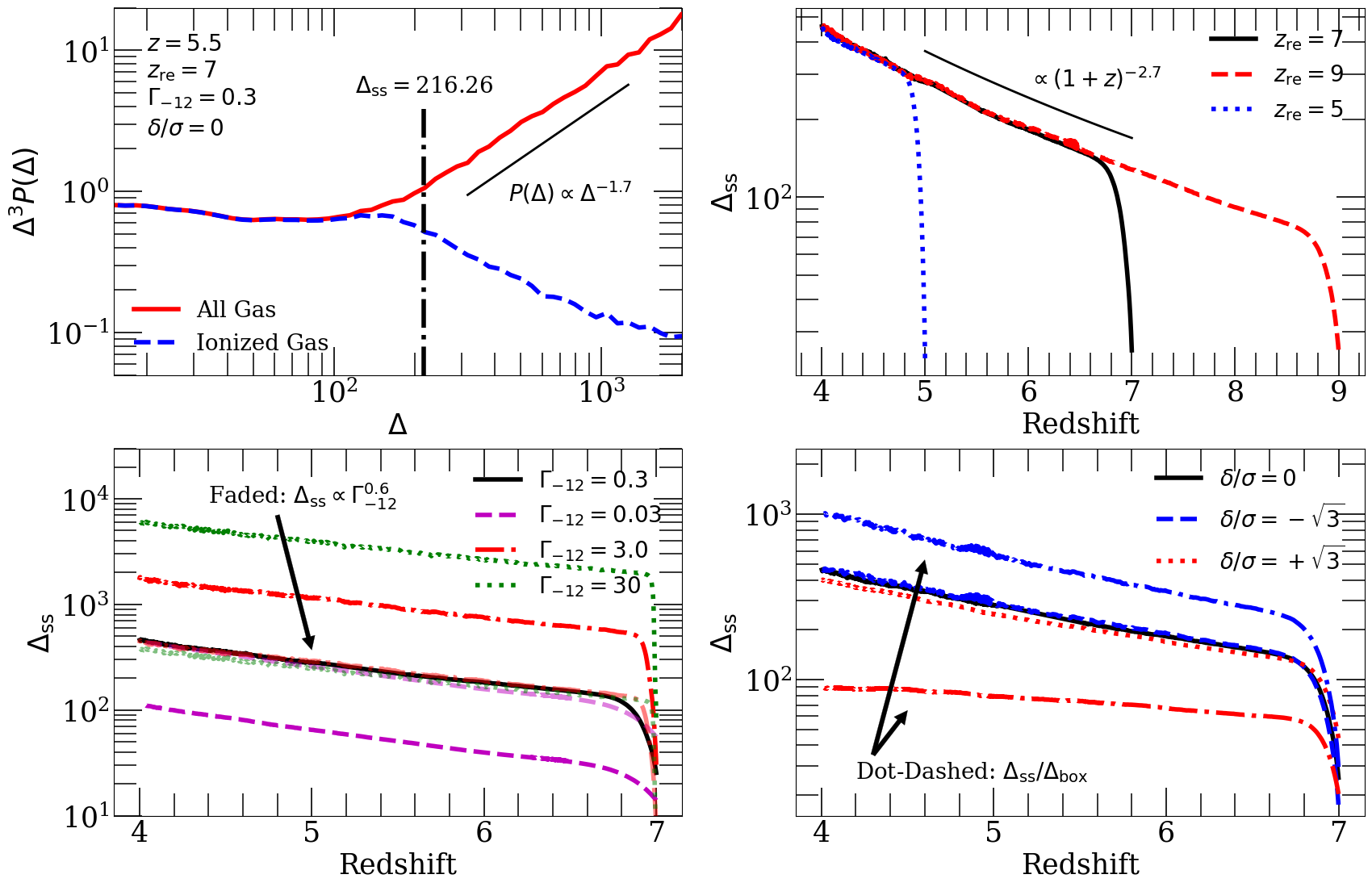}
    \caption{Dependence of the self-shielding density on IGM environment.  {\bf Top Left}: an example density PDF from our fiducial run that shows how $\Delta_{\rm ss}$ is defined.  {\bf Top Right}: dependence on $z_{\rm re}$.  For a short time after $z_{\rm re}$, $\Delta_{\rm ss}$ evolves rapidly as small self-shielding halos photo-evaporate, but quickly asymptotes to a $\propto (1+z)^{-2.7}$ scaling independent of $z_{\rm re}$.  {\bf Bottom Left}: Dependence on $\Gamma_{-12}$.  The faded curves, which closely over-lap the black curve, show $\Delta_{\rm ss}$ for each case re-scaled by a factor of $(\Gamma_{-12}/0.3)^{-0.6}$. 
    The fact that they all agree with the black curve demonstrates that $\Delta_{\rm ss} \propto \Gamma_{-12}^{0.6}$ is true across $3$ orders of magnitude.  {\bf Bottom Right}: dependence on $\delta/\sigma$.  The dot-dashed curves show the ratio of $\Delta_{\rm ss}/\Delta_{\rm box}$, which decreases with $\Delta_{\rm box}$ - however, $\Delta_{\rm ss}$ itself is only mildly sensitive to the more advanced structure formation in the boxes with higher density.  }
    \label{fig:density_pdf_example}
\end{figure}

In this section, we study self-shielding and the ionizing photon opacity of the IGM in \textsc{saguaro}, beginning in this section with the PDF of gas densities.  In Figure~\ref{fig:density_pdf_example}, the top left panel shows the gas density PDF, $P(\Delta)$, multiplied by $\Delta^3$, at $z = 5.5$ in our fiducial Core simulation\footnote{The quantity $\Delta^3 P(\Delta)$ for ionized gas is proportional to the recombination rate per log interval in $\Delta$ for isothermal gas.  }.  The solid red curve shows the PDF for all gas, and the dashed blue curve shows that of ionized gas only.  The latter is defined, as in Ref.~\cite{McQuinn2011}, by multiplying $\Delta^3 P(\Delta)$ by the square of the mean ionized fraction for gas at that $\Delta$, $x_{\rm HII}^2(\Delta)$ (see also \S4.3 of Ref.~\cite{DAloisio2020}).  At around $\Delta \sim 200$, the ionized PDF quickly drops below the total, marking the transition from highly ionized to self-shielded gas.  We define the self-shielding density, $\Delta_{\rm ss}$ such that $x_{\rm HII}^2(\Delta_{\rm ss}) \equiv 0.5$, such that the recombination rate is half what it would be if the gas was fully ionized.  Close to $\Delta_{\rm ss}$, the PDF changes shape from $P(\Delta) \propto \Delta^{-3}$ at $\Delta \lesssim \Delta_{\rm ss}$ to $P(\Delta) \propto \Delta^{-1.7}$ for $\Delta \gtrsim \Delta_{\rm ss}$.  Ref.~\cite{DAloisio2020} pointed out that this is a result of the gas below $\Delta_{\rm ss}$ being heated and evacuated to the lower-density IGM, while the gas above this threshold remains neutral, relatively cold, and embedded within halos.  

In the remaining panels, we show how $\Delta_{\rm ss}$ evolves with redshift in simulations with different parameters.  The top right shows the dependence on $z_{\rm re}$, with $\Gamma_{-12} = 0.3$ and $\delta/\sigma = 0$.  At $z_{\rm re}$, $\Delta_{\rm ss}$ rises sharply as neutral gas at moderate over-densities is ionized, and within a few 10s of Myr settles on a steady $\propto (1+z)^{-2.7}$ scaling that is independent of $z_{\rm re}$.  This redshift scaling is slightly shallower than that of constant physical density ($\propto (1+z)^{-3}$), probably reflecting the evolution of gas temperature with redshift. The transition to this redshift scaling happens much more quickly than the transition to pressure equilibrium ($10$s vs.\ $100$s of Myr), reflecting the fact that dense gas photo-ionizes on a much shorter timescale than it can respond dynamically.  This result also suggests that $\Delta_{\rm ss}$ is mostly independent of the detailed structure and dynamics of the clumps, the latter of which retain memory of $z_{\rm re}$ for much longer (Figures~\ref{fig:example_density} and~\ref{fig:phase_diagrams_L2}).  This suggests that self-shielding corrections in simulations without RT can be safely assume that the self-shielding density does not depend on $z_{\rm re}$.  

The bottom left shows the effect of $\Gamma_{\rm HI}$ on $\Delta_{\rm ss}$ with $z_{\rm re}$ and $\delta/\sigma$ held fixed to their fiducial values.  We see that higher $\Gamma_{\rm HI}$ increases $\Delta_{\rm ss}$ without changing its redshift evolution.  The faded green and magenta curves show the $\Gamma_{-12} = 0.03$, $3.0$ and $30$ cases re-scaled by a factor of $(0.3/\Gamma_{-12})^{0.6}$, which tightly overlap the black curve.  This indicates that \textsc{saguaro} predicts a scaling of $\Delta_{\rm ss} \propto \Gamma_{\rm HI}^{0.6}$ across the full range of $\Gamma_{\rm HI}$ that we simulate.  This scaling is fairly close to the commonly-assumed scaling of $\Delta_{\rm ss} \propto \Gamma_{\rm HI}^{2/3}$~\cite{Furlanetto2005,Rahmati2013,Chardin2018,Nasir2021}.  Using the analytic argument in Ref.~\cite{McQuinn2011} based on the self-shielding model of Ref.~\cite{MiraldaEscude2000}, a scaling of $0.6$ implies\footnote{Specifically, in the Ref.~\cite{MiraldaEscude2000} model (which assumes a sharp density cutoff between ionized and shielded gas), $\Delta_{\rm ss} \propto \Gamma_{\rm HI}^{3/(7-\gamma)}$, where $\gamma$ is the slope of $P(\Delta)$.  } $P(\Delta) \propto \Delta^{-2}$.  This is reasonably close to the $-1.7$ scaling seen in the top left of Figure~\ref{fig:density_pdf_example}.  In the bottom right, we show the dependence on $\delta/\sigma$.  The dashed and dotted curves show $\Delta_{\rm ss}$ defined relative to the cosmic mean density, while the dot-dashed curves show it as a fraction of the  the box-scale mean density, $\Delta_{\rm box}$ (see annotation).  We see that $\Delta_{\rm ss}$ varies little as a function of the box-scale density.  This is perhaps a bit surprising in the over-dense case, since its thermal history differs significantly from the mean-density case (Figure~\ref{fig:fractions_T}).  The dot-dashed curves show that the over (under)-dense run has a much lower (higher) $\Delta_{\rm ss}/\Delta_{\rm box}$, reflecting the fact that the fraction of self-shielded gas increases with box-scale over-density (also see Figure~\ref{fig:fractions_T}).  

\begin{figure}[h!]
    \centering
    \includegraphics[scale=0.19]{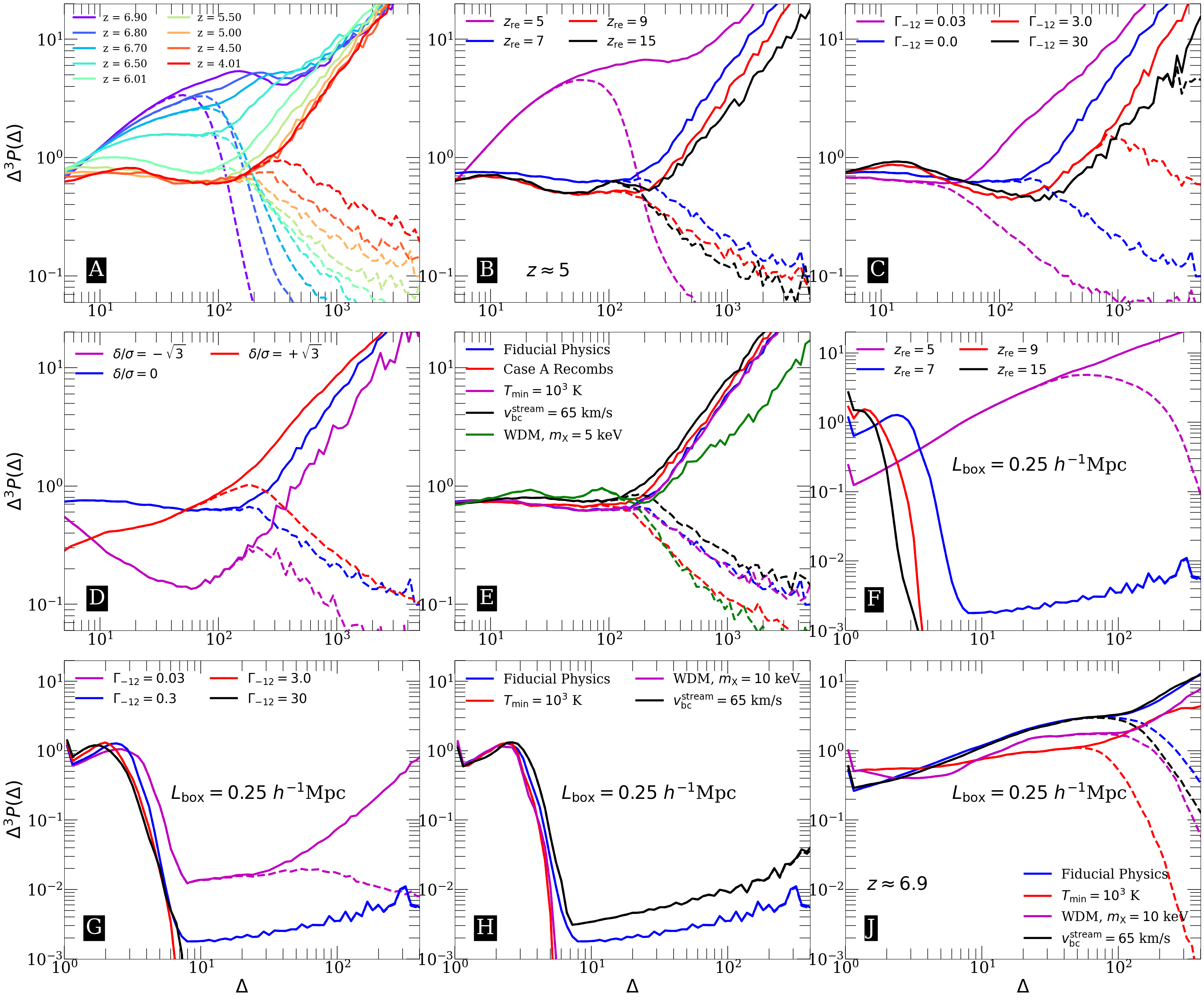}
    \caption{Gas density PDFs in a subset of the \textsc{saguaro} suite.  Note that all results are shown at $z = 5$, except where otherwise denoted (that is, panels A and J).  Panel A shows the redshift evolution in the fiducial Core run.  Panels B, C, and D show the dependence of $P(\Delta)$ on $z_{\rm re}$, $\Gamma_{\rm HI}$, and $\delta/\sigma$, respectively, at fixed $z \approx 5$.  In Panel E, we show the effects of alternative physical parameters (see \S~\ref{subsec:additional_parameters}).  The remaining panels show results from the HR suite.  Panels F and G show the effects of $z_{\rm re}$ and $\Gamma_{\rm HI}$, while H and J shows the effect of alternative physical parameters.  Panel H shows results at $z = 5$ (well past $z_{\rm re} = 7$) while Panel J shows $z = 6.9$ (shortly after $z_{\rm re}$).  }
    \label{fig:PDelta_parameter_plot}
\end{figure}

In Figure~\ref{fig:PDelta_parameter_plot}, we show $\Delta^3 P(\Delta)$ for a subset of \textsc{saguaro} simulations.  Solid curves show the PDF for all gas, and dashed curves show that of ionized gas only (as in the top left of Figure~\ref{fig:density_pdf_example}).  Panel A shows the evolution of $\Delta^3 P(\Delta)$ in our fiducial $2$ $h^{-1}$Mpc box at $9$ redshifts between $z_{\rm re} = 7$ and $4$ (see legend).  Near $z_{\rm re}$, the full and ionized PDFs begin to differ at $\Delta \sim 50$, indicating where the gas self-shields.  This cutoff increases steadily with redshift and reaches $\Delta \sim 300$ by $z = 4$.  At the same time, gas initially at densities near this threshold is evacuated to lower densities as the IGM responds to photo-heating.  The PDF flattens below $\Delta_{\rm ss}$, and steepens above it - the latter indicating that neutral gas at $\Delta \gtrsim 500$ is being compressed as lower-density gas is stripped away from halos.  The total gas PDF does not evolve much at $z < 5$, indicating that the IGM has reached a new equilibrium.  

Panel B shows the effect of $z_{\rm re}$ at fixed\footnote{For the $z_{\rm re} = 5$ case, we show $z = 4.95$.  } $z\approx 5$.  The $z_{\rm re} = 5$ case, which has just reionized, has a shape similar to the $z = 6.9$ case in Panel A, while the others are all similar to the ``relaxed limit'' in Panel A.  There are subtle differences, however, between the $z_{\rm re} = 7$, $9$, and $15$ cases.  Most notably, the slope of $P(\Delta)$ is slightly steeper at $\Delta > \Delta_{\rm ss}$ for simulations with lower $z_{\rm re}$.  This suggests that structures dense enough to survive pressure smoothing form more easily in parts of the IGM that re-ionize later, a result seen qualitatively in Figure~\ref{fig:example_density}.  In Panel C, we see that the shape of $P(\Delta)$ is very sensitive to $\Gamma_{\rm HI}$.  In general, the transition from ionized to neutral shifts to higher densities, as already seen in Figure~\ref{fig:density_pdf_example}, and the characteristic steepening of the PDF also shifts to the right.  At $\Gamma_{-12} = 0.03$ and $0.3$, these two features track each other - however, at higher values of $\Gamma_{-12}$, the shape transition stalls at around $\Delta \approx 300$ while $\Delta_{\rm ss}$ continues to increase.  This reflects the fact that, at high enough $\Gamma_{\rm HI}$, the halos hosting gas near $\Delta_{\rm ss}$ are massive enough to retain photo-ionized, $\sim 10^4$ K gas.  Above this point, $P(\Delta)$ stops being sensitive to $\Gamma_{\rm HI}$.  

In Panel D, we see a sharp contrast in $P(\Delta)$ between runs with different $\delta/\sigma$.  In the under-dense case, the evacuation of gas just below $\Delta_{\rm ss}$ is much more dramatic, indicating that gas near this threshold resides on average in less massive halos that are more easily disrupted by reionization heating.  The opposite is true in the over-dense case, for which this feature is less pronounced.  The slopes of the PDFs well above $\Delta_{\rm ss}$ are similar, although more self-shielded gas is retained for higher $\delta/\sigma$.  In Panel E, we show results assuming fiducial box-scale parameters assuming different IGM physics models.  We see that the modeling choices we consider all have a modest effect on the relaxed PDF shape, with the exception of the WDM model with $m_{\rm X} = 5$ keV (green curve).  In that case, the slope at $\Delta > \Delta_{\rm ss}$ is appreciably shallower than the other cases, reflecting the relative rarity of halos capable of hosting self-shielding gas.  

In the remaining panels (F-J), we show similar results for our HR simulations.  Panels F and G are formatted the same way as B and C, respectively.  We see that in the $z_{\rm re} = 5$ case, which has just re-ionized, the PDF follows roughly a power law with a self-shielding cutoff around $\Delta_{\rm ss} \approx 50$.  For other values of $z_{\rm re}$, which have been ionized for hundreds of Myr, nearly all the gas above $\Delta$ of a few has been evacuated to lower density.   In the $z_{\rm re} = 7$ case, there is still a small amount of gas above this threshold, and in the other cases, nearly all density fluctuations have been erased.  Indeed, this box size is small enough that it does not contain any halos massive enough to retain their gas after pressure smoothing is finished.  This holds true in panel G for all values of $\Gamma_{-12}$ except the lowest, $0.03$.  In that case, $\Delta_{\rm ss}$ is low enough that a few of the most massive halos are able to self-shield (see Figure~\ref{fig:example_tau}).  

Panels H and J show the effect of varying IGM physics, at $z = 5$ and $6.9$ (for $z_{\rm re} = 7$) respectively.  In H, the pre-heated and WDM models lose their over-densities more quickly than the fiducial physics case.  Surprisingly, the case with high stream velocity retains slightly more over-dense gas - despite the tendency of $v_{\rm bc}$ to erase dense structures.  This suggests that the different initial conditions assumed (starting at $z = 1080$ vs.\ $300$, see \S\ref{subsubsec:vbc}) may play a role.  The differences between models are more interesting at $z = 6.9$, before neutral gas has been evaporated.  The preheated and WDM models have slightly flatter $P(\Delta)$ with less gas at high densities, and the preheated case has a lower $\Delta_{\rm ss}$.  

\subsection{The mean free path}
\label{subsec:mfp}

\begin{figure}[h!]
    \centering
    \includegraphics[scale=0.22]{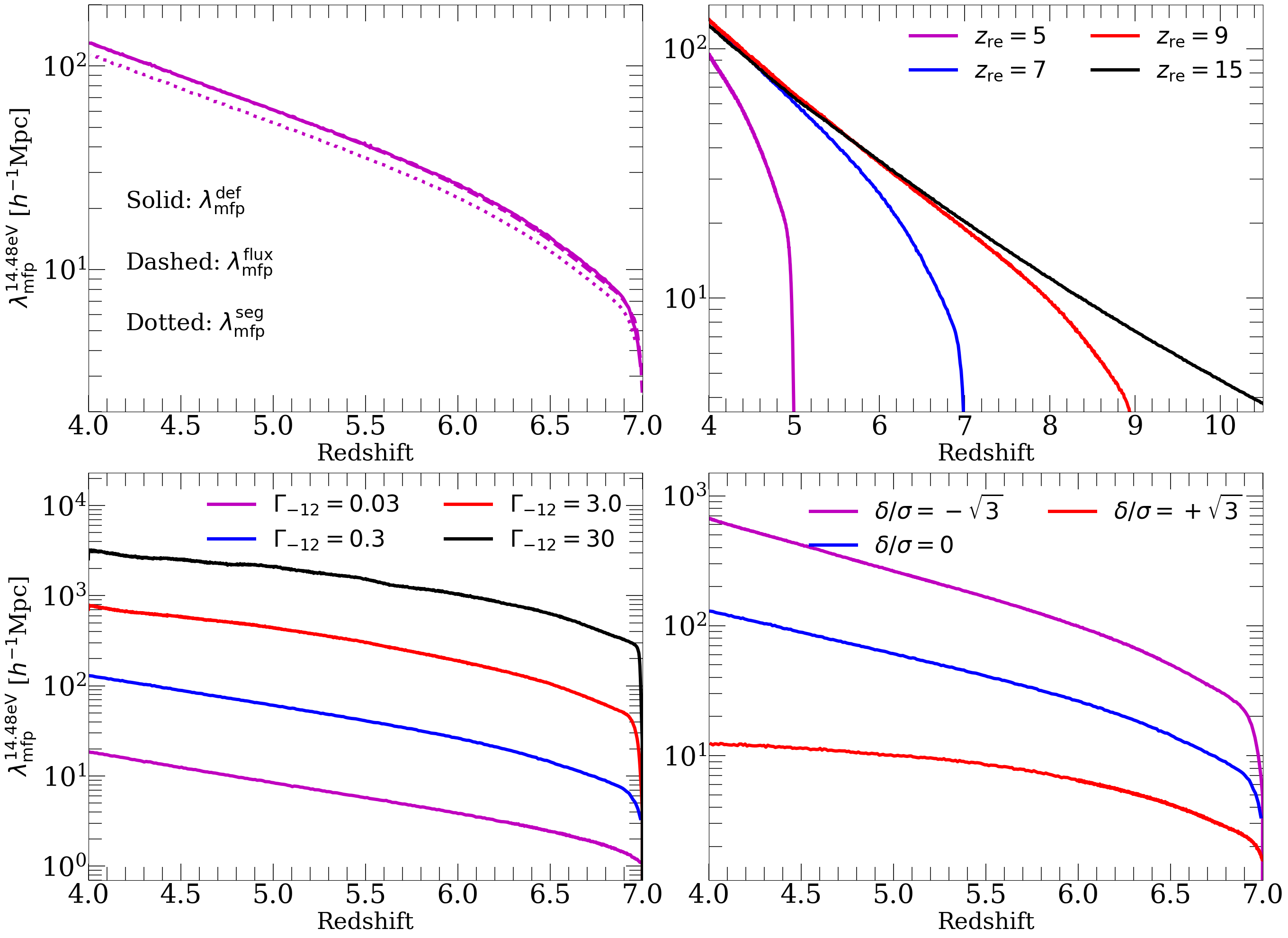}
    \caption{Evolution of $\lambda_{\rm mfp}$ in the $14.48$ eV energy bin in \textsc{saguaro}. {\bf Top Left:} the reference Core run using the three definitions of $\lambda_{\rm mfp}$ described in \S\ref{subsubsec:est_mfp}.   $\lambda_{\rm mfp}^{\rm def}$ and $\lambda_{\rm mfp}^{\rm flux}$ (solid and dashed) agree at percent=level, while $\lambda_{\rm mfp}^{\rm seg}$ (with $L_{\rm seg} = 0.25$ $h^{-1}$Mpc) is $\approx 10\%$ lower (dotted curve).  {\bf Top Right:} dependence on $z_{\rm re}$.  $\lambda_{\rm mfp}$ grows rapidly just after $z_{\rm re}$, and in all cases approaches a limiting power law evolution of $\propto (1+z)^{-4.4}$, similar to the findings of Ref.~\cite{DAloisio2020}.  {\bf Bottom Left:} dependence of $\lambda_{\rm mfp}$ with $\Gamma_{\rm HI}$, which is roughly independent of redshift and becomes weaker as $\Gamma_{\rm HI}$ increases.  {\bf Bottom right:} dependence on $\delta/\sigma$.  We find $\lambda \propto \Delta_{\rm box}^{-1.5 }$ ($\propto \Delta_{\rm box}^{-2}$) between the mean and over (under)-dense runs.  }
    \label{fig:mfp_summary}
\end{figure}

In this section, we briefly study the MFP in \textsc{saguaro}, including its dependence on different MFP definitions (see \S\ref{subsubsec:est_mfp}).  In Figure~\ref{fig:mfp_summary}, we show the MFP in our lowest-energy bin, $14.48$ eV, as a function of redshift in a selection of our simulations ($\lambda_{\rm mfp}^{14.48 {\rm eV}}$).  The top left panel shows the results for our reference run using three definitions given in \S\ref{subsec:mfp} (Eq.~\ref{eq:lambda_flux_def}-\ref{eq:lambda_theory_def}, solid, dashed, and dotted lines, respectively).  For the latter definition, we assume a segment length of $L_{\rm seg} = 0.25$ $h^{-1}$Mpc.  We see that $\lambda_{\rm mfp}$ is, encouragingly, fairly insensitive to its definition, especially for $\lambda_{\rm mfp}^{\rm flux}$ and $\lambda_{\rm mfp}^{\rm def}$, which are nearly on top of each other.  We see that $\lambda_{\rm mfp}^{\rm seg}$ is $\sim 10\%$ below the others, suggesting some sensitivity to the definition (and the choice of $L_{\rm seg}$).  This sensitivity suggests that $\lambda_{\rm mfp}^{\rm def}$ or $\lambda_{\rm mfp}^{\rm flux}$ do a better job of estimating the MFP in simulations. 

We show the dependence of $\lambda_{\rm mfp}$ on $z_{\rm re}$, $\Gamma_{\rm HI}$, and $\delta/\sigma$ top right, bottom left, and bottom right panels, respectively.  In the top right, we find (similar to Ref.~\cite{DAloisio2020}) that $\lambda_{\rm mfp}$ increases rapidly just after $z_{\rm re}$, reflecting the photo-evaporation of dense systems and pressure-smoothing of the IGM, and evolves more slowly afterwards.  The MFP in our simulations with $z_{\rm re} = 15$, $9$, and $7$ converge to a common power-law scaling of roughly $\propto (1+z)^{-4.4}$ by $z \approx 5$, while the $z_{\rm re} = 5$ run has not yet caught up with this limiting behavior by $z = 4$.  In the bottom left, we see that $\lambda_{\rm mfp}^{912}$ scales with $\Gamma_{\rm HI}$ in a way that is mostly redshift-independent, except near $z_{\rm re}$.  The scaling between $\Gamma_{-12} = 0.03$ and $0.3$ is close to $\propto \Gamma_{\rm HI}^{0.9}$, while it becomes closer to $\propto \Gamma_{\rm HI}^{0.5}$ between $\Gamma_{-12} = 3.0$ and $30$.  This roughly brackets the $\propto \Gamma_{\rm HI}^{2/3}$ scaling commonly assumed in the literature~\citep{Davies2016,Becker2021}, but highlights the fact that there is considerable sensitivity to the ionizing background strength.  Lastly, we see that $\lambda_{\rm mfp}$ scales with density like $\propto \Delta^{-1.5}$ between the mean density and over-dense runs and $\propto \Delta^{-2}$ between the mean density and under-dense runs.  We note that all these scalings assume a constant value of $\Gamma_{\rm HI}$ throughout the simulation runtime.  The scaling of $\lambda_{\rm mfp}$ with $\Gamma_{\rm HI}$ may be different when $\Gamma_{\rm HI}$ is changing over a short timescale (see Appendix D of Ref.~\cite{Cain2022b} and Figure 7 of ~\cite{Tohfa2026}).  We plan to return to this point in a future study.  

\subsection{The HI column density distribution}
\label{subsec:HI_col_dist}

The HI column density distribution function (CDDF, $f(N_{\rm HI})$) determines the IGM opacity and its sensitivity to ionizing photon energy.  If the CDDF is dominated by high-column systems containing significant neutral gas, the IGM opacity will be high (short MFP) and independent of the energy of ionizing photons (since neutral gas is highly opaque at all UV frequencies).  If instead the CDDF is dominated by low-column, highly ionized gas, the opacity will be low and will scale with photon energy in the same way as the HI cross-section, $\sigma_{\rm HI} \propto \nu^{-2.75}$.  The shape of the CDDF is an input in measurements of the ionizing photon output of the galaxy population based on high-redshift quasar spectra~\citep[][]{Bolton2007,Becker2013,Gaikwad2023,Bosman2024,Cain2025}.

\begin{figure}[h!]
    \centering
    \includegraphics[scale=0.34]{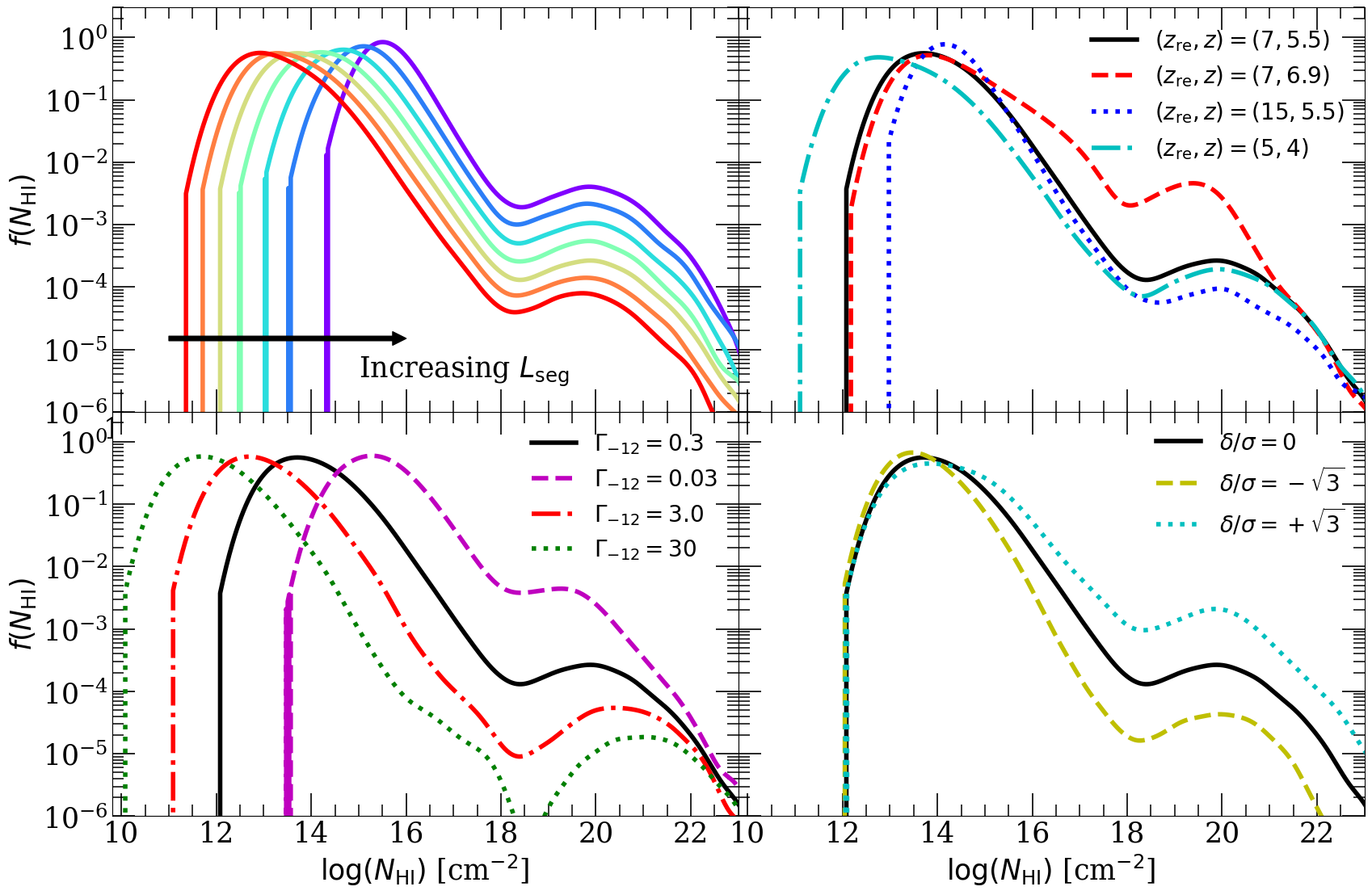}
    \caption{HI column density distribution in Core \textsc{saguaro} runs.  {\bf Top Left:} results from our $z_{\rm re} = 7$, $\Gamma_{-12} = 0.3$, $\delta/\sigma = 0$ run for all $7$ values of $L_{\rm seg}$.  At the low-$N_{\rm HI}$ end, $f(N_{\rm HI})$ shifts to the right (see annotation) with increasing $L_{\rm seg}$, since the average $N_{\rm HI}$ through a low-density patch of the IGM is proportional to $L_{\rm seg}$.  The second peak at high $N_{\rm HI}$ is set by the distribution of self-shielded systems, and its location is independent of $L_{\rm seg}$.  {\bf Top Right:} dependence on $z$ and $z_{\rm re}$ for $L_{\rm seg} = 0.25$ $h^{-1}$Mpc.  The high-column peak is largest when reionizaiton happened recently ($z_{\rm re} = 7$, $z = 6.9$), when most of the small-scale structure is as-yet unaffected by pressure smoothing.  The high-column peak is similar between runs at lower redshift, while the low-column end shifts depending on redshift and how recently the gas was ionized.  {\bf Lower Left:} Dependence on $\Gamma_{-12}$.  Lower ionizing background strengths shift the distribution to the right and enhance the amplitude of the high-column peak.  {\bf Lower Right:} effect of the box-scale density.  The amplitude of the high-column peak increases significantly with box-scale density, reflecting the larger role that self-shielded systems play in setting the IGM opacity in over-dense regions. }
    \label{fig:column_density_example} 
\end{figure}

The top left panel of Figure~\ref{fig:column_density_example} shows $f(N_{\rm HI})$ defined for $7$ values of $L_{\rm seg}$ (see discussion surrounding Eq.~\ref{eq:lambda_seg_def}) for the reference Core run at the same redshifts shown in the top left of Figure~\ref{fig:density_pdf_example}.  The annotation indicates that curves that start further to the right on the plot have higher values of $L_{\rm seg}$, increasing in powers of $2$ between adjacent curves.  The CDDF is bimodal, peaking at low $N_{\rm HI}$ close to the minimum $N_{\rm HI}$, and again at $\log(N_{\rm HI}/[{\rm cm}^{-2}]) \sim 20$.  At low column densities, $N_{\rm HI} \propto L_{\rm seg}$, and the left edge of $f(N_{\rm HI})$ shifts to the right with increasing $L_{\rm seg}$.  Much higher columns correspond to segments intersecting a self-shielded system.  The peak in $f(N_{\rm HI})$ occurs at $\log(N_{\rm HI}/[{\rm cm}^{-2}]) \sim 20$, independent of $L_{\rm seg}$.  The normalization of $f(N_{\rm NHI})$ at high $N_{\rm HI}$ is $\propto L_{\rm seg}$, since the likelihood that a given segment will intersect at least one self-shielding system is proportional to the length of the segment.  The insensitivity of the shape of the CDDF at $\log(N_{\rm HI}) > 16$, which sets the frequency dependence of the MFP, suggests that our results in this section are likely insensitive to the choice of $L_{\rm seg}$, which we take to be $0.25$ $h^{-1}$Mpc in what follows.  

The top right panel shows how $f(N_{\rm NHI})$ depends on $z$ and $z_{\rm re}$.  The black solid and red dashed curves show $z_{\rm re} = 7$ at $z = 5.5$ and $6.9$, respectively.  The abundance of systems with $\log(N_{\rm HI}/[{\rm cm}^{-2}]) \sim 16-21$ is elevated at $z = 6.9$, especially at $\log(N_{\rm HI}/[{\rm cm}^{-2}]) \sim 18-20$ where the systems become self-shielding.  This owes to the higher fraction of gas in self-shielded systems shortly after the gas is ionized (see top left of Figure~\ref{fig:PDelta_parameter_plot}).  The blue dotted and cyan curves show runs with $z_{\rm re} = 15$ and $5$ at $z = 5.5$ and $4$, respectively.  At fixed $z = 5.5$, the $z_{\rm re} = 15$ run has lower $f(N_{\rm NHI})$ near the self-shielding threshold than the $z_{\rm re} = 7$ case.  In the former, the gas ionized when there were fewer dense structures, and pressure smoothing has had more time to erase those structures.  We also see that the cutoff at low $N_{\rm HI}$ is slightly higher in the $z_{\rm re} = 15$ run, due to the most rarefied voids having been ``filled in'' by gas escaping from the smallest halos.  The run with $(z_{\rm re},z) = (5,4)$ is similar to the $(z_{\rm re},z) = (7,5.5)$ run at the highest columns, but is more extended at lower columns.  This comparison isolates the effect of changing both $z_{\rm re}$ and $z$ while keeping the time since reionization approximately fixed\footnote{The time interval between $z = 7$ and $5.5$ is $\Delta t = 279$ Myr, and for $z = 5$ to $4$ this is $369$ Myr.}.

The bottom left panel shows the effect of $\Gamma_{\rm HI}$ on the CDDF.  We see that increasing $\Gamma_{\rm HI}$ makes the CDDF extend to lower $N_{\rm HI}$, with the cutoff $N_{\rm HI}$ scaling linearly with $\Gamma_{\rm HI}$, and reduces the abundance of high-$N_{\rm HI}$ systems.  This is due to the increase in the self-shielding density seen in Figure~\ref{fig:density_pdf_example}-\ref{fig:PDelta_parameter_plot}.  The bimodality in $f(N_{\rm NHI})$ also becomes more prominent, as $\Gamma_{\rm HI}$ increases and self-shielded systems become rarer and more isolated.  Lastly, the bottom right shows the effect of box-scale density.  The shape and cutoff of $f(N_{\rm NHI})$ at low $N_{\rm HI}$ is remarkably similar, suggesting that the distribution and structure of low-density voids in $2$ $h^{-1}$Mpc patches of the IGM is reasonably insensitive to the density contrast at the scale of the patch.  There is a striking difference at high columns, with the over(under)-dense runs showing a factor of $5-10$ more (fewer) systems with $\log(N_{\rm HI}/[{\rm cm}^{-2}]) \gtrsim 18$.  Runs at a higher box-scale density form more dense structures, and contain more halos with masses $\gtrsim 10^8$ $M_{\odot}$ that are able to self-shield and retain their gas despite heating by reionization.  Note that the bi-modality of $f(N_{\rm HI})$ seen in our simulations is driven entirely by self-shielding of dense gas structures and is distinct from the effect of still neutral regions of the IGM during reionization, which also generate a bimodality on larger spatial scales~\citep[see Figure 7 of Ref.][]{Feron2024}. 

We pause here to note the important caveat that, since our simulations model only primordial H and He, and ignore the effects of potential star formation in the most massive halos, our simulations likely over-estimate the very high-column ($\log(N_{\rm HI}/[{\rm cm}^{-2}]) \gtrsim 21$) of $f(N_{\rm HI})$.  Correcting for self-ionization by stellar sources within these halos would likely push the high-column peak in $f(N_{\rm HI})$ to modestly lower columns, especially in the over-dense boxes (where star forming halos are more common) and at high $\Gamma_{-12}$ (where the peak occurs at very high columns).  Other effects, such as corrections arising from molecular H2 formation, could have a similar effect~\citep[e.g. Figure 6 of][]{Altay2013}.  Given these considerations, we caution against over-interpreting the exact placement of the peak in our simulations or the shape of $f(N_{\rm HI})$ on the high-column end of it.  Fortunately, these details should have minimal effect on our subsequent analysis of the ionizing photon opacity, since gas at somewhat lower columns dominates absorption in the IGM at these redshifts~\citep{Rahmati2017,Nasir2021}.  

\begin{figure}[h!]
    \centering
    \includegraphics[scale=0.275]{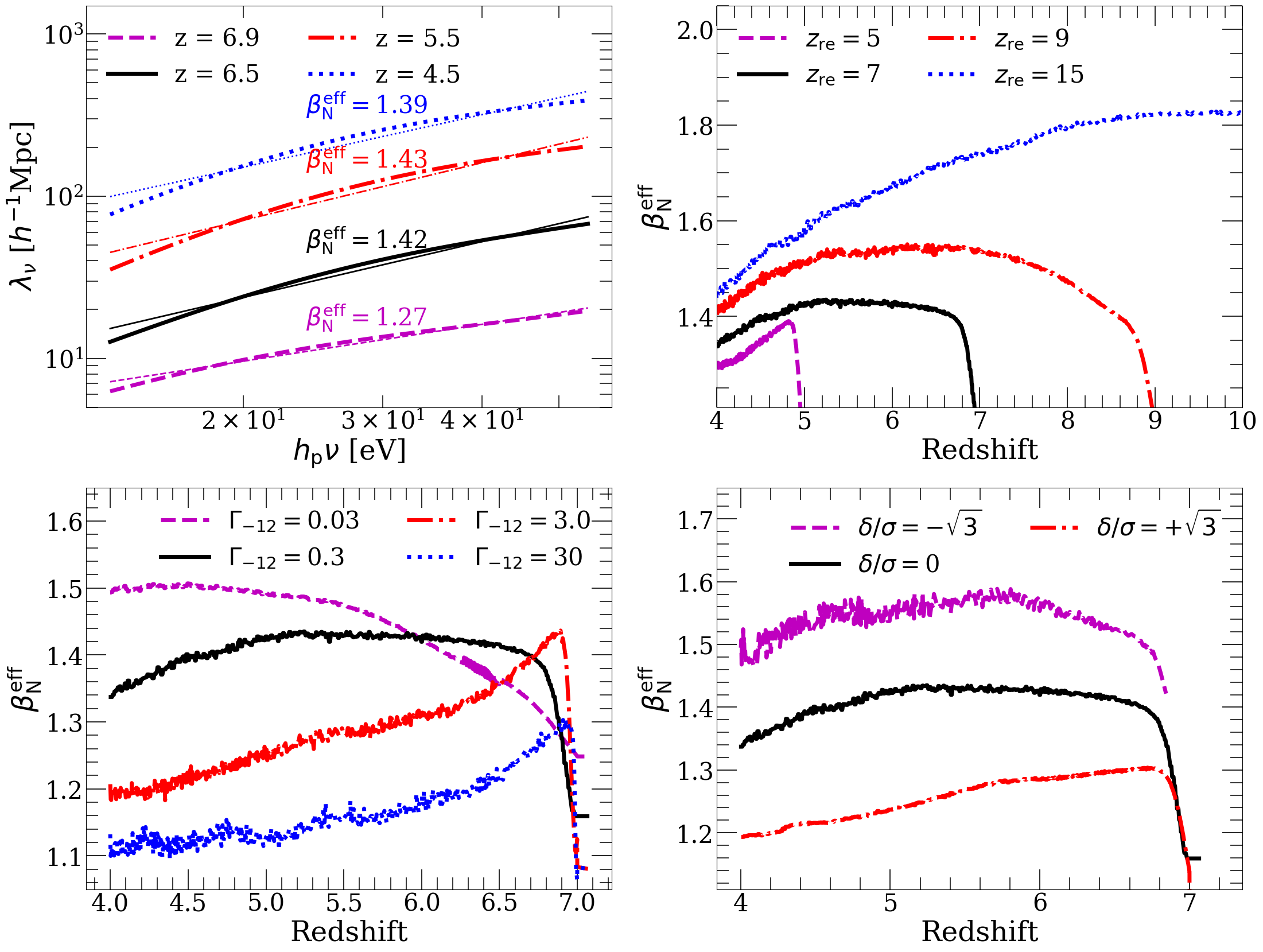}
    \caption{Dependence of the MFP on photon frequency in different large-scale IGM environments.  {\bf Top Left:} dependence of the MFP on photon energy at several redshifts, with the thin lines denoting best-fit power laws.  The values of $\beta_{\rm N}^{\rm eff}$ are the best-fit effective column density slopes recovered from the fits.  {\bf Remaining panels:} $\beta_{\rm N}^{\rm eff}$ vs.\ redshift for simulations with varying $z_{\rm re}$, $\Gamma_{-12}$, and $\delta/\sigma$.  See text for details.  }
    \label{fig:column_density_mfp_example}
\end{figure}

We next use $f(N_{\rm NHI})$ to study the frequency dependence of $\lambda_{\rm mfp}$ in \textsc{saguaro}.  In the limit that the IGM consists of a discrete set of HI absorbers, we can write the average absorption coefficient at frequency $\nu$ as~\citep{Meiksin1993,Rahmati2017}
\begin{equation}
    \label{eq:kappa_nu}
    \kappa_{\nu} = \int dN_{\rm HI} f(N_{\rm HI}) (1-\exp[-N_{\rm HI}\sigma_{\rm HI}^{\nu}])
\end{equation}
where the integral runs over all columns and $\sigma_{\rm HI}^{\nu}$ is the HI-ionizing cross-section at frequency $\nu$.  Ref.~\cite{Nasir2021} showed that Eq.~\ref{eq:kappa_nu} is equivalent to our segment-based MFP definition (Eq.~\ref{eq:lambda_seg_def}) assuming $f(N_{\rm NHI})$ is estimated using the same set of segments used to calculate $\lambda_{\rm mfp}^{\rm seg}$.  

In Figure~\ref{fig:column_density_mfp_example}, we use Eq.~\ref{eq:kappa_nu} to explore the frequency dependence of the MFP in \textsc{saguaro}.  In the top left panel, we show $\lambda_{\rm mfp}^{\nu}$ vs.\ photon energy ($E = h_{\rm p}\nu$) between $E = 13.6$ and $54.4$ eV at several redshifts in our fiducial Core run.  We see that the frequency dependence is almost (but not quite) a power law.  Indeed, it can be shown using Eq.~\ref{eq:kappa_nu} that if $f(N_{\rm NHI})$ is a power law in $N_{\rm HI}$ with slope $\beta_{\rm N}$, then the MFP should be exactly a power law in frequency with scaling $\lambda_{\rm mfp}^{\nu} \propto \sigma_{\rm HI}(\nu)^{(\beta_{\rm N}-1)} \propto \nu^{2.75(\beta_{\rm N}-1)}$.  Thus, deviations from a power law reflect the more complicated shape of $f(N_{\rm NHI})$.  Nevertheless, we can fit $\lambda_{\rm mfp}^{\nu}$ to a power law (thin lines) and extract an effective column density slope that would produce the recovered best-fit, which we denote $\beta_{\rm N}^{\rm eff}$. We report this value for each curve in the top left. 

In the remaining panels, we plot $\beta_{\rm N}^{\rm eff}$ vs.\ redshift and explore the dependence on \textsc{saguaro} parameters.  In the top right, we see that for a short time after $z_{\rm re}$, $\beta_{\rm N}^{\rm eff}$ increases rapidly, peaking between $1.4$ and $1.5$ for $5 \leq z_{\rm re} \leq 9$ (given fiducial values of other parameters), and close to $2$ for $z_{\rm re} = 15$.  After several hundred Myr have passed, however, $\beta_{\rm N}^{\rm eff}$ begins decreasing again, reaching between $1.45$ and $1.3$ for all models by $z = 4$.  We can understand this behavior by considering the physics that sets $\beta_{\rm N}$.  In the limit of a highly-ionized, optically thin IGM, the opacity everywhere is proportional to $\sigma_{\rm HI}(\nu)$, such that $\beta_{\rm N} = 2$.  In the opposite limit, in which the opacity is dominated by self-shielded neutral gas, the opacity is independent of frequency, such that $\beta_{\rm N} = 1$.  
A short time after $z_{\rm re}$, these systems dominate the opacity because most of them have not photo-evaporated yet, such that $\beta_{\rm N}$ starts out close to $1$.  As these systems evaporate over the next few hundred Myr, $\beta_{\rm N}$ increases as self-shielded systems contribute less to the total opacity.  At still lower redshifts, accretion onto $\gtrsim 10^8$ $M_{\odot}$ halos, which can retain neutral gas in the photo-heated IGM, causes $\beta_{\rm N}$ to decrease again.  The peak in $\beta_{\rm N}$ vs. redshift occurs when self-shielding plays the smallest role in setting the IGM opacity.  Lower $z_{\rm re}$ reduces $\beta_{\rm N}^{\rm eff}$ at fixed redshift, since more structures capable of self-shielding were able to form while the IGM was still cold and neutral.  

The bottom left panel shows that the effect of $\Gamma_{\rm HI}$ on $\beta_{\rm N}^{\rm eff}$ is somewhat complicated.  Simulations with higher $\Gamma_{\rm HI}$ see a faster jump in $\beta_{\rm N}^{\rm eff}$ shortly after $z_{\rm re}$, owing to the more rapid evaporation of small self-shielded systems.  However, at lower redshifts, $\beta_{\rm N}^{\rm eff}$ is considerably lower in systems with higher $\Gamma_{\rm HI}$ - the opposite trend.  The reason for this is that the handful of rare systems dense enough to self-shield contribute a larger fraction to the {\it total} opacity when $\Gamma_{\rm HI}$ is very high, since their effective cross-sections decrease sub-linearly with $\Gamma_{\rm HI}$.  In the bottom right, we see the effect of box-scale density.  Unsurprisingly, boxes with larger densities have a higher abundance of self-shielded systems, and thus smaller $\beta_{\rm N}^{\rm eff}$ at all redshifts.  On average, the over-dense box has $\beta_{\rm N}^{\rm eff} \approx 1.2$, while under-dense boxes have $\beta_{\rm N}^{\rm eff} \approx 1.6$, and with the mean-density box in the middle.  

While there is significant variation in our recovered $\beta_{\rm N}$ values, our results tend to be somewhere between $\beta_{\rm N} = 1.2$ and $1.5$, except near the ``edges'' of reasonable parameter space.  The commonly adopted value of $1.3$~\citep[e.g.][]{Becker2013,Davies2024} is comfortably within this range.  This finding suggests that adopting a tight prior on $\beta_{\rm N}$ within this range when measuring quantities such as the ionizing emissivity of galaxies~\citep{Bolton2007,Becker2013,Cain2025} and other related observables such as the clumping factor~\citep{Davies2024} is justified\footnote{See the bottom right panel of Figure 11 in Ref.~\cite{Cain2025} for an assessment of how much uncertainty in $\beta_{\rm N}$ affects emissivity measurements.  }.  For example, in their recent observational measurement of the clumping factor, Ref.~\cite{Davies2024} assumed $\lambda \propto \nu$, which corresponds to $\beta_{\rm N} \approx 1.35$, roughly the middle of our range.  An uncertainty of $\pm 0.15$ in $\beta_{\rm N}$ would map to about $\pm 0.4$ in the $\lambda$-$\nu$ index $\alpha_{\lambda}$, or about a $20\%$ uncertainty in them measured clumping factor (see their Figure 6).  This is less than the shift of $\sim +2$ shift in $\alpha_{\lambda}$ required to reconcile their unexpectedly high clumping factor measurements with cosmological radiative transfer simulations~\citep{Asthana2025}.  

Our findings in this section yield several interesting conclusions.  The first is that the shape of the HI column density distribution is much more complicated than the oft-assumed power law form (as previously noted by Ref.~\cite{Nasir2021}), which causes the MFP to deviate modestly from the commonly-assumed power-law form.  Both of these assumptions are commonly made in the literature (e.g. Refs.~\cite{Bolton2007,Becker2013,Davies2024}).  It is also interesting that our results for $\beta_{\rm N}$ indicated that this quantity is modestly higher than the oft-assumed $\beta_{\rm N} = 1.3$ during most of reionization, but approach this value towards lower redshifts.  Higher values of $\beta_{\rm N}$ require fewer ionizing photons from galaxies at fixed $\lambda_{\rm mfp}^{912}$ to complete reionization~\cite{Bolton2007,Becker2013}, which may affect the interpretation of measurements of the MFP at the end of reionization~\citep{Becker2021,Gaikwad2023,Zhu2023}.  

\section{Effect of the IGM on reionization}
\label{sec:global_reion}

\subsection{The Clumping Factor}
\label{subsec:clumping} 

The clumping factor (Eq.~\ref{eq:clump_density}) can be roughly defined as the boost in the recombination rate produced by fluctuations in density, and perhaps temperature, relative to that expected in a uniform-density, highly-ionized, isothermal IGM.  It has been employed as a description of the recombination rate in the IGM as a whole~\citep[e.g.][]{Madau1999,Pawlik2009}, and as a way to quantify missing recombinations in simulations that cannot resolve small-scale IGM density fluctuations~\citep[e.g.][]{Mao2019,Bianco2021}, such as those studied here.  One problem with the clumping factor is that it is, in general, a poorly defined quantity.  Counting the number of recombinations in the IGM requires a criterion for which recombinations to include, which itself requires a definition of the IGM.  This should exclude, for instance, recombinations in the ISM and CGM of halos, which are driven by the halo's own ionizing output - such absorptions are already accounted for in the definition of the halo's ionizing escape fraction.  
A further complication arises from the fact that not every recombination in the IGM is immediately balanced by an ionization.  In high-density regions near the self-shielding threshold, non-equilibrium effects can become important, rendering the local recombination rate much different from the ionization rate.  

We explore several definitions of the clumping factor.  Following Ref.~\cite{Kohler2007,DAloisio2020,Chen2020}, we define the recombination clumping factor to be
\begin{equation}
    \label{eq:C_recomb}
    C_{\rm R} \equiv \frac{\langle \alpha_{\rm B}(T) n_e n_{\rm HII} \rangle}{\alpha_{\rm B}(T_{\rm ref}) \langle n_e \rangle \langle n_{\rm HII} \rangle}
\end{equation}
where $\alpha_{\rm B}$ is the case B recombination coefficient of hydrogen and $T_{\rm ref} = 10^4$ K.  The averages in the denominator are always taken to be those for a mean-density IGM (even in over/under-dense boxes), while the average in the numerator is over the simulation volume.  Following Ref.~\cite{DAloisio2020}, we exclude gas at extremely high densities (much of which is within the virial radius of halos that would be star-forming) by averaging the numerator of Eq.~\ref{eq:C_recomb} only over gas with $\Gamma_{\rm HI}/\langle \Gamma_{\rm HI} \rangle > 0.01$, which we denote $C_{\rm R}^{\Gamma/\langle \Gamma \rangle > 0.01}$.  We note that the total recombination rate clumping factor, $C_{\rm R}^{\rm total}$, is noisy due to the effect of collisional ionizations (see Appendices~\ref{app:RSLA} \&~\ref{app:subcycle}), so we do not show it here.  

We also use a second definition of the clumping factor that is based on the ionization (or absorption) rate of HI.  We define this to be the ``absorption clumping factor'', defined as
\begin{equation}
    \label{eq:C_ion}
    C_{\rm A,n_{\rm HI}} \equiv \frac{\langle \Gamma_{\rm HI} n_{\rm HI} \rangle}{\alpha_{\rm B}(T_{\rm ref}) \langle n_e \rangle \langle n_{\rm HII} \rangle}
\end{equation}
where the denominator is the same as in Eq.~\ref{eq:C_recomb}, but the numerator is now the volume-averaged absorption rate of ionizing photons \footnote{Note that this is different from the ``ionization clumping factor'' defined in Ref.~\cite{Kohler2007} (see also Ref.~\cite{Chen2020}), in that the former has $\langle n_{\rm HI}\rangle \langle \Gamma_{\rm HI} \rangle$ in the denominator instead of the average recombination rate.  We use Eq.~\ref{eq:C_ion} instead because it is more easily comparable to $C_{\rm R}$, and indeed is equal to it in the limit that ionizations and recombinations exactly balance.}.  We further define $C_{\rm A,n_{\rm HI}}^{\rm total}$ and $C_{\rm A,n_{\rm HI}}^{\Gamma/\langle \Gamma \rangle > 0.01}$ in the same way as above.  This definition will differ from Eq.~\ref{eq:C_recomb} to the extent that non-equilibrium absorptions and recombinations in self-shielding systems are important.  

\begin{figure}[h!]
    \centering
    \includegraphics[scale=0.225]{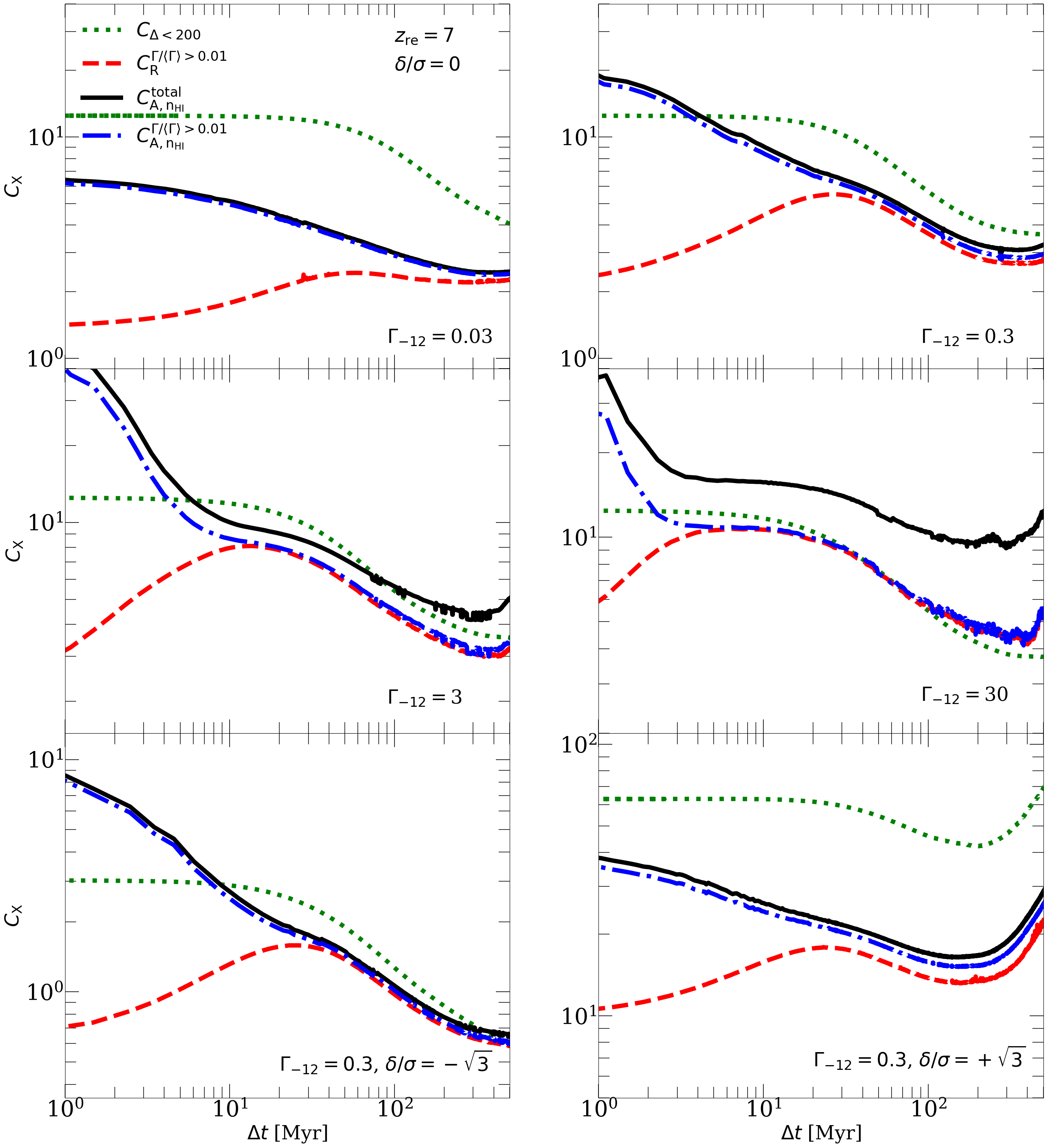}
    \caption{Evolution of the clumping factor in several \textsc{saguaro} simulations.  Each panel shows  $C_{\rm R}$ (red dashed), $C_{\rm A,n_{\rm HI}}$ (solid black and blue dashed), and $C_{\Delta < 200}$ (green dotted).  All simulations have $z_{\rm re} = 7$, and we show the clumping factors vs. time since reionization.  The top four panels show results for mean density simulations with different values of $\Gamma_{\rm HI}$, and the bottom two show results for different box-scale densities. 
 See text for details.  }
    \label{fig:ion_vs_recomb_clumping_example}
\end{figure}

Figure~\ref{fig:ion_vs_recomb_clumping_example} shows the evolution of each of the clumping factors defined above, a definition based on only thresholding the density field, $C_{\Delta < 200}$ (Eq.~\ref{eq:clump_density}), as a function of time since reionization ($\Delta t$).  All six panels have $z_{\rm re} = 7$.  In the top four, we hold $\delta/\sigma = 0$ and show results for all four values of $\Gamma_{-12}$ in the Core suite (see bottom right of each panel).  The bottom two panels hold $\Gamma_{-12} = 0.3$ fixed and show $\delta/\sigma = \pm \sqrt{3}$.  Remarkably, we see that none of the four definitions of $C$ used here agree with each other!  In general, $C_{\Delta < 200}$ (green dashed) starts out roughly constant during the first $10$ Myr or so, then starts gradually decreasing as the IGM dynamically relaxes.  $C_{\rm R}$ initially increases as D-type I-fronts work their way into self-shielded clumps, raising the recombination rate, and then decreases later as the gas becomes less clumpy.  By contrast, $C_{\rm A}$ starts out elevated due to the high absorption rate required to evaporate these clumps, and monotonically decreases over the next several hundred Myr.  

We also see considerable differences as a function of both $\Gamma_{-12}$ and $\delta/\sigma$.  In the top left panel, which has $\Gamma_{-12} = 0.03$, $C_{\rm R}$ and $C_{\rm A}$ differ considerably at small $\Delta t$, with the latter being larger, and $C_{\Delta < 200}$ is still larger than both.  The difference between $C_{\rm A,n_{\rm HI}}$ and $C_{\rm R}$ occurs because it takes a long time for D-type I-fronts to photo-ionized moderately over-dense systems when $\Gamma_{-12}$ is low, such that these systems (which dominate the ionization rate in the ionized IGM) remain out of photo-ionization equilibrium for $\sim 100$ Myr.  Furthermore, $\Delta_{\rm ss}$ is well below $200$ in this run (see Figure~\ref{fig:density_pdf_example}), such that $C_{\Delta > 200}$ includes a significant amount of neutral gas where the absorption and recombination rates are both $0$.  The two definitions of $C_{\rm A,n_{\rm HI}}$ are in good agreement with each other, since very few absorptions occur at densities where $\Gamma_{\rm HI}$ is below $1\%$ of its volume-average.  

As $\Gamma_{\rm HI}$ increases, we see a few key trends.  First, $C_{\rm R}$ and $C_{\rm A,n_{\rm HI}}$ both increase as gas becomes ionized at higher densities, while $C_{\Delta < 200}$ changes very little, since it is sensitive to a fixed range of densities.  We also see that $C_{\rm R}^{\Gamma/\langle \Gamma \rangle > 0.01}$ and $C_{\rm A,n_{\rm HI}}^{\Gamma/\langle \Gamma \rangle > 0.01}$ start agreeing at smaller $\Delta t$, as the absorption and recombination rates in mostly ionized gas reaches equilibrium more quickly, due to the shorter photo-evaporation time-scales.  However, at the highest $\Gamma_{-12}$, a significant difference emerges between $C_{\rm A,n_{\rm HI}}^{\Gamma/\langle \Gamma \rangle > 0.01}$ and $C_{\rm A,n_{\rm HI}}^{\rm total}$.  This is because the higher mean value of $\Gamma_{\rm HI}$ in these runs means that gas at $1\%$ of average can still have a significant ionized fraction and absorption rate.  In the bottom panels, we see that similar trends as in the top right panel, but with all clumping factors reduced (increased) in the under (over)-dense run.  In the under-dense case, the absorption and combination rates reach equilibrium more quickly than for mean density, while in the over-dense there is a modest difference between them even after several hundred Myr.  Importantly, we find no situation in which all four definitions of the clumping factor shown here all mutually agree!  

Our results beg an important question - if different commonly-used definitions of the clumping factor disagree, which one is ``correct'' (if any)?  We can answer this by considering the purpose that the clumping factor is intended to serve - that is, to count the number of ionizing photons that are consumed in ionized regions of the IGM.  More specifically, we wish to count how many photons are consumed by recombinations and self-shielded absorbers {\it after they escape the ISM/CGM of the galaxies that produced them}.  In \textsc{saguaro}, all photons are treated as if they have escaped their host galaxies (thanks to our RT domain setup) and the IGM is highly ionized everywhere in the volume (except within over-dense self-shielded systems) thanks to our freezing procedure.  As such, {\it every} photon absorbed in our simulations at $z < z_{\rm re}$ meets the above criteria (even those absorbed at very high densities that may be associated with the CGM of a halo that would be star-forming).  As such, we conclude that $C_{\rm A,n_{\rm HI}}^{\rm total}$ is the appropriate clumping factor to use in our simulations.  Note that, from Eq.~\ref{eq:lambda_flux_def} and~\ref{eq:C_ion}, we see that $C_{\rm A,n_{\rm HI}} \propto \langle \Gamma_{\rm HI}\rangle/\lambda_{\rm mfp}^{\rm flux}$, such that using this clumping factor is equivalent to using the ionizing photon mean free path to count absorptions (see further discussion in Ref.~\cite{Davies2024}).  

\subsection{Ionizing photon budget}
\label{subsec:ion_budget}

In this section, we turn to the implications of our findings in the previous section for the number of ionizing photons required to re-ionize the universe.  Following Ref.~\cite{DAloisio2020}, we define the globally-averaged clumping factor, $C_{\rm global}$, to be
\begin{equation}
    \label{eq:C_global}
    C_{\rm global}(z) = \int dz_{\rm re} P(z_{\rm re}) C(z_{\rm re},z)
\end{equation}
where $P(z_{\rm re})$ is the PDF of reionization redshifts (that is, the redshift derivative of the reionization history\footnote{Note that this is because, if the reionization history is $x_i(z)$, $dx_i/dz$ counts the differential fraction of the IGM that re-ionizes at redshift $z$ per unit redshift.  This quantity can be interpreted as a PDF of reionization redshifts.  }).  Here, $C(z_{\rm re},z)$ can be any of the clumping factors described in the previous section.  We then use the same approach described in Refs.~\cite{DAloisio2020,Long2022}, based on the ``ionization accounting'' equation of Ref.~\cite{Madau1999} (their Eq. 20), to calculate the reionization history for a given ionizing emissivity, $\dot{N}_{\rm ion}$.  In what follows, we use the $\dot{N}_{\rm ion}$ for the \textsc{late start/late end} and \textsc{early start/late end} models from Ref.~\cite{Cain2024b} (see their Figure 4).  Note that when we use $C_{\rm A,n_{\rm HI}}$ in this equation, we are directly counting the number of ionizations in the gas, thus relaxing the assumption of Ref.~\cite{Madau1999} that ionizations are balanced by recombinations.  
In these calculations, we use three definitions of $C$.  Two of these are $C_{\rm R}^{\Gamma/\langle \Gamma \rangle > 0.01}$ and $C_{\rm A,n_{\rm HI}}^{\rm total}$, taken from our suite of mean density ($\delta/\sigma = 0$) simulations.  We also adopt a second estimation of $C_{\rm A,n_{\rm HI}}$, where we first perform a Gaussian quadrature average over $\delta/\sigma = 0$ and $\pm \sqrt{3}$ before taking the average over $z_{\rm re}$ in Eq.~\ref{eq:C_global}.  This procedure, which is explained in the main text and Appendix B of Ref.~\cite{DAloisio2020}, approximately accounts for the effect of density fluctuations above the box scale, giving a more accurate estimate of the total clumping factor for the entire IGM.  We denote this averaged quantity by $C_{\rm A,n_{\rm HI}}^{\rm total,DC}$.  

\begin{figure}[h!]
    \centering
    \includegraphics[scale=0.21]{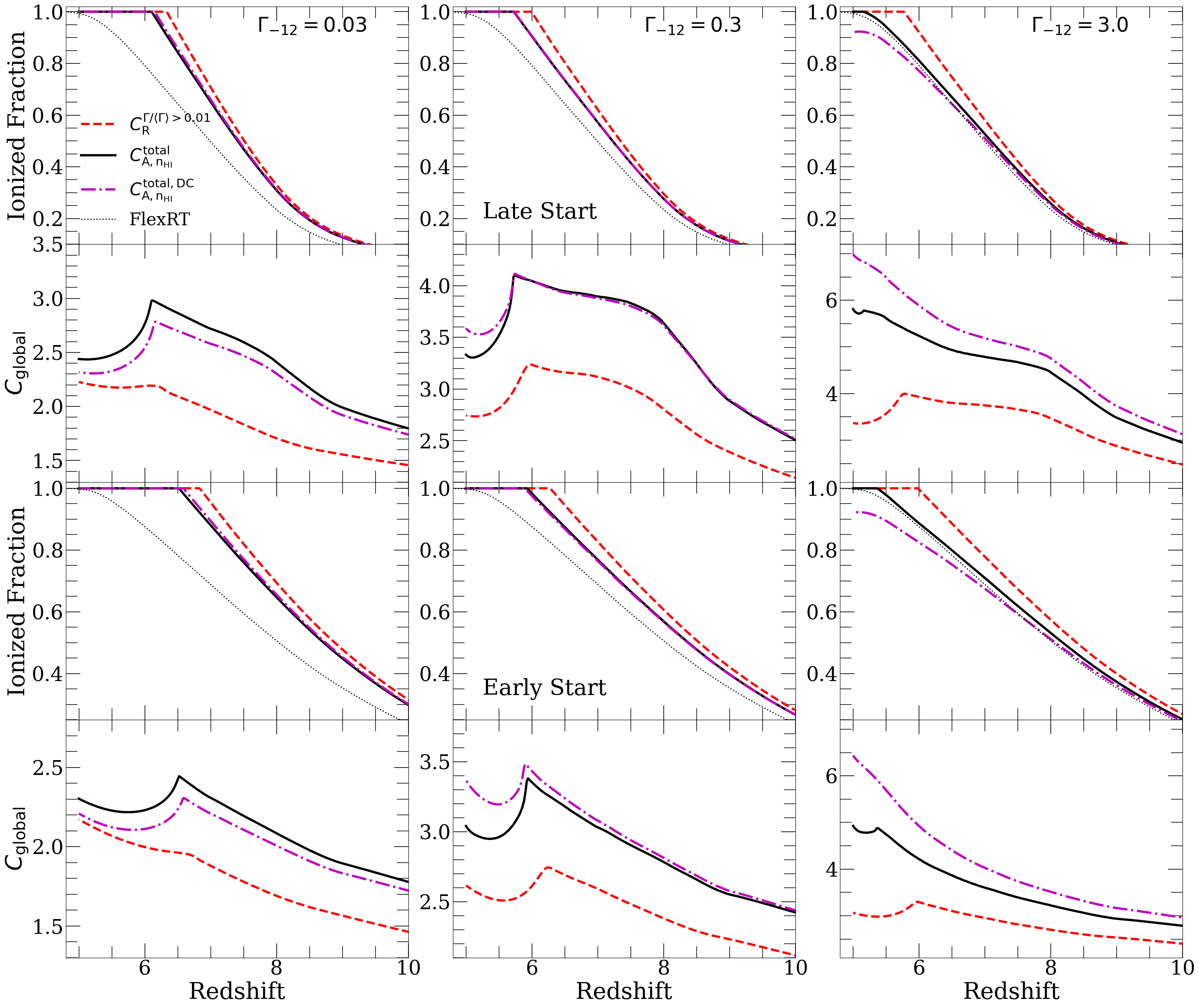}
    \caption{Reionization histories using different definitions of the clumping factor, estimated using the ionization counting equation from Ref.~\cite{Madau1999}.  The top two rows show results for the \textsc{late start/late end} model of Ref.~\cite{Cain2024b}, and the bottom two rows show results for the \textsc{early start/late end} case.  The first and third rows show the reionization history, and the second and fourth show $C_{\rm global}$, for three values of $\Gamma_{\rm HI}$ (increasing left to right).  The red dashed, black solid, and magenta dot-dashed lines denote results obtained using $C_{\rm R}^{\Gamma/\langle \Gamma \rangle > 0.01}$, $C_{\rm ion,n_{\rm HI}}^{\rm total}$, and $C_{\rm ion,n_{\rm HI}}^{\rm total,DC}$, respectively.  The thin dotted line in the reionization history panels shows the result from the corresponding FlexRT simulation (see Figure 1 of Ref.~\cite{Cain2024b}).  }
    \label{fig:reion_histories_clumping}
\end{figure}

We show the results of this exercise in Figure~\ref{fig:reion_histories_clumping}.  The top two rows show results for the \textsc{late start/late end} emissivity model from Ref.~\cite{Cain2024b}, with the first row showing the reionization history and the second showing $C_{\rm global}$.  The columns show results for different values of $\Gamma_{\rm HI}$, increasing from left to right.  The red dashed, black solid, and magenta dot-dashed curves denote $C_{\rm R}^{\Gamma/\langle \Gamma \rangle > 0.01}$, $C_{\rm A,n_{\rm HI}}^{\rm total}$, and $C_{\rm A,n_{\rm HI}}^{\rm total,DC}$ clumping factor definitions, respectively.  In the top row, the thin dotted line shows the reionization history from FlexRT\footnote{Note that the IGM opacity in FlexRT is calibrated using a suite of high-resolution IGM simulations with a setup similar to \textsc{saguaro}~\citep{Cain2021,Cain2022b}.  }, as shown in Figure 1 of Ref.~\cite{Cain2024b}.  The bottom two rows are identical to the top two, but for the \textsc{early start/late end} model.  For models assuming $\Gamma_{-12} = 0.03$ and $0.3$, our clumping factor-based models always finish reionization earlier than the corresponding FlexRT simulation.  For $\Gamma_{-12} = 0.03$, this difference is $\Delta z \approx 1.25$ ($1.5$), and $\Delta z \approx 0.75$ ($1$) for $\Gamma_{-12} = 0.3$ for the \textsc{Late Start/Late End} (\textsc{early Start/Late End}) case.  The spread in reionization end time between different clumping factors is smaller - about $0.25$ in redshift.  However, for $\Gamma_{-12} = 3$, reionization finishes around the same time as in FlexRT in the $C_{\rm A,n_{\rm HI}}^{\rm total}$, slightly later with $C_{\rm A,n_{\rm HI}}^{\rm total,DC}$, and still earlier when using $C_{\rm R}^{\Gamma/\langle \Gamma \rangle > 0.01}$.   
 
We see the origin of the differences between clumping factor models in the second row.  As we would expect from Figure~\ref{fig:ion_vs_recomb_clumping_example}, $C_{\rm A,n_{\rm HI}}^{\rm total}$ is always higher than $C_{\rm R}^{\Gamma/\langle \Gamma \rangle > 0.01}$, resulting in a later end to reionization.  Consistent with the findings of Ref.~\cite{DAloisio2020}, we find that $C_{\rm A,n_{\rm HI}}^{\rm total,DC}$ is reasonably similar to the mean density case, although not to the same degree for all values of $\Gamma_{-12}$.  Specifically, we find very close agreement between the two for $\Gamma_{-12} = 0.3$, while for $\Gamma_{-12} = 0.03$ ($3$), the DC mode average comes in slightly below (above) the mean-density case.  We can understand this behavior by considering the role of over-dense gas in setting the clumping factor.  At higher $\Gamma_{-12}$, the self-shielding density is higher, and thus the highest density peaks contribute more to the total absorption rate.  When $\Gamma_{\rm HI}$ is low, over-dense patches of the IGM play a smaller role in setting the absorption rate.  In this case, the under-dense patches ``win'' out over the over-dense ones in the DC mode average, bringing the total clumping factor lower.  The opposite occurs when $\Gamma_{-12}$ is high.  We see that $\Gamma_{-12} = 0.3$ is the ``sweet spot'' where the contribution of over and under-dense regions to the average approximately cancel out.   

We find fairly similar results for the \textsc{early start/late end} model.  In this case, all clumping factors are slightly lower, since the reionization history is more extended and hence ionized regions have more time to dynamically relax before reionization ends globally.  The delay in the reionization history in FlexRT relative to our models is slightly longer in the $\Gamma_{-12} = 0.03$ case, but is similar for the other two $\Gamma_{-12}$ to the \textsc{late start/late end} scenario.  The differences between clumping factors are also similar, suggesting that our findings in this section are robust to modest differences in the duration of reionization.  We also see a similar trend in how the DC-mode average clumping factor relates to the mean density one.  

There are several potential contributors to the overall delay in reionization in the FlexRT simulation relative to most of the models shown in Figure~\ref{fig:reion_histories_clumping}.  First, FlexRT solves the radiative transfer equation self-consistently, which accounts for finite speed of light effects that slightly delay the reionization history as compared to our analytic model~\citep{Deparis2019,Cain2024d}.  Second, the simulations upon which FlexRT is based have a factor of $2^3$ higher spatial resolution than the Core \textsc{saguaro} runs, so they may capture more absorption due to small-scale structure.  Lastly, FlexRT accounts for correlations between fluctuations in $\Gamma_{\rm HI}$, ionized fraction, and density that is not accounted for in our simple model.  Since the recombination rate increases with both $\Gamma_{\rm HI}$ and density, our simple model is likely under-predicting the recombination rate for a fixed mean $\Gamma_{\rm HI}$.  These issues highlight the limitations of simple analytic approaches like the one we employ here, demonstrating the need for self-consistent RT simulations.

\section{The dark matter halo mass function}
\label{sec:dm_halos}

In this section, we briefly study the dark matter halo mass function (DM-HMF) in \textsc{saguaro}.  We identify halos from the DM particle distributions using the publicly available \textsc{rockstar} phase-space halo-finding code~\citep{Behroozi2013}.  Our DM particle mass is $5 \times 10^2$ $h^{-1}M_{\odot}$ for the $2$ $h^{-1}$Mpc boxes, and $\approx 1$ $h^{-1}M_{\odot}$ in the $=0.25$ $h^{-1}$Mpc boxes.  In our results, we show halos of all masses identified by \textsc{rockstar} down to a minimum particle count of $20$, but note that Ref.~\cite{Leroy2020} found that around $50-100$ particles are required for robust halo identification in \textsc{rockstar}.  As such, we expect the minimum robust halo mass for these two cases to be at least $M_{\rm halo}^{\min} \approx 5\times10^4$ and $100$ $h^{-1}M_{\odot}$, respectively.  
In what follows, we show spherical over-density halo masses estimated such that the mean density within the halo is $200\times$ the critical density (that is, $M_{200}^c$), although in Appendix~\ref{app:halos} we show how things change if we use the friends-of-friends virial mass.  
One of the input parameters in \textsc{rockstar} is the force resolution, which sets the minimum size of structures that the code can identify as halos.  Since the force resolution is not well-defined in PM gravity solvers, we have set this parameter equal to a small fraction of our grid size ($0.1$ and $0.01$ $h^{-1}$kpc for our Core and Core-HR runs, respectively), and have checked that further lowering it does not change our results (see Appendix~\ref{app:halos}).

In Figure~\ref{fig:dm_density}, we show slices through the DM density at $z = 4$.  We show the fiducial Core model (panel A), over-dense run with $\delta/\sigma = +\sqrt{3}$ (B), our WDM model with $m_{\rm X} = 5$ keV (C), and our fiducial Core-HR run (D).  The \textsc{rockstar} halos with centers within the $2$ $h^{-1}$kpc slice are overlaid as red dots, with size proportional to the halo virial radius.  As expected, we see that halos are found along filaments and nodes in the cosmic web.  The halos are more concentrated in the over-dense run than in the mean density case, reflecting the fact that more of the mass is collapsed onto the most massive filaments and nodes.  The WDM model has only a few halos, concentrated at the highest density peaks.  In this model, dark matter free-streaming prevents bottom-up structure formation in smaller filaments, resulting in a dearth of halos at masses below the free-streaming scale.  The DM and halo distribution in our small volume looks remarkably similar to the others, especially the over-dense $2$ $h^{-1}$Mpc box.  This is because the two box sizes were initialized with the same Gaussian random field, and because dark matter formation is self-similar down to earth-mass scales in CDM.  

\begin{figure*}
    \centering
    \includegraphics[scale=0.37]{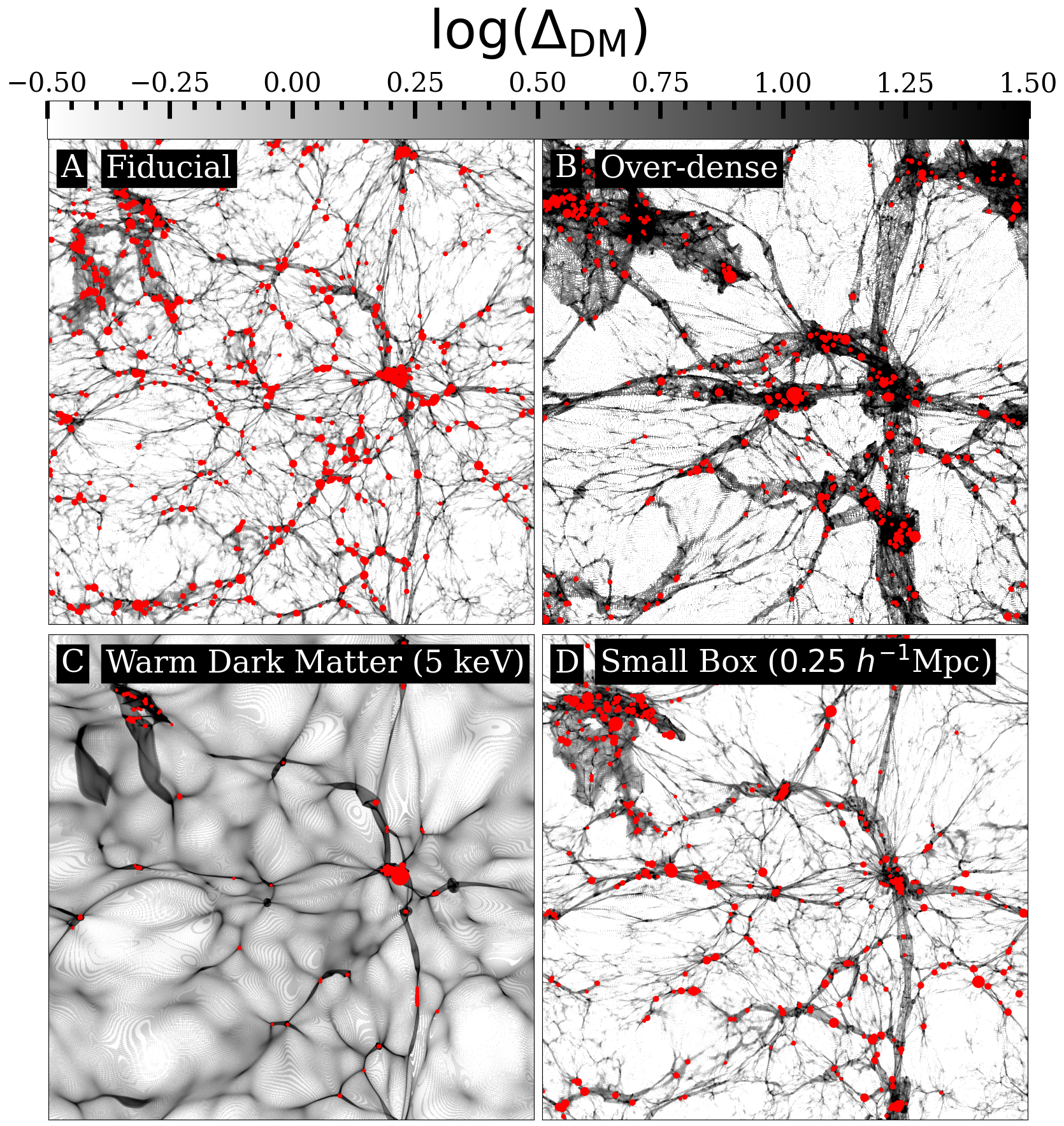}
    \caption{Slices through the dark matter density field at $z = 4$ in our fiducial Core simulation (A), over-dense ($\delta/\sigma = +\sqrt{3}$, B), the warm dark matter model with $m_{\rm X} = 5$ keV (C), and our fiducial small-box run ($L_{\rm box} = 0.25$ $h^{-1}$Mpc, D).  Red points denote the locations of dark matter halos identified with \textsc{rockstar}, with point sizes scaled by the virial radius of the halo.  }
    \label{fig:dm_density}
\end{figure*}

\begin{figure*}
    \centering
    \includegraphics[scale=0.21]{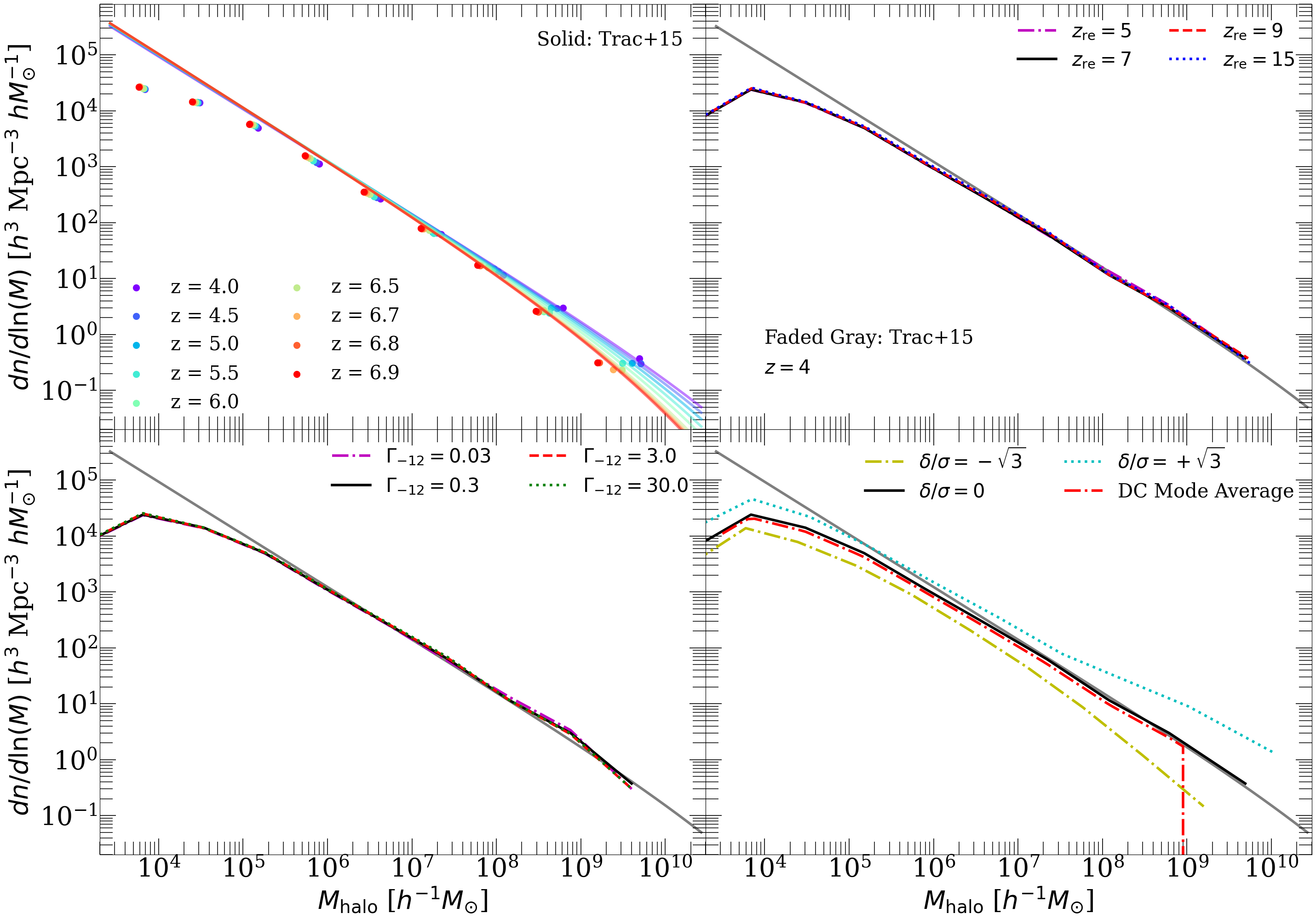}
    \caption{Dark matter halo mass function in Core \textsc{saguaro} runs.  {\bf Top Left:} Evolution of the DM-HMF with redshift in our fiducial Core run.  Points indicate simulation data, and the solid curves denote the DM-HMF from Ref.~\cite{Trac2015}.  Colors denote different redshifts (see legend). {\bf Remaining Panels:} The DM-HMF at $z = 4$ as a function of different box-scale parameters.  There is little dependence on $\Gamma_{-12}$ or $z_{\rm re}$, indicating that reionization has a minimal effect on dark matter structure at the level of the DM-HMF.  }
    \label{fig:HMF}
\end{figure*}

Figure~\ref{fig:HMF} shows the DM-HMF in several \textsc{saguaro} simulations.  The dots in the top left panel show the DM-HMF in our fiducial Core run, with different color points denoting different redshifts (see legend).  The solid lines show the DM-HMF at those redshifts calibrated from N-body simulations\footnote{These simulations were DM-only and thus their particle masses differ from ours by a factor of $\Omega_{\rm m}/\Omega_{\rm dm}$ (at fixed cosmology).  We correct for this offset when plotting the Ref.~\cite{Trac2015} DM-HMF.  } by Ref.~\cite{Trac2015}.  The agreement between the simulation and the Ref.~\cite{Trac2015} expectation is reasonably good at $M_{\rm halo} \gtrsim 10^6$ $h^{-1}M_{\odot}$, with the simulated DM-HMF falling off at smaller masses due to lack of mass resolution.  Interestingly, in agreement with the Ref.~\cite{Trac2015} expectation, the DM-HMF evolves very little with redshift at $M_{\rm halo} \lesssim 10^8$ $h^{-1}M_{\odot}$, indicating that halos in this mass range form at a similar rate that they are consumed by mergers with larger halos.  
The remaining panels show the DM-HMF as a function of different box-scale parameters at $z = 4$, with the gray line showing the Ref.~\cite{Trac2015} expectation.  The top right and bottom left panels show that $z_{\rm re}$ and $\Gamma_{\rm HI}$ have a small effect on the DM-HMF, suggesting that the differences in baryonic structure caused by these parameters does not substantially change dark matter structure at the level of the DM-HMF (although it may affect the internal structures of halos, which we do not consider here).  This is true even for the smallest halos we simulate, which have nearly all their baryonic content removed by reionization.  

In the bottom right, we see that $\delta/\sigma$ has a significant effect on the DM-HMF, which we expect since it modulates the total matter density.  Boxes with larger $\delta/\sigma$ form more massive halos, with the most massive halo in the over-dense box being $1.7 \times 10^{10}$ $h^{-1}M_{\odot}$ at $z = 4$ (compared to $9.1 \times 10^{9}$ and $4.1 \times 10^{9}$ $h^{-1}M_{\odot}$ for $\delta/\sigma = 0$ and $-\sqrt{3}$, respectively).  The red dot-dashed line shows the Gaussian quadrature average using all three simulations, which agrees well with the mean density case and the Ref.~\cite{Trac2015} expectation across most of the resolved mass range.  Note that we cut off the DC mode average at $10^9$ $h^{-1}M_{\odot}$, since the under-dense volume lacks halos at higher masses.  

\begin{figure*}
    \centering
    \includegraphics[scale=0.21]{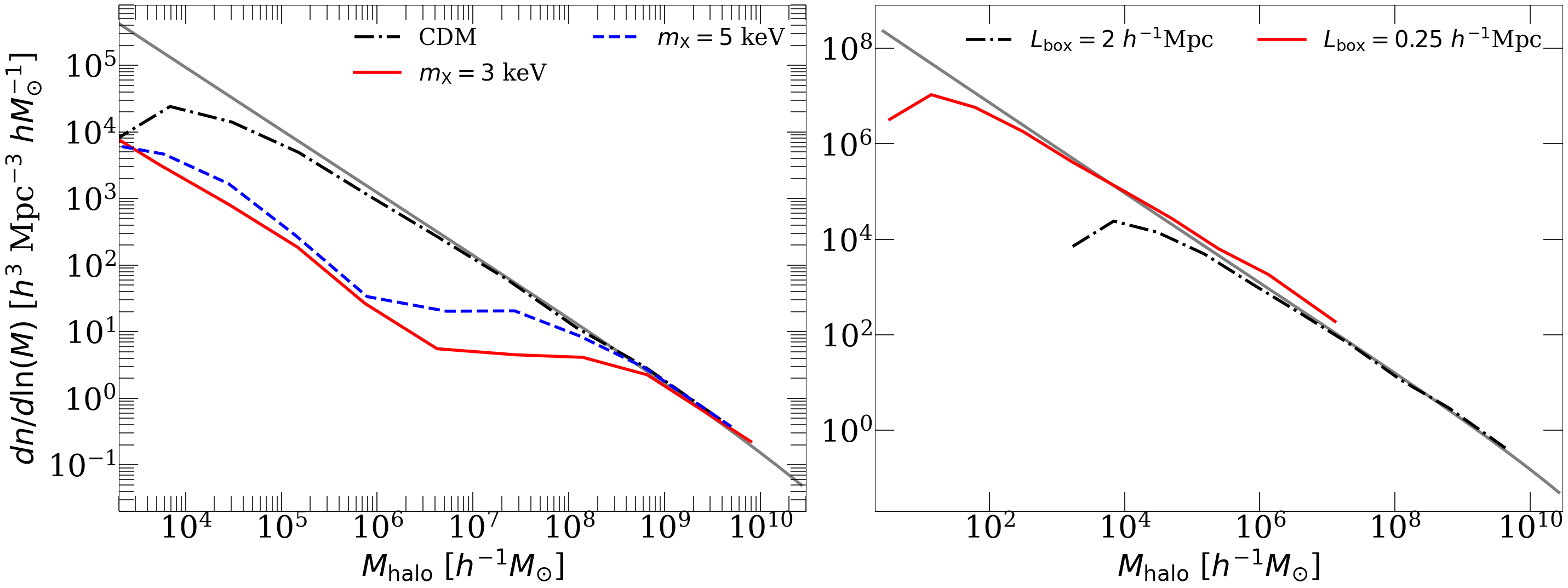}
    \caption{Dependence of DM-HMF on the assumed dark matter model ({\bf left}) and on box size ({\bf right}).  See text for details.  }
    \label{fig:HMF_WDM_small}
\end{figure*}

In Figure~\ref{fig:HMF_WDM_small}, we show the DM-HMF for two more cases of interest.  The left panel compares our fiducial Core sim to WDM runs with $m_{\rm X} = 3$ and $5$ keV.  We see that the WDM runs only agree with the fiducial run (and the Ref.~\cite{Trac2015} result) at the highest halo masses, and they fall below it by over and order of magnitude at lower masses.  This reflects the effect free streaming at small scales in WDM, which suppresses the formation of the smallest halos.  The run with higher $m_{\rm X}$ cuts off at a slightly lower mass scale, reflecting its smaller free-streaming scale~\citep{Viel2005}.  
The right panel compares the DM-HMF from the fiducial Core and Core-HR runs.  These curves have similar shapes, but are offset in halo mass by about a factor of $10^3$, which is expected given their different box sizes.  The HR box does not reach the $\sim 10^{8}$ $h^{-1}M_{\odot}$ threshold where halos are able to retain their gas after reionization, explaining why no self-shielding gas survives in that case.  Unlike the Core run, however, it is complete to masses well below $\sim 10^{4}$ $h^{-1}M_{\odot}$, thus fully resolving the Jeans scale of the cold, pre-ionized IGM~\citep{DAloisio2020}.  It is interesting that the small box shows an excess of halos (by a factor of $\approx 1.7$) relative to the Ref.~\cite{Trac2015} expectation (and our $2$ $h^{-1}$Mpc DM-HMF) at $M_{\rm halo} > 10^4$ $h^{-1} M_{\odot}$.  We have checked that this high-mass excess becomes less pronounced (factor of $\approx 1.3$) at $z = 7.2$.  

One possible reason for the excess in halos at the high-mass end, and disagreement between the two box sizes where there mass functions overlap, is the fact that structure formation at or above the box scale begins to go non-linear long before $z = 4$ in the HR runs.  As a result, it is reasonable to expect that missing mode coupling with scales greater than $L_{\rm box}$ may lead to inaccuracies in the formation of small-scale structure.  
At $z = 4$, the non-linear scale, $k_{\rm NL}$, is $\sim 1$ $h^{-1}$Mpc, close to our Core box size and a factor of a few larger than our HR box size.  Another possible culprit could be that Rockstar is over-linking particles, and thus identifying too many halos at the high-mass end.  In Appendix~\ref{app:halos}, we investigate the effects of several numerical parameters in \textsc{rockstar}, which may affect over-linking, on our recovered DM-HMFs.  We find that while some parameter choices do result in over-linking, the DM-HMF is reasonably insensitive to these, and that our fiducial parameter choices do not show evidence of over-linking.  Furthermore, the offset between our fiducial Core and Core-HR DM-HMFs at $10^{5} \lesssim M_{\rm halo}/[h^{-1}M_{\odot}] \lesssim 10^7$ is robust to these parameter choices, reinforcing our hypothesis non-linear structure formation above our HR box scale is the main culprit.  Appendix~\ref{app:halos} provides further details.  

In the post-EoR IGM, HI is found almost entirely within self-shielded dark matter halos.  The 21$\,$cm emission from HI in these halos is being targeted by a number of current and forthcoming intensity mapping experiments, including CHIME~\cite{CHIMEoverview}, CHORD~\cite{CHORDwhitepaper}, the SKA Observatory~\citep{Santos2015}, HIRAX~\cite{crichton2022-hirax}, and FAST~\citep{BigotSazy2016}. CHIME has provided the first direct measurement of the 21$\,$cm auto spectrum at $z \approx 1$~\cite{CHIME2025} and can access up to $z=2.5$, while CHORD and SKA-Low will enable measurements at higher redshifts, approaching the end of reionization. 
At such redshifts, the influence of reionization of the post-EoR 21$\,$cm signal must be fully characterized, for two reasons: to ensure that it does not bias cosmological inference from intensity mapping measurements, and to reveal whether we can extract information about reionization from the post-EoR signal.
This influence has been studied in Refs.~\cite{uptonsanderbeck2019-UVB21cm,modi2019-HVsims,long2022-streaming,long2023-EoR21cm,giri2024-endofEoRsims,incley2026-EoR21cm}, but is more commonly ignored in currently-available modeling approaches, including 
perturbation theory~\cite{pourtsidou2023-forecasts,obuljen2023-fieldlevelHI,foreman2024-HIstoch}), hydrodynamical simulations~\cite{villaescusa-navarro2018}, semi-analytical models~\cite{wolz2016-sam,spinelli2020-sam,li2024-sam}, and HI halo occupation distribution models~\cite{padmanabhan2017-halomodel,wolz2019-halomodel,chen2021-halomodel}. 
\textsc{saguaro} has the potential to shed new light on the relationship between reionization properties and the HI-rich halos that contribute to the 21$\,$cm signal, with higher spatial resolution than previous studies,
including the contribution from halos much smaller than $10^{8}\,M_{\odot}$, which are typically un-resolved in previous simulations. We plan to explore this in future work.

\section{Ly$\alpha$ transmission properties}
\label{sec:Lya}

The transmission of Ly$\alpha$ photons is an important probe of conditions in the IGM~\cite{Becker2012,Wilson2022}, including reionization~\citep{Kulkarni2019,Mason2020,Bosman2021,Park2021b}, the physics of dark matter~\citep{Irsic2019}, and cosmology.  Ly$\alpha$ damping wing absorption, due to both the neutral IGM and self-shielded systems, has also been used to study the IGM conditions deep into the EoR~\citep{Mason2018}.  We make mock Ly$\alpha$ transmission spectra by integrating the Ly$\alpha$ opacity along $N = 1000$ sightlines with random starting points and orientations in our boxes.  Our Ly$\alpha$ forest calculations span a total path length of over $7 \times 10^6$ km/s, more than enough to calculate a converged 1-D flux power spectrum.  In our Ly$\alpha$ damping wing calculations, we integrate for a co-moving distance of $80$ $h^{-1}$Mpc, sufficient to converge on the damping wing optical depth.  In both cases, our sightlines wrap multiple times through the box, leveraging the simulation's periodic boundary conditions.  The resulting Ly$\alpha$ forest and damping wing transmission properties we recover from this exercise are idealized in the only probe the average transmission properties of what is, in reality, a small patch of the IGM.  As such, our calculations are not intended to produce realistic {\it global} IGM Ly$\alpha$ transmission properties, but rather to assess the differences between small patches of the IGM embedded in different large-scale environments.  

\subsection{Ly$\alpha$ forest flux power spectrum}

\begin{figure}
    \centering
    \includegraphics[scale=0.24]{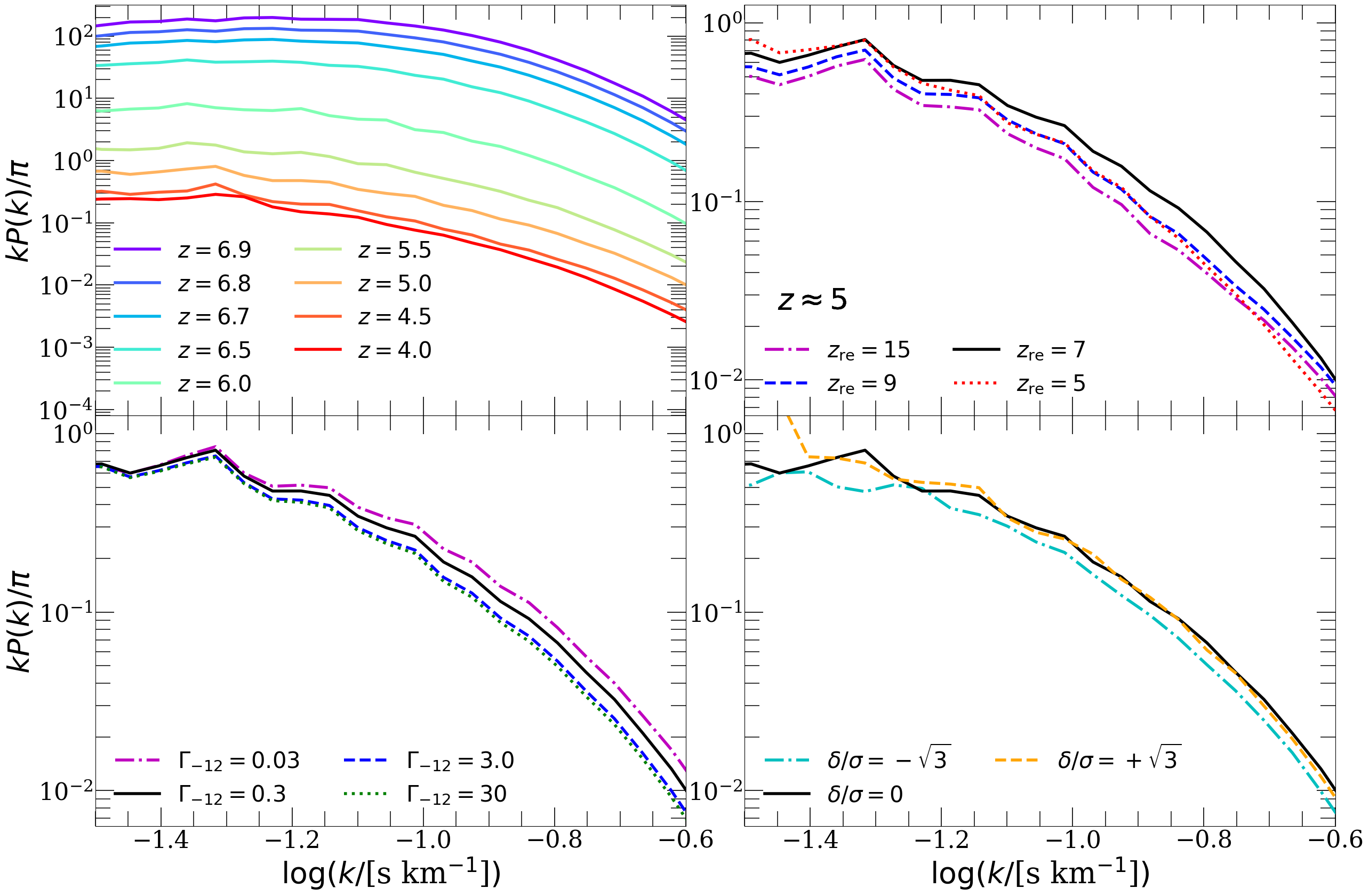}
    \caption{Ly$\alpha$ forest power spectrum in a subset of Core simulations.  The top left panel shows redshift evolution, the remaining panels show different values of $z_{\rm re}$ (top right), $\Gamma_{-12}$ (bottom left), and $\delta/\sigma$ (bottom right) at $z = 5$.  In all cases, optical depths in highly ionized cells are re-scaled such that the mean transmission matches recent observations at each redshift.  See text for further details.  }
    \label{fig:power_spectrum_plot}
\end{figure}

The Ly$\alpha$ forest power spectrum is a commonly-used statistic in studies seeking to use the forest to constrain IGM properties, including its the thermal structure of the IGM~\citep{Wilson2022} and the nature of dark matter~\citep{Irsic2024}.  When we calculate transmission spectra, we re-scale the optical depths in highly ionized ($x_{\rm HI} < 0.1$) cells such that we match, at each redshift, the mean transmission measured from Refs.~\cite{Becker2012,Bosman2021}.  We show the results in Figure~\ref{fig:power_spectrum_plot} as a function of wavenumber in velocity space. The top-left panel shows the redshift evolution from $z = 6.9$ to $4$ for our fiducial Core sim, and the remaining panels show the dependence on the box-scale quantities.  

In the top left panel, we see that the redshift evolution of the flux power spectrum is dominated by evolution in the mean flux (note that we extrapolate measurements from Ref.~\cite{Bosman2021} for $z > 6.2$).  By contrast, the other parameters introduce interesting shape differences at fixed $z$.  At fixed $z$, higher $z_{\rm re}$ (top right) generally results in less small-scale flux power, due to two effects: (1) the density field has less small-scale structure when it has been re-ionized for longer (Jeans smoothing, see Ref~\cite{Nasir2016}) and (2) small-scale temperature fluctuations left over from relaxation have had more time to disappear~\citep{Hirata2018,Cain2024a}.  The exception is the red dotted curve, for which reionization has {\it just} happened ($z_{\rm re} = 5$, $z = 4.95$).  In this case, there is slightly more power at the low end of our $k$ range, but the flux power spectrum declines more steeply at high $k$.  The faster fall-off at high $k$ probably owes to the temperature being nearly homogeneous (panel E of Figure~\ref{fig:example_temperature}) and higher on average in transmissive voids, resulting fewer $T$ fluctuations and more thermal line broadening.  At the high-$k$ end, the enhanced clumpiness of the gas likely offsets these effects, resulting in more power.  

In the bottom left, we see that $\Gamma_{-12}$ has a modest, but interesting effect on small-scale flux power.  
Recall that our procedure of re-scaling optical depths to match mean flux measurements effectively re-scales $\Gamma_{\rm HI}$ in ionized gas, meaning that the only differences we should see are due the effect of $\Gamma_{-12}$ on gas temperature, self-shielding and structure formation.  
At higher $\Gamma_{-12}$, we expect higher temperatures, less small-scale structure in the density field, and less self-shielded gas, all of which should suppress small-scale power.  This is exactly what we see.  While there is little difference between runs with $\Gamma_{-12} = 30$ and $3$, small-scale power grows by a factor of $\approx 1.5$ between $3$ and $0.03$ at $\log(k/[{\rm s\text{ }km}^{-1}]) > -1$.  
We have investigated the origin of these differences and found that the dominant effect is that of temperature.  Simulations that re-ionize at higher $\Gamma_{-12}$ end up hotter after ionization due to their faster I-front speeds (see bottom-middle of Figure~\ref{fig:fractions_T} and Appendix~\ref{app:treion}.)  Thus, at higher $\Gamma_{-12}$, thermal broadening smooths out more of the small-scale features in the Ly$\alpha$ forest, suppressing the flux power spectrum.  A sub-dominant but non-negligible effect also arises from damping wing features imprinted by self-shielded gas on the forest~\citep{Park2024} and differences in the structure of the density field arising from the effect of $\Gamma_{-12}$ on mini-halo photo-evaporation. These latter effects may be partially or entirely missed in Ly$\alpha$ forest simulations that have low resolution and/or do not model self-shielding.   In the bottom right, we see that local over-density has a modest impact on flux power.  There is slightly less power in the under-dense case, probably reflecting the reduction in small-scale structure.  Interestingly, the over-dense case is similar to the mean density run, suggesting that it has a similar distribution of small-scale voids to the mean-density case.  It may also be due to a cancellation of competing effects - namely, the over-dense run has more damping wing absorption, but also higher temperatures.  We defer a more through investigation to future work.  Our findings in this section reflect those of earlier studies noting the importance of small-scale structure evolution for the forest flux power, and its sensitivity to reionization at $z \lesssim 5$.  In future work, we plan to study these effects in more detail in \textsc{saguaro}, and their possible implications for constraints on alternative DM cosmologies from the Ly$\alpha$ forest~\citep[e.g.][]{Irsic2024,GarciaGallego2026}.  

\subsection{Ly$\alpha$ damping wing absorption}
\label{subsec:lyadamping}

\begin{figure}
    \centering
    \includegraphics[scale=0.295]{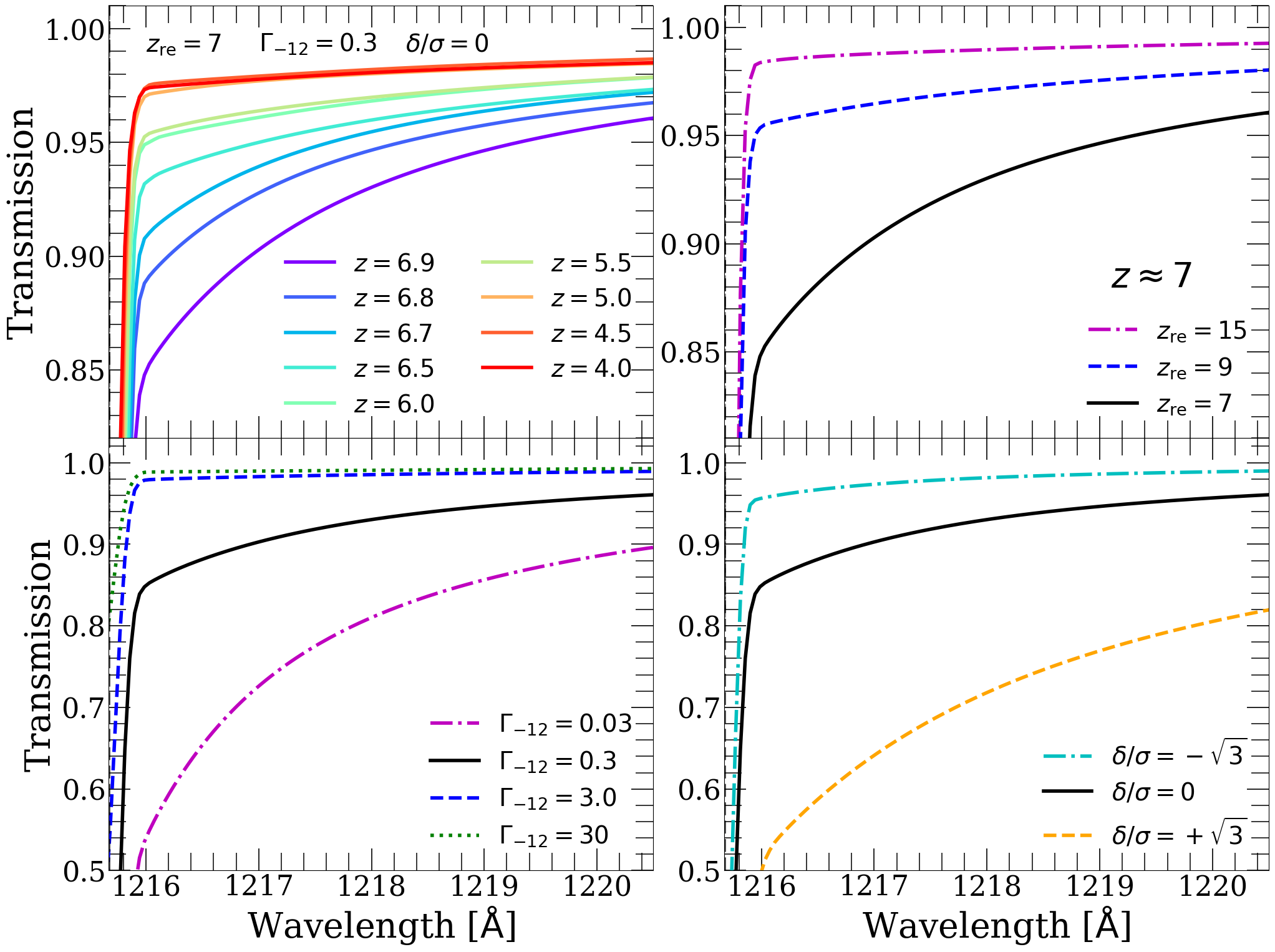}
    \caption{Ly$\alpha$ damping wing transmission in \textsc{saguaro}.  The layout and curves are in the same format as Figure~\ref{fig:power_spectrum_plot}, except that the fix-redshift panels are shown at $z \approx 7$, close to when the universe is expected to be half ionized.  {\bf Top Left}: Redshift evolution in our fiducial Core simulation.  Damping wing absorption is significant at $z = 6.9$, and mostly disappears by $z = 4$.  {\bf Top Right}: sensitivity to $z_{\rm re}$ at $z \approx 7$, close to reionization's likely midpoint.   {\bf Bottom Left}: dependence on $\Gamma_{-12}$, at the same redshift.  {\bf Bottom Right}: Dependence on the box-scale density parameter. See text for details}
    \label{fig:Lya_transmission}
\end{figure}

Self-shielding systems can also contribute non-negligibly to Ly$\alpha$ damping wing absorption in galaxy and quasar spectra during reionization~\citep{Choudhury2015,Mesinger2015,Park2021b}.  In \textsc{saguaro}, the diffuse IGM is highly ionized at $z < z_{\rm re}$,  so any damping-wing absorption come from self-shielded systems.  We show the Ly$\alpha$ damping wing transmission averaged over our $1000$ sightlines through our fiducial Core simulation in the top-left panel of Figure~\ref{fig:Lya_transmission} between $z = 6.9$ and $4$.  We see significant damping-wing absorption at $z = 6.9$, with transmission fractions of $\approx 85\%$ and $\approx 95\%$ at $1216\text{\AA}$ and $1220\text{\AA}$, respectively.  Most of this absorption disappears by $z = 5$, indicating that it was sourced by short-lived self-shielded systems that were photo-evaporated within a few hundred Myr.  A small amount of absorption ($\sim 2\%$) persists to $z = 4$ and is sourced by the largest halos in the box.  The fact that most of this absorption disappears shortly after reionization suggests it is sourced by self-shielded structures below the atomic cooling limit - that is, metal-free DLAs.  These may be missing or under-represented in simulations that probe much larger volumes that are used to model damping wing absorption towards quasars and LAEs~\citep[e.g.][]{Keating2024,Maitra2025}, which often lack radiative transfer and/or sufficient resolution to fully capture self-shielding.  

The remaining panels show the effects of varying different \textsc{saguaro} parameters.  The top right panel shows several $z_{\rm re}$ values at $z \approx 7$\footnote{We use $z = 6.9$ for the $z_{\rm re} = 7$ case.  }, close to the midpoint of reionization when Ly$\alpha$ damping wing statistics are most constraining on the IGM neutral fraction~\citep{Mason2018a,Davies2018,Durovcikova2024,Tang2024b}.  We see a considerable difference, reflecting the larger fraction of self-shielding gas remaining in recently-ionized gas.  The ionizing background strength (bottom left) has an even larger effect.  For our two highest $\Gamma_{\rm HI}$ values, most of the self-shielding structures have been ionized already by $z = 6.9$, but for $\Gamma_{-12} = 0.03$, the damping wing optical depth is a factor of a few higher than the fiducial case.  In the bottom right, we see that increasing $\delta/\sigma$ has a similar effect.  These findings hint that damping wing absorption in the spectra of high-$z$ galaxies and quasars by proximate metal-free DLAs could be important during reionization.  Suites of high-resolution simulations like \textsc{saguaro} could prove useful in correcting for this and related effects, which will become more important as observations become more precise\citep[see also e.g.][]{Choudhury2015}.  

\section{Conclusions}
\label{sec:conc}

We have run a large suite of fully coupled hydrodynamical and radiative transfer simulations, 
\textsc{saguaro}, designed to capture the dynamics of intergalactic structures near the Jeans scale of the cold, pre-ionized IGM.  \textsc{saguaro} includes two box sizes that together span nearly four orders of magnitude in spatial scale, and sample a grid spanning representative values of reionization redshift, photo-ionizing background strength, and large-scale over-density at the box scale.  Furthermore, \textsc{saguaro} captures the dynamics of self-shielding systems via an optimized ray-tracing RT treatment that is computationally efficient enough to allow us to run over $200$ simulations.  \textsc{saguaro} has the potential to facilitate detailed studies of how small-scale structure in the IGM affects both the processes that drive reionization and the observables used to probe it. These include the distribution of densities in the universe, the opacity of the intergalactic medium to H-ionizing photons, the absorption and recombination rates in the IGM, the halo mass function, and the transmission properties of Ly$\alpha$.  Our results have the potential to inform simulations of reionization at larger scales and allow for the development of improved forward-modeling methods that take small-scale structure into account.  

In addition to laying out the methodology and scope of \textsc{saguaro}, our initial investigations also revealed a number of interesting findings that deserve further follow-up.  These are summarized below:

\begin{itemize}

    \item Similar to several prior studies, we find that small-scale structure in the IGM are very sensitive to heating from reionization, with gas in halos less than $10^8$ $M_{\odot}$ evaporating on a timescale of a few hundred Myr.  For a fixed reionization history and large-scale density, we find that the details of this process depend most strongly on the assumed dark matter cosmology and pre-heating of the IGM, and less sensitively on the ionizing background strength, spectral index of the radiation field, baryon-dark matter streaming velocity, and assumptions about Case B vs. Case A recombination rates.  

    \item Echoing the findings of Refs.~\cite{Hirata2018,Cain2024a,Cain2025b}, we find that the hydrodynamic response of the IGM drives considerable temperature fluctuations at $5-10$ kpc scales, and turbulence at smaller scales under some conditions.  We characterize how these effects depend on reionization history, ionizing background, local over-density, and/or various modeling assumptions.  

    \item We find that the characteristic density at which gas self-shields is fairly insensitive to the local redshift of reionization and the local over-density of the IGM.  It scales with photo-ionization rate as $\propto \Gamma_{\rm HI}^{0.6}$ across several orders of magnitude in $\Gamma_{\rm HI}$.  These results could provide helpful inputs for treatments of self-shielding effects in analytical models and simulations.  

    \item The PDF of gas densities in the IGM evolves dramatically in response to reionization, in a way that depends sensitively on large-scale environment and the assumed physics.  Most notably, the structure of the density PDF is sensitive to the self-shielding density, as this determines the thermal structure and resulting hydrodynamics inside dense clumps.  

    \item Echoing previous work, we find that the HI column density distribution during reionization is much more complicated than a power law, having a complex double-peaked shape that is sensitive to the local properties of the IGM.  As a result, the dependence of the HI-ionizing mean free path depends on photon energy in a way that varies with both the local IGM environment and redshift.  

    \item We find that the clumping factor is sensitive not only to IGM properties, but also on how it is defined.  We find that differences in the definition of the clumping factor can lead to shifts in the ending of reionization of up to a few tenths of a redshift for observationally-motivated ionizing photon emissivity histories.  

    \item The dark matter halo mass function is (almost) completely insensitive to reionization.  

    \item The shape of the Ly$\alpha$ forest flux power spectrum on small scales (high $k$) is locally sensitive to the reionization redshift and the strength of the local ionizing background, owing to the effect these have on small-scale structure.  Such effects may be missed, to some degree, in simulations that lack sufficient resolution to capture this structure and its evolution.  

    \item In recently-ionized gas, self-shielding clumps of gas can contribute significant Ly$\alpha$ damping wing absorption at redshifts redward of systemic.  This effect is enhanced in over-dense parts of the universe and those with low ionizing background strengths.  
    
\end{itemize}

In the near future, we plan to leverage \textsc{saguaro} to significantly improve the subgrid treatment of the intergalactic opacity in the FlexRT code of Ref.~\cite{Cain2024c} using an emulation approach similar to that recently developed by Ref.~\cite{Tohfa2026}.  Crucial improvements over the existing treatment include: (1) the wider parameter space surveyed, (2) the larger box size and increased dynamic range, (3) the completion of all simulations down to $z = 4$, and (4) more comprehensive characterization of the intergalactic opacity.  These improvements will give valuable and timely new insights into the relationship between the ionizing output of galaxies and the IGM during reionization, which has important implications for the interpretation of observations of the first galaxies and AGN at high redshift with experiments such as the James Webb Space Telescope, and broadly for cosmology.  \textsc{saguaro} will also be able to help correct for inaccuracies in modeling Ly$\alpha$ forest transmission and damping wing absorption, both of which have emerged as crucial probes of the ionization state of the early universe.  

\acknowledgments

The authors thank Matthew McQuinn for helpful input on simulation setup and comments on the draft version of this manuscript.  CC thanks the Beus Center for Cosmic Foundations for support throughout the completion of this work. AD acknowledges support from NSF grant AST-2045600. ES acknowledges support from NASA grants 80NSSC22K1265, 80NSSC23K0646, and 80NSSC25K7299. 
SF acknowledges
support from the U.S.\ Department of Energy, Office of Science, Office of High Energy Physics under Award Number DE-SC0024309.  HT acknowledges support from NASA grant 80NSSC22K0821.
Some of the simulations in this work were run on the Sol supercomputer~\cite{10.1145/3569951.3597573} at Arizona State University, and the authors acknowledge ASU Research Computing for support.
Additional computations were run using computational resources distributed under NSF ACCESS allocations TG-PHY230158, TG-PHY240332, and TG-PHY250179.

\bibliography{references.bib}
\bibliographystyle{JHEP}

\appendix

\section{Sensitivity to $t_{\rm freeze}$}
\label{app:tfreeze}

\begin{figure}[h!]
    \centering
    \includegraphics[scale=0.19]{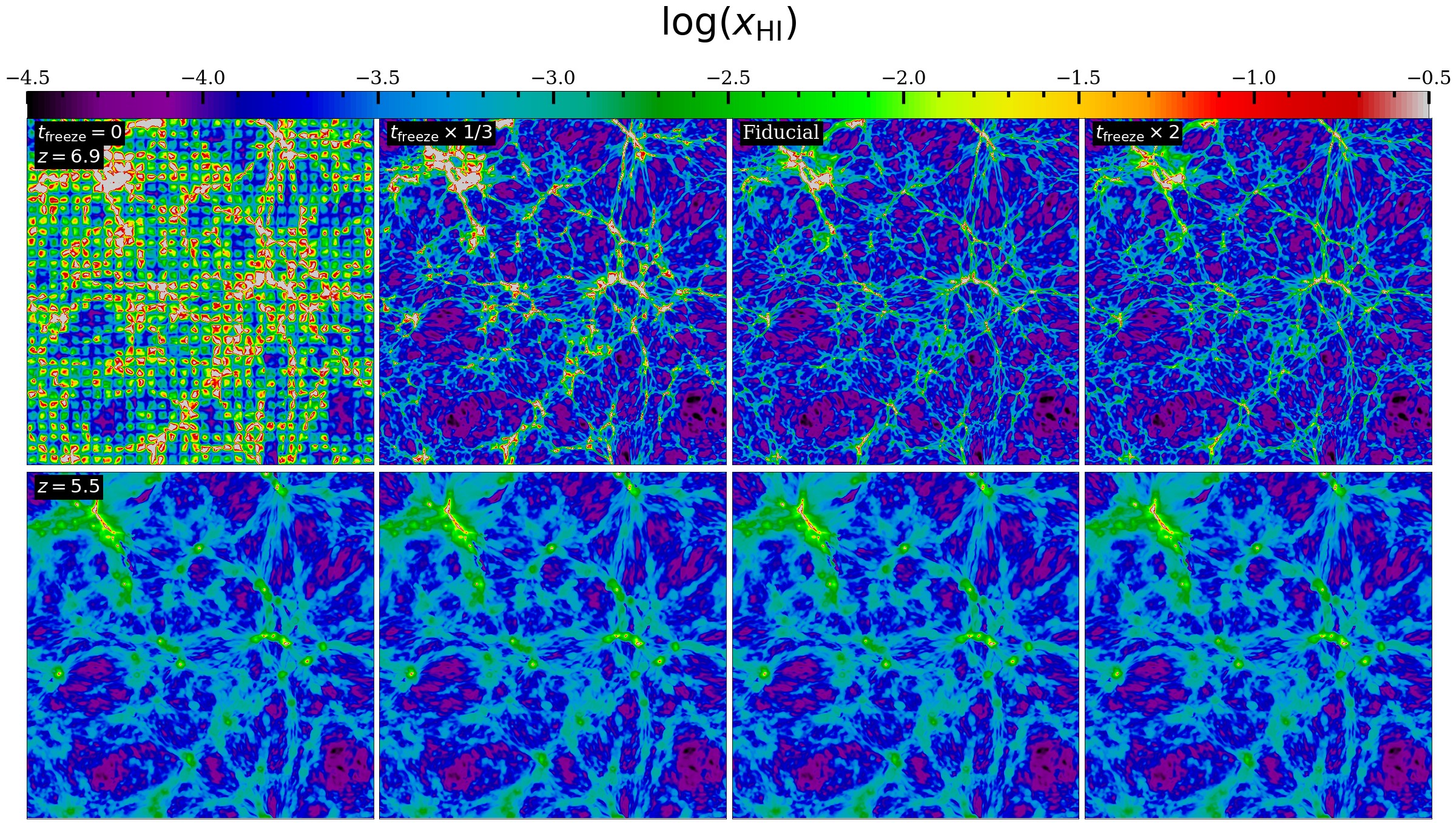}
    \caption{Maps of the neutral fraction in test runs with different values of $t_{\rm freeze}$ (increasing from left to right) at $z = 6.9$ (just after reionization) and $z = 5.5$.  See text for details.  }
    \label{fig:tfreeze_vis}
\end{figure}

In this appendix, we quantify the effect of choosing different time intervals, $t_{\rm freeze}$, during which to freeze the time evolution of the density field during initial I-front passage (see \S~\ref{subsubsec:IF}, Eq.~\ref{eq:tfreeze}).  The choice of $t_{\rm freeze}$ has the most effect on the ionization state of the gas in scenarios where I-fronts move the slowest - therefore, we run our test with $\Gamma_{-12} = 0.03$ to maximize differences.  In Figure~\ref{fig:tfreeze_vis}, we show maps of the HI fraction for tests with $t_{\rm freeze} = 0$, and $1/3\times$, $1\times$, and $2\times$ the fiducial value (Eq.~\ref{eq:tfreeze}), respectively, for our fiducial $z_{\rm re}$ and $\delta/\sigma$ values.  The top row shows maps at $z = 6.9$, and the bottom shows $z = 5.5$.

In the top-left panel, we see that failing to apply our freezing procedure results in significant artifacts in the $x_{\rm HI}$ map at $z = 6.9$, since I-fronts have not had enough time to cross the RT domains.  In this case, the neutral fraction is significantly over-estimated during this initial I-front crossing period.  The remaining panels in the top row show that as long as $t_{\rm freeze}$ is at least $1/3$ of it's fiducial value (that is, at least one I-front crossing time in mean-density neutral gas), the $x_{\rm HI}$ distribution is reasonably insensitive to this choice at $z = 6.9$.  While there is a small difference between the $1/3$ and fiducial cases, this disappears entirely by $z = 5.5$ (bottom row).  Indeed, even the $t_{\rm freeze} = 0$ case does not differ appreciably from the other three at this redshift, suggesting that our freezing procedure only important for gas properties within the first couple hundred Myr.  This test confirms the robustness of our results to the choice of $t_{\rm freeze}$.  

\section{Accuracy of $T_{\rm reion}$}
\label{app:treion}

\begin{figure}[h!]
    \centering
    \includegraphics[scale=0.38]{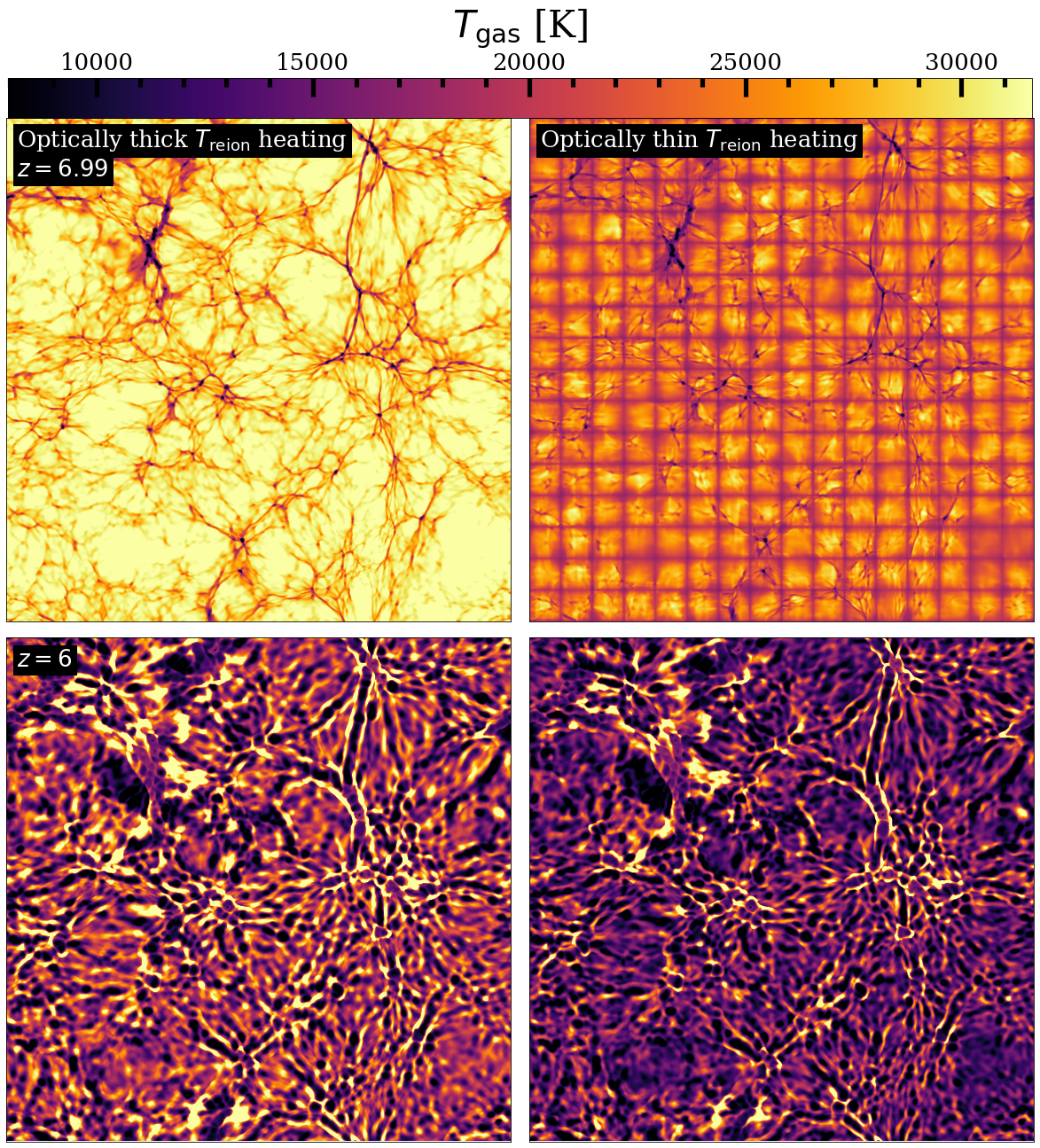}
    \caption{Map of $T_{\rm reion}$ with (top left) and without (top right) our optically thick heating approximation during I-front passage.  This procedure eliminates artifacts in the $T_{\rm reion}$ distribution that trace the structure of the RT grids.  These artifacts disappear by $z = 6$ (bottom panels), but a significant difference in the average temperature remains evident.  }
    \label{fig:treion_artifacts}
\end{figure}

We next examine the accuracy of the IGM temperature in the wake of I-fronts, $T_{\rm reion}$ (see \S~\ref{subsubsec:heating_Ifronts}) in \textsc{saguaro}.  A key change to the setup of Ref.~\cite{DAloisio2020} was to calculate heating rates during I-front passage in the optically thick limit - that is, assuming that the spectrum of absorbed photons at each point matches that of the spectrum incident at the boundaries of the RT domains.  This is needed because the domains are, in general, comparable to or smaller than the width of the I-fronts themselves, which introduces inaccuracies in the heating rates.  In Figure~\ref{fig:treion_artifacts}, we show maps of the IGM temperature $\sim 1$ Myr after I-front passage using our setup (including density freezing and optically thick $T_{\rm reion}$ heating, top left panel) and using optically thin $T_{\rm reion}$ heating as in Ref.~\cite{DAloisio2020} (top right panel).  We see that our procedure produces a smooth, nearly homogeneous $T$ distribution free of noticeable artifacts, with temperature fluctuations tracing the structure of dense filaments, as expected.  However, optically thin $T_{\rm reion}$ heating causes artifacts that clearly trace the boundaries of RT domains and under-estimates the mean $T_{\rm reion}$. (Note that the cells within these boundaries were excluded from the clumping factor and mean free path calculations in Ref. \cite{DAloisio2020} to mitigate the effects of such artifacts.)  These structures are a result of the RT domains being too small to capture the integrated heat injection across the entire I-front structure.  In the bottom row, we show that after $\sim 150$ Myr ($z \sim 6$), the domain artifacts have disappeared and leave no significant effects on the temperature fluctuations.  However, the average temperature remains lower than the optically thick case.  

\begin{figure}
    \centering
    \includegraphics[scale=0.37]{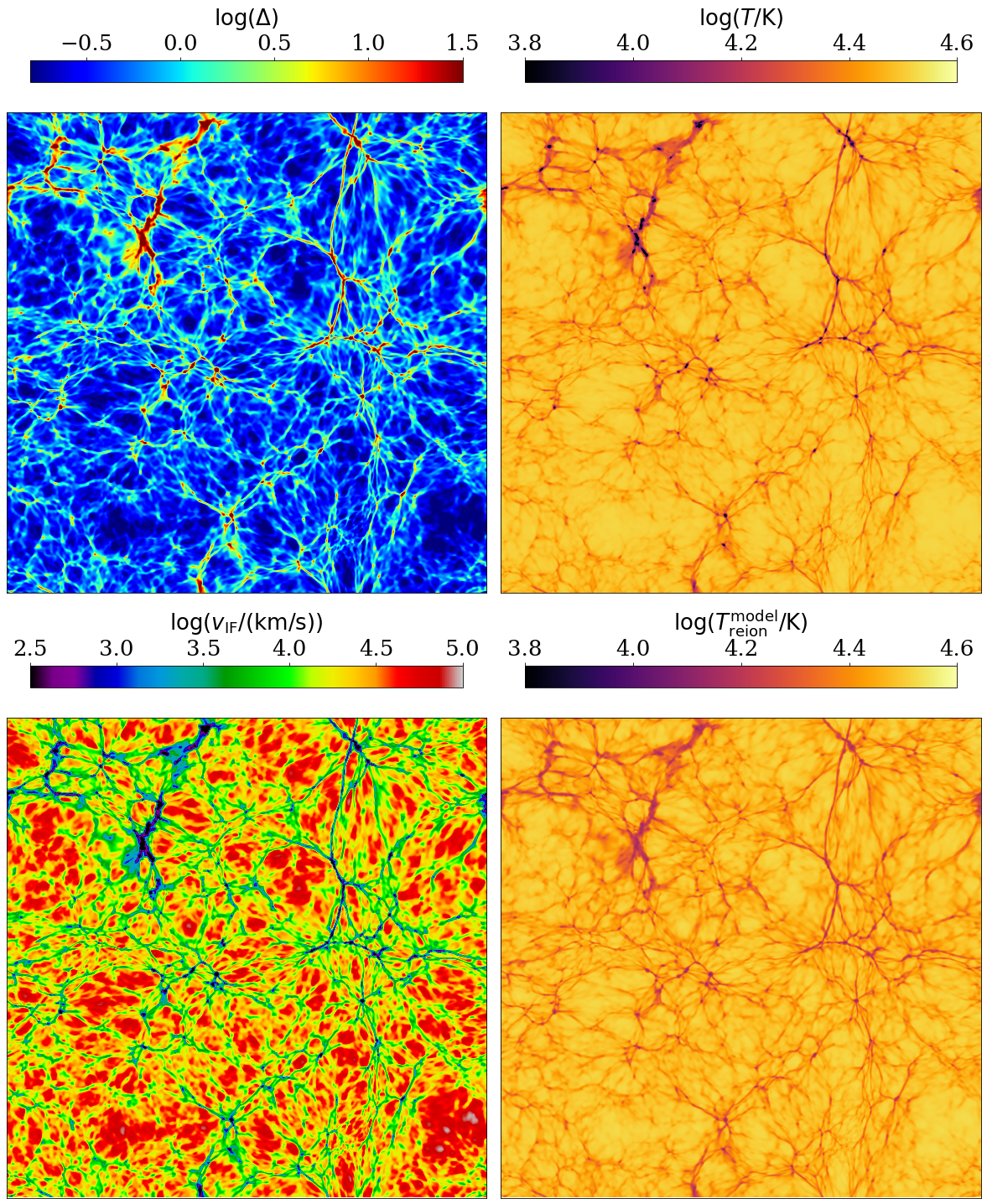}
    \caption{Agreement of $T_{\rm reion}$ in our fiducial Core run with the prediction from Ref.~\cite{DAloisio2019}.  {\bf Top Left}: A slice through the density field in our fiducial Core simulation at $z = 6.99$, immediately after reionization.  {\bf Top Right}: Gas temperature at the same redshift. I-fronts slow down in the dense filaments, causing $T_{\rm reion}$ to be lower there.  {\bf Bottom Left}: Map of $v_{\rm IF}$ estimated using Eq.~\ref{eq:vIF}.  {\bf Bottom Right}: predicted $T_{\rm reion}$ using the $v_{\rm IF}$ in the bottom left and the model of Ref.~\cite{DAloisio2019}.  We see very good agreement except in the densest filaments and halos, where self-shielding likely keeps the gas cooler than the Ref.~\cite{DAloisio2019} prediction.  }
    \label{fig:Treion_appendix_plot}
\end{figure}

One may be rightly concerned that assuming optically thick $T_{\rm reion}$ causes incorrect modeling collisional cooling inside the I-front (which is the dominant cooling channel during $T_{\rm reion}$ heating,~\citep{DAloisio2019,Zeng2021}), which may themselves cause inaccuracies in $T_{\rm reion}$.  We examine this possibility in Figure~\ref{fig:Treion_appendix_plot}.  The top left panel shows a slice through the density field shortly after reionization, and the top right shows $T_{\rm reion}$ (the same as the left panel of Figure~\ref{fig:treion_artifacts}, but shown in a log color scale).  In the bottom panels, we assess how accurate our $T_{\rm reion}$ field is using the $T_{\rm reion}$ model of Ref.~\cite{DAloisio2019}.  This model uses high-resolution 1D radiative transfer simulations to estimate $T_{\rm reion}$ as a function of $v_{\rm IF}$ and the spectral shape of the incident radiation.  The bottom-left panel shows an estimate of $v_{\rm IF}$ in the slice, where we have evaluated Eq.~\ref{eq:vIF} using $\Gamma_{-12}$ to estimate incident flux and using the local density at each point to calculate $n_{\rm H}$.  We then evaluate Eq. 3 from Ref.~\cite{DAloisio2019} at each cell using the estimated $v_{\rm IF}$ assuming a power law spectrum of $\alpha = 1.5$, consistent with the fiducial spectrum used in \textsc{saguaro}. 

We see by comparing the top-right and bottom-right panels that our simulations match well the prediction of the Ref.~\cite{DAloisio2019} $T_{\rm reion}$ model.  The average value of $T_{\rm reion}$ differs only by a few percent, and the small-scale fluctuations trace the density fluctuations in a similar way.  The only difference is that $T_{\rm reion}$ is lower in the densest filaments in the simulation than expected from the model.  This is reasonable, because the Ref.~\cite{DAloisio2019} cannot be applied to gas that remains partially or fully self-shielded.  Such gas, which is found at high densities in the simulation, should have $T_{\rm reion}$ lower than the Ref.~\cite{DAloisio2019} prediction. This result provides encouraging confirmation that $T_{\rm reion}$ and its spatial fluctuations are being treated accurately in \textsc{saguaro}.  

\section{Reduced speed of light}
\label{app:RSLA}

\begin{figure}
    \centering
    \includegraphics[scale=0.27]{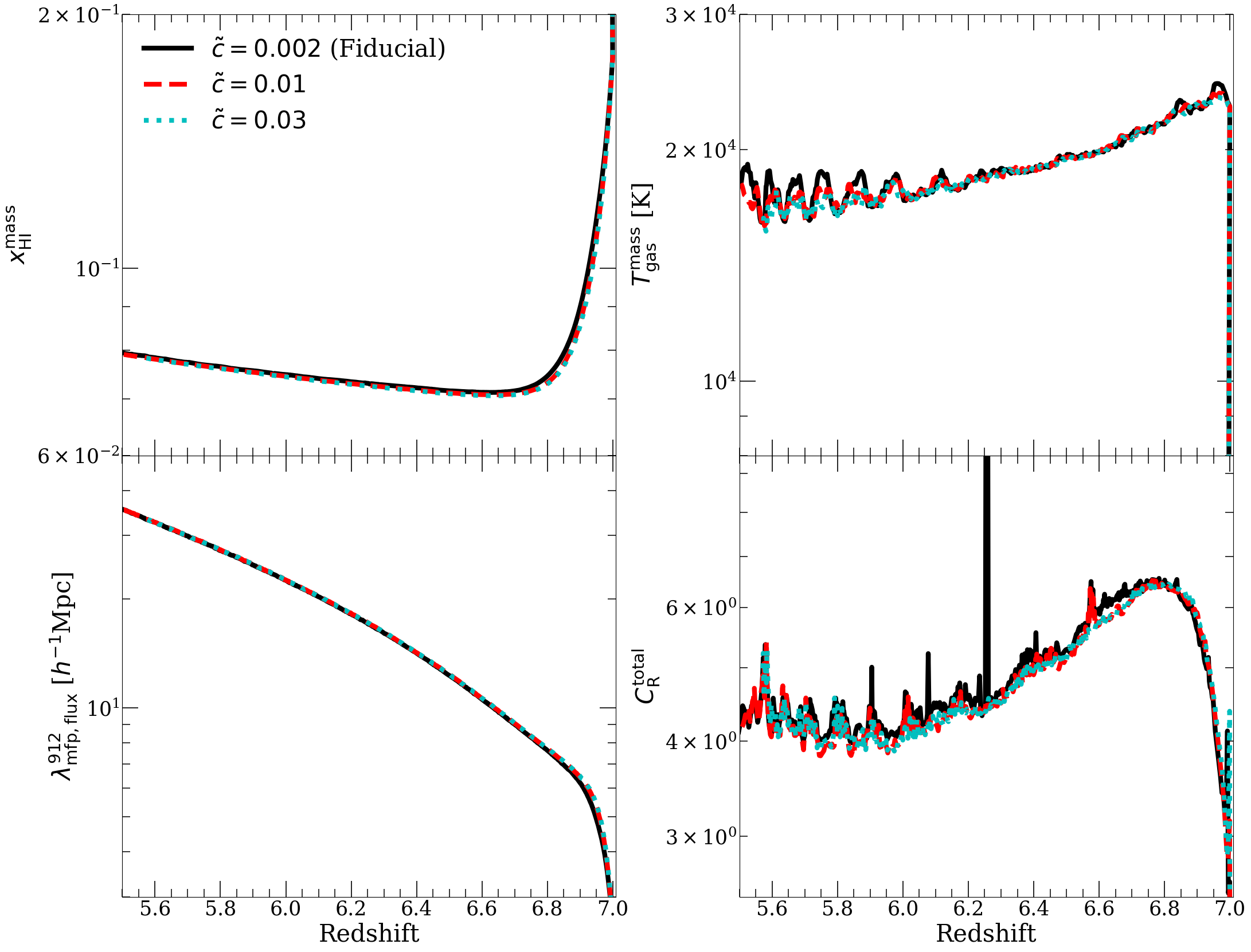}
    \caption{Test of how our choice of $\tilde{c} = 0.002$ after I-front passage affects the accuracy of our results.  {\bf Clockwise from top left:} the mass-averaged HI fraction, mass-averaged gas temperature, total recombination clumping factor, and ionizing photon mean free path for $\tilde{c} = 0.002$, $0.01$, and $0.03$.  }
    \label{fig:reduced_c}
\end{figure}

We next test whether our choice of the reduced speed of light after I-front passage, $\tilde{c} = 0.002$ (see \S\ref{subsubsec:RSLA}) is sufficiently large to accurately capture the coupling between the gas chemistry and radiation field.  In Figure~\ref{fig:reduced_c}, we show results from simulations with $\tilde{c} = 0.002$, $0.01$, and $0.03$.  Clockwise from top left, we plot the mass-averaged HI fraction, the mass-averaged gas temperature, the {\it total} recombination clumping factor, $C_{\rm R}^{\rm total}$ (defined analogously to $C_{\rm A,n_{\rm HI}}^{\rm total}$, but for recombinations), and the ionizing photon MFP at $912\text{\AA}$.  The second and third quantities are intrinsically noisy and sensitive to short-timescale processes occurring at the highest densities, so they should be most sensitive to the choice of $\tilde{c}$.  We see first that the average HI mass and MFP are insensitive to the choice of $\tilde{c}$, demonstrating convergence in self-shielding and opacity for our fiducial choice.  Scatter in $T_{\rm gas}^{\rm mass}$ and $C_{\rm R}^{\rm total}$ are both modestly reduced at higher values of $\tilde{c}$.  As we will show in the next appendix, these quantities are extremely sensitive to short-time-scale physics in a handful of highly over-dense cells ($\Delta > 10^3$), and the noise seen here is due to the time step being too large (in spite of our sub-cycling thermo-chemistry time step).  Since larger $\tilde{c}$ gives shorter $\Delta t_{\rm RT}$, the noise is reduced for higher $\tilde{c}$.  Fortunately, this handful of problematic cells has very little effect on the most quantities of interest, as we show below.  Thus, we conclude that \textsc{saguaro} is converged in the choice of $\tilde{c}$.  

\section{Sub-cycling time steps}
\label{app:subcycle}

\begin{figure}
    \centering
    \includegraphics[scale=0.27]{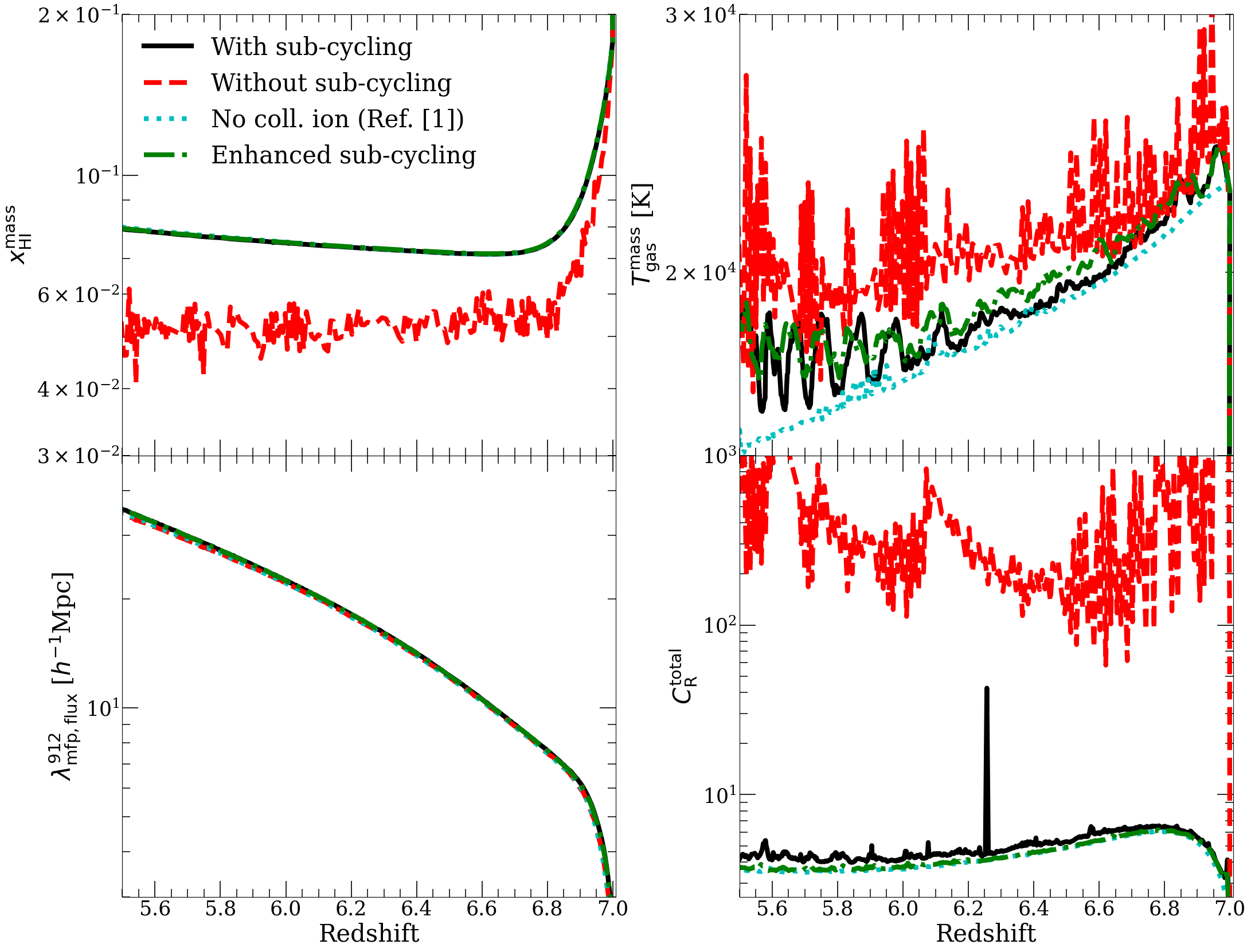}
    \caption{Effect of including the sub-cycling procedure described in \S\ref{subsubsec:timestep}, showing the same quantities as in Figure~\ref{fig:reduced_c}.  The black solid line shows our fiducial case, the red dashed turns off sub-cycling, the blue dotted line also turns off collisional ionizations (as in Ref.~\citep{DAloisio2020}), and the green dot-dashed curve includes a more stringent sub-cycling criterion than our fiducial choice.  }
    \label{fig:subcycle}
\end{figure}

Here, we quantify the effect of including a sub-cycling time step to solve for the evolution of chemistry and temperature in high-density cells, as described in \S\ref{subsubsec:timestep}.  Figure~\ref{fig:subcycle} shows the same quantities as in Figure~\ref{fig:reduced_c} for four cases - our fiducial Core run (black), a run with sub-cycling turned off (red), one with no sub-cycling and with collisional ionizations turned off (cyan, mimicking the Ref.~\cite{DAloisio2020} setup), and a test with enhanced sub-cycling.  For the last case, we replace $0.1 t_{\rm therm}$ with $0.01 t_{\rm therm}$ in Equation~\ref{eq:deltat_cycle}, and define $t_{\rm chem} \equiv \min(\min(n_{\rm HI},n_{\rm HII})/|dn_{\rm HI}/dt|)$.  The former makes $\Delta t_{\rm cycle}$ more sensitive to abrupt temperature changes, and the latter allows the code to better track changes in the ionized fraction in cells that are mostly ($> 90\%$) neutral.  

Turning off sub-cycling dramatically worsens the noise in both the temperature and recombination rate, and boosts the latter by over an order of magnitude. Most importantly, it also reduces the HI fraction by $20-30\%$.  We have checked and found that this is because gas at the highest densities, well above the self-shielding threshold, becomes spuriously ionized by collisional ionizations.  Turning  off collisional ionizations (as was done in Ref.~\cite{DAloisio2020}) brings things back into agreement with the fiducial case.  The run with enhanced sub-cycling has much less noise in $C_{\rm R}^{\rm total}$ and agrees better with the case without collisional ionizations.  This suggests that our fiducial choice of time-stepping parameter is not quite stringent enough to properly capture sub-percent level ionized fractions in very dense cells above the self-shielding threshold.  However, they are sufficient to avoid the catastrophic over-ionization of clumps seen in the model without sub-cycling, which is sufficient for our purposes.  Encouragingly, all the models except the red-dashed curve agree in $x_{\rm HI}^{\rm mass}$ and $\lambda_{\rm mfp}^{912}$, confirming that self-shielding is being treated well enough for our purposes.

\section{Convergence of the halo mass function}
\label{app:halos}

There are several numerical parameters in \textsc{Rockstar} that can affect the identification and masses of halos.  There are two parameters that control the adaptive friends-of-friends halo identification in \textsc{rockstar}: an initial linking length, LL, and a ``friends-of-friends fraction', or foff, which controls for the identification of particle sub-groups in the halo-finding algorithm.   There is also the force resolution, discussed in \S\ref{sec:dm_halos}.  Lastly, \textsc{rockstar} computes both friends-of-friends virial masses ($M_{\rm vir}$) and spherical over-density (SO) masses (e.g. $M_{200}^{c}$), the latter of which should provide a more apples-to-apples comparison to DM-HMFs calibrated using SO halo finders (such as that of Ref.~\cite{Trac2015}).  Below, we investigate the importance of these considerations for our DM-HMF calculations.  

\begin{figure}
    \centering
    \includegraphics[scale=0.21]{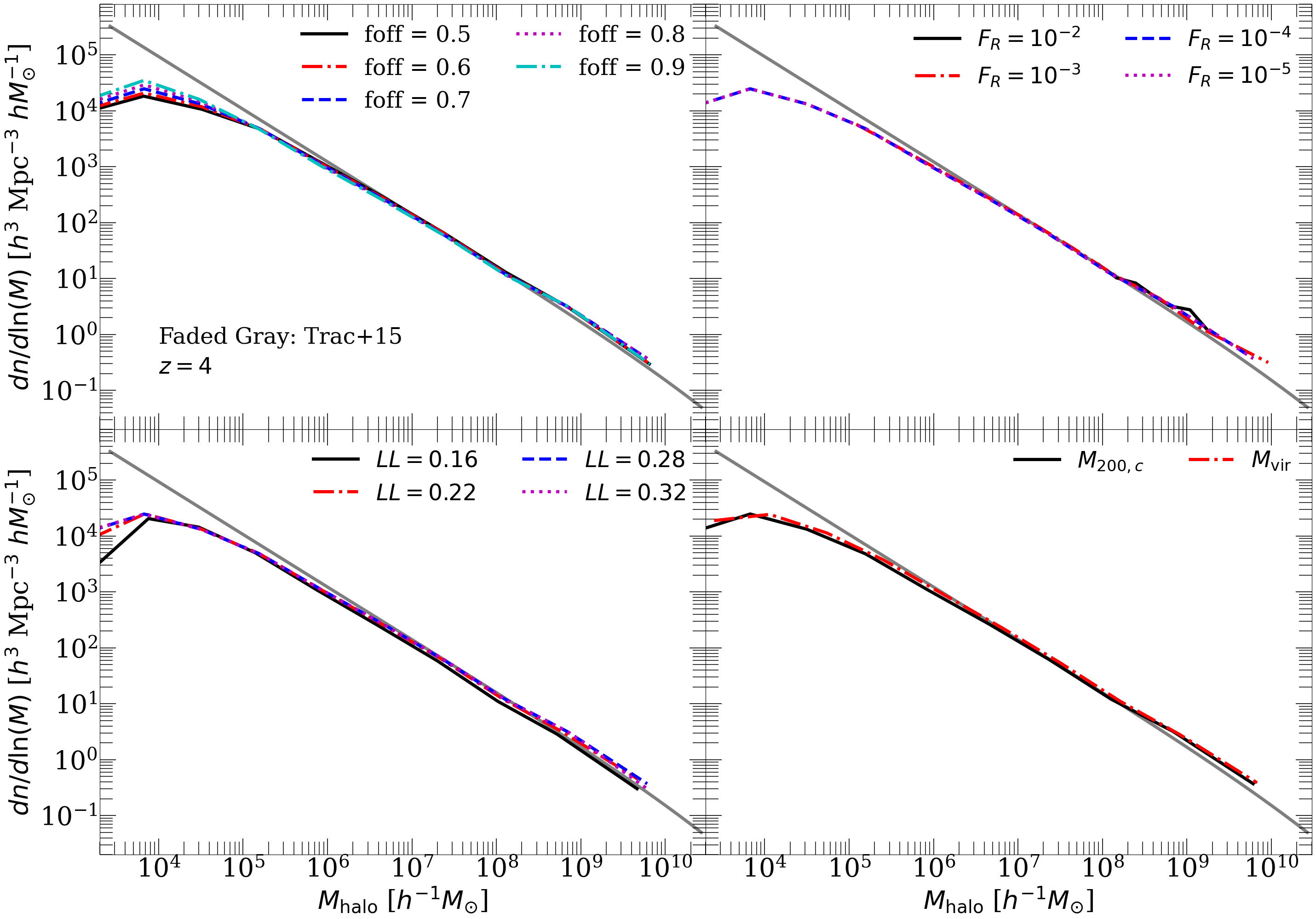}
    \caption{Convergence of the our \textsc{rockstar} halo mass function with respect to numerical parameters.  From top left going clockwise, we show the effect of varying the friends-of-friends fraction, force resolution, the effect of using friends-of-friends virial masses vs. spherical over-density masses, and initial linking length.  See text for details.  }
    \label{fig:HMF_test}
\end{figure}

Figure~\ref{fig:HMF_test} shows how each of these parameters affects our recovered DM-HMF in the fiducial Core run at $z = 4$.  In each panel, the gray solid curve shows the Ref.~\cite{Trac2015} prediction, as in Figures~\ref{fig:HMF}-\ref{fig:HMF_WDM_small}.  From top left going clockwise, the panels compare different values of the foff, force resolution ($F_R$), the effect of using $M_{\rm vir}$ vs. $M_{200}^{c}$, and the initial linking length.  We see that none of these parameters/definitions have a dramatic effect on the DM-HMF - in all cases, our results are within $30\%$ or less of the Ref.~\cite{Trac2015} prediction at $M_{\rm halo} > 10^{6}$ $h^{-1}M_{\odot}$.  In the top left, we see that the foff has little effect on the high-mass end, and mainly modulates the number of halos at $M_{\rm halo} \lesssim 10^5$ $h^{-1}M_{\odot}$, where the identified objects may not even be real halos.  The $F_R$ parameter (shown in units of $h^{-1}$Mpc in the legend) has a similar effect (top right).  The high-mass end of the DM-HMF is reasonably insensitive to it, and more halos appear at lower masses as $F_R$ increases.  Fortunately, values of $F_R = 10^{-4}$ and $10^{-5}$ $h^{-1}$Mpc (blue dashed and magenta dotted curves) are identical, justifying our assertion that the DM-HMF converges at small $F_R$, and that our DM-HMF at the high-mass end is converged in this parameter.

\begin{figure}
    \centering
    \includegraphics[scale=0.75]{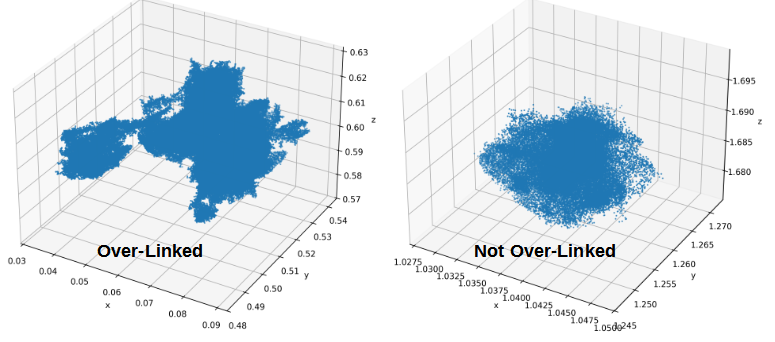}
    \caption{Examples of over-linked (left) and not over-linked (right) halos.  We see the former commonly in our tests if $LL = 0.28$ or $0.32$.  }
    \label{fig:overlinking}
\end{figure}

In the bottom-left panel, we see a modest effect at the high-mass end for the initial linking length, with high (low) linking lengths over (under)-predicting the DM-HMF by $\sim 20\%$.  We have carefully examined individual halos and find that values of $LL = 0.28$ and $0.32$ result in being identified that are clearly ``over-linked'' - that is, they are clearly a combination of more than one distinct halo.  For $LL \leq 0.22$, we are unable to identify any clearly over-linked halos.  Figure~\ref{fig:overlinking} shows examples of halos that are clearly over-linked (left) and not over-linked (right), the former of which we find are common in the $LL = 0.28$ and $0.32$ cases.  On the flip side, we see that for $LL = 0.16$, the DM-HMF begins to fall somewhat below the Ref.~\cite{Trac2015} prediction, suggesting that \textsc{rockstar} may be starting to miss particles near the edges of halos.  As such, we chose $LL = 0.2$ for our analysis, in order to avoid over-linking whilst preserving reasonable agreement with the high-mass end of the Ref.~\cite{Trac2015} at $z = 4$.  We note that values equal or close to $0.2$ have been commonly-used in the literature~\citep{Davis1985,White2001,More2011}.  Lastly, we see that using $M_{200}^{c}$ gives a slightly lower DM-HMF than $M_{\rm vir}$, but both are still reasonably consistent with the Ref.~\cite{Trac2015} result.  Since Ref.~\cite{Trac2015} used a spherical over-density halo finder, we show results using $M_{200}^{c}$ in the main text.  

\section{Testing the DC Mode Average}
\label{app:DCmode}

In this appendix, we have tested the DC mode averaging procedure discussed in \S\ref{subsec:code} and laid out in Appendix B of Ref.~\cite{DAloisio2020}.  
To do this, we have run $5$ simulations with $N = 256^3$ for our fiducial box size of $2$ $h^{-1}$Mpc for each of the $\delta/\sigma$ values reported in this work, and a single $8$ $h^{-1}$Mpc box ($64\times$ the volume of the smaller runs) at the same spatial resolution ($N = 1024^3$,  assuming fiducial $\Gamma_{-12}$ and $z_{\rm re}$ values of $0.3$ and $7$, respectively).  
The runs assume the fiducial setup  and physics of the CORE suite.  
This exercise can give us a sense of whether the effects of missing density fluctuations at and above the box scale are accounted for correctly, although it will under-estimate the effect of un-resolved small-scale structure that is computationally infeasible to capture in the $8$ $h^{-1}$Mpc run.  

\begin{figure*}[h!]
    \centering
    \includegraphics[width=\linewidth]{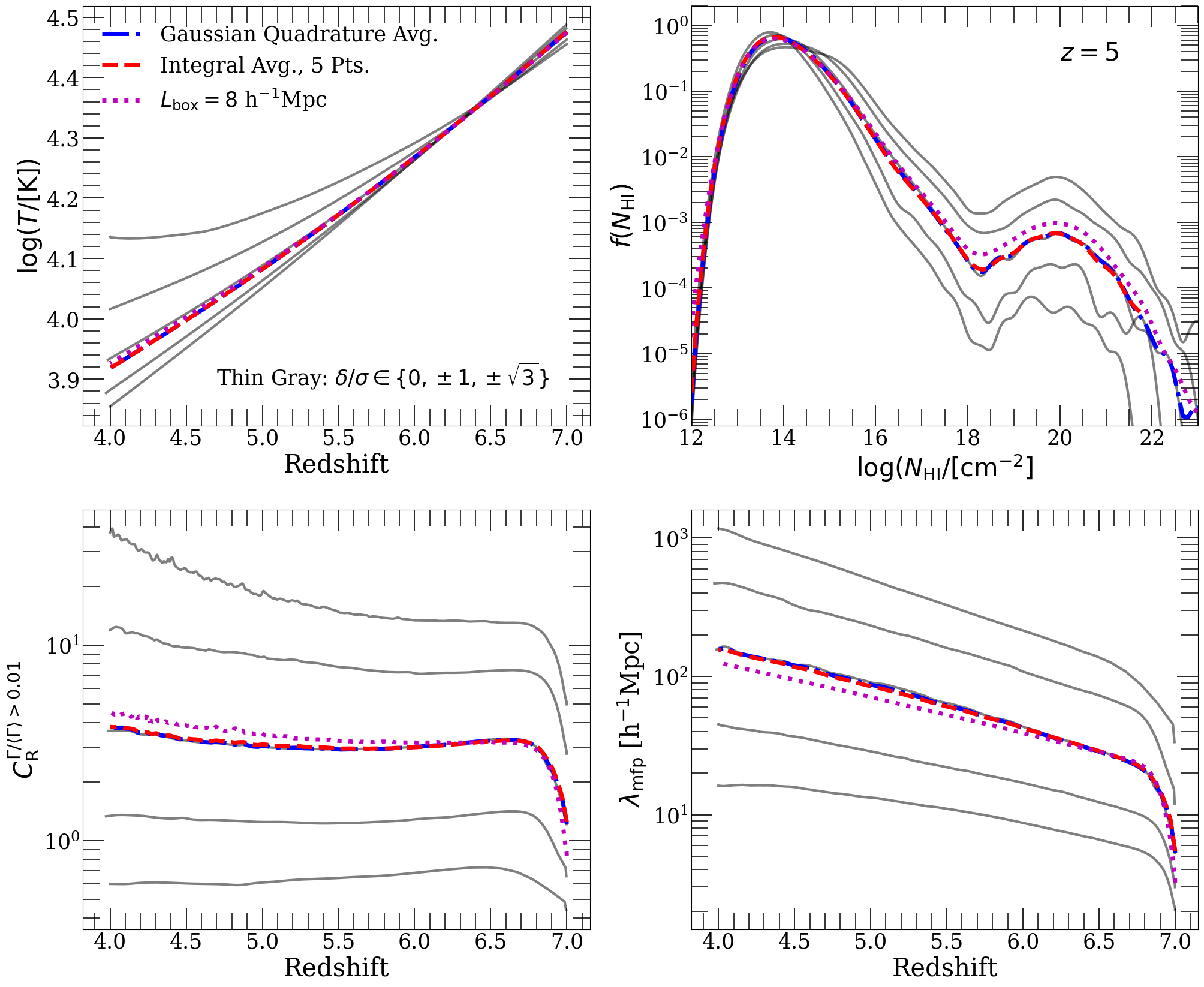}
    \caption{Test of our procedure for averaging over DC modes.  The thin gray lines show results for five $2$ $h^{-1}$Mpc runs with differing $\delta/\sigma$, and the blue and red lines denote two different methods of averaging them (see text).  The dotted magenta line denotes an $8$ $h^{-1}$Mpc run with $4^3 \times$ the volume of the smaller runs.  We find that both averaging procedures give results that are in excellent agreement with each other and the $\delta/\sigma = 0$ $2$ $h^{-1}$Mpc simulation.  The $8$ $h^{-1}$Mpc box is in reasonable agreement, but displays modest but non-negligible ($10-30\%$) differences for some quantities.  See text for details.  }
    \label{fig:DC_mode_test}
\end{figure*}

We show the results of this exercise in Figure~\ref{fig:DC_mode_test}.  
From top left going clockwise, the panels show the volume-weighted gas temperature, HI column density distribution $f(N_{\rm HI})$ at $z = 5$, the ``flux-based'' definition of the MFP (averaged over frequencies), and the recombination clumping factor averaged over cells with $\Gamma_{\rm HI} > 0.01\langle \Gamma_{\rm HI}\rangle$.  
The thin grey lines denote the individual values of each quantity for our five $2$ $h^{-1}$Mpc runs with different box-scale densities (indicated in top left panel).  
The blue dot-dashed line indicates the Gaussian quadrature average (which uses only $\delta/\sigma = 0$ and $\pm \sqrt{3}$ - Eq. B3 of Ref.~\cite{DAloisio2020}).  
The red dashed line denotes an evaluation of the average in Eq. B1 of Ref.~\cite{DAloisio2020}, assuming power-law interpolation in $\Delta_{\rm box}$ between all $5$ values of $\delta/\sigma$.  
The magenta dotted curve denotes the result of the $L_{\rm box} = 8$ $h^{-1}$Mpc simulation.

The Gaussian quadrature averaging procedure and the direct integration are in remarkable agreement with both each other and the $\delta/\sigma = 0$ $2$ $h^{-1}$Mpc simulation for all the quantities shown in Figure~\ref{fig:DC_mode_test}.  
This is consistent with our finding in Figure~\ref{fig:reion_histories_clumping} that the DC-mode averaged clumping factor is quite close to the mean density case, especially for our fiducial $\Gamma_{-12} = 0.3$.  
In the case of the volume-averaged temperature, the $L = 8$ $h^{-1}$Mpc box is also in few-percent level agreement with both averages.  
The other quantities show slightly more appreciable differences.  
The HI CDDF is higher by $\sim 30\%$ at the high-column end of the distribution, while the clumping factor and MFP differ by $\sim 10\%$ and $20\%$, respectively, by $z = 4$. 
These results suggest that our approach to estimating the globally-averaged IGM opacity and absorption rate from an ensemble of small-volume simulations may be biased high by $10-20\%$, especially at the lowest redshifts when non-linear effects near the box-scale become most important.  
We plan to investigate in future work whether this offset is expected to similarly affect the accuracy of sub-grid opacity models based on \textsc{saguaro}.  
Still, the broad agreement of our averaging method with the larger volume is encouraging, and suggests that our conclusions drawn in this work from this procedure are reasonably robust.

\end{document}